\documentclass{emulateapj}
\pdfoutput=1
\usepackage{longtable}
\usepackage{amsmath}
\usepackage{amssymb}
\usepackage{epsfig}
\usepackage{natbib}

\renewcommand\plotone[1]{%
\typeout{Plotone included the file #1}
\centering
\leavevmode
\includegraphics[clip=true,width=7.5cm]{#1}%
}%
\renewcommand\plottwo[2]{{%
\typeout{Plottwo included the files #1 #2}
\centering
\leavevmode
\columnwidth=.45\columnwidth
\includegraphics*[clip=true,width=7.5cm]{#1}%
\hfil
\includegraphics*[clip=true,width=7.5cm]{#2}%
}}%

\newcommand\plotfour[4]{{%
\begin{center}
\includegraphics*[clip=true,width=7cm]{#1}%
\hfill
\includegraphics*[clip=true,width=7cm]{#2}\\
\includegraphics*[clip=true,width=7cm]{#3}%
\hfill
\includegraphics*[clip=true,width=7cm]{#4}%
\end{center}
}}%

\newcommand\plottwolarge[2]{{%
\typeout{Plottwo included the files #1 #2}
\centering
\includegraphics*[clip=true,width=12cm]{#1}\\
\includegraphics*[clip=true,width=12cm]{#2}%
}}%

\def\iso#1#2{\mbox{${}^{#2}{\rm #1}$}}

\newcommand{\beq}{\begin{equation}}
\newcommand{\eeq}{\end{equation}}
\newcommand{\beqar}{\begin{eqnarray}}
\newcommand{\eeqar}{\end{eqnarray}}
\def\fun#1#2{\lower3.6pt\vbox{\baselineskip0pt\lineskip.9pt
  \ialign{$\mathsurround=0pt#1\hfil##\hfil$\crcr#2\crcr\sim\crcr}}}

\newcommand{\Kepler}{\textsc{Kepler}}

\begin{document}

\title{Dependence of X-Ray Burst Models on Nuclear Reaction Rates}

\author{R.
H. Cyburt\altaffilmark{1,2}, A. M. Amthor\altaffilmark{3},  A. Heger\altaffilmark{2,4,5,6}, E. Johnson\altaffilmark{7},  L. Keek\altaffilmark{1,2,7,10}, Z. Meisel\altaffilmark{2,8}, H. Schatz\altaffilmark{1,2,7}, and K. Smith\altaffilmark{2,9}
}

\altaffiltext{1}{National Superconducting Cyclotron Laboratory,
Michigan State University, East Lansing, MI 48824}

\altaffiltext{2}{Joint Institute for Nuclear Astrophysics (JINA),
http://www.jinaweb.org}

\altaffiltext{3}{Department of Physics and Astronomy,
Bucknell University, Lewisburg, PA 17837}

\altaffiltext{4}{Monash Centre for Astrophysics, School of Physics and Astronomy, Monash University, Victoria 3800 Australia}

\altaffiltext{5}{School of Physics \& Astronomy,
  University of Minnesota, Minneapolis, MN 55455, U.S.A.}

\altaffiltext{6}{Center for Nuclear Astrophysics, Department of
  Physics and Astronomy, Shanghai Jiao-Tong University, Shanghai
  200240, P. R. China.}

\altaffiltext{7}{Department of Physics and Astronomy, Michigan
State University, East Lansing, MI 48824}

\altaffiltext{8}{Department of Physics, University of Notre Dame,
  Notre Dame, IN 46556}

\altaffiltext{9}{Current Address: Oak Ridge National Laboratory}

\altaffiltext{10}{Current Address: CRESST and X-ray Astrophysics Laboratory NASA/GSFC, Greenbelt, MD 20771}

\begin{abstract}
X-ray bursts are thermonuclear flashes on the surface of accreting
neutron stars and reliable burst models are needed to interpret
observations in terms of properties of the neutron star and the binary system.  
We investigate the dependence of X-ray burst models on
uncertainties in (p,$\gamma$), ($\alpha$,$\gamma$), and ($\alpha$,p) nuclear reaction rates
using fully self-consistent burst models that account for the
feedbacks between changes in nuclear energy generation and changes in
astrophysical conditions. A two-step approach first identified
sensitive nuclear reaction rates in a single-zone model with 
ignition conditions chosen to match 
calculations with a
state-of-the-art 1D multi-zone model based on the {\Kepler} stellar
evolution code. All relevant reaction rates on neutron deficient 
isotopes up to mass 106 were individually varied by a factor of 100 up and down. 
Calculations of the 84
highest impact reaction rate changes were then repeated in the 1D multi-zone model. 
 We find a number of uncertain reaction rates that
affect predictions of light curves and burst ashes significantly. The
results provide insights into the nuclear processes that shape X-ray
burst observables and guidance for future nuclear physics work to
reduce nuclear uncertainties in X-ray burst models.
\end{abstract}

\maketitle

\section{Introduction}
\label{sect:introduction}

Type I X-ray bursts are the most frequently observed thermonuclear
explosions in nature
\citep{Schatz2006a,Strohmayer2006,Lewin93,Parikh2013}.  They take
place on the surface of accreting neutron stars in low-mass X-ray
binary systems for a certain range of mass transfer rates, generally
within two orders of magnitude below the Eddington mass accretion rate
($\dot{M}_\mathrm{Edd}\approx2\times10^{-8}\,\mathrm{M}_{\odot}/\textrm{yr}$;
\citealt{Woosley1976,Joss1977,Fujimoto1981,Strohmayer2006}).  Since
these events are not cataclysmic, bursts will repeat with recurrence
times ranging from hours to days.  The importance of understanding
X-ray bursts as a probe of neutron star properties and the underlying
physics has been discussed
extensively~\citep{Lewin93,Steiner2010,Zamfir2012,Ozel2013,Guver2013}.

The bursts are powered by the triple-$\alpha$ reaction, the
$\alpha$\textsl{p}-process and the rapid proton capture process
(\textsl{rp}-process; \citealt{Wallace1981,
  VanWormer1994,Schatz1998,Schatz2001,Fisker2008,Woosley2004a,Jose2010}).
These nuclear processes involve hundreds of nuclear species from
stable isotopes to the proton drip line. 
Models predicting burst light curves and the composition of the burst
ashes are therefore sensitive to a broad range of uncertain nuclear
structure and reaction properties near and beyond the current frontier
of experimental knowledge. This limits interpretation of the vast body
of observational data that has been accumulated \citep{Galloway2008},
for example in the MINBAR data base which will eventually contain
X-ray light curve data of over 5000 X-ray bursts
\citep{Galloway2010}. 

Models with reliable nuclear physics are needed 
 to validate the astrophysical model assumptions through comparison with 
 observations, to guide 
 future model developments towards an understanding of the full variety 
 of  observed bursting behavior, and to constrain parameters such as
distance, accretion rate, accreted composition, and neutron star
properties \citep{Heger2007,Galloway2004}. An example for the latter
are recent attempts to constrain the neutron star surface gravity by
matching a set of model bursts to observed light curves
\citep{Zamfir2012}.  

Whereas the X-ray light curve is the main direct observable of X-ray
bursts, accurate nuclear physics is also needed to predict the
composition of the burst ashes. Reliable calculations of this
composition are required to predict potential spectroscopic signatures
in the X-ray burst light curve from small amounts of ejected material
\citep{Weinberg2006} and to predict the composition of the neutron
star crust, which in mass accreting systems is made in part or
entirely out of X-ray burst ashes. Of particular importance is the
amount of $^{12}$C, which may reignite at greater depth
\citep{Cumming2001,Strohmayer2002,Cumming2006} and explain the origin
of occasionally observed superbursts \citep{Keek2008,Keek2011}. The
thermal transport properties of the neutron star crust, as well as the
amount of heating and cooling through weak interaction processes
\citep{Haensel2008,Gupta2007, Schatz2014} also depend sensitively on
composition.

The goal of this paper is to
identify important reaction rates that affect observables and nucleosynthesis.
Besides
nuclear reaction rates, burst models also depend on 
 $\beta$-decay rates and masses of neutron deficient nuclei.  
 All relevant $\beta$-decay rates have been
determined experimentally \citep{Schatz2006a}, and corrections due to
the high densities and temperatures reached in X-ray bursts are
predicted to be small in most cases \citep{FFN1982,Pruet2003}. However,
the validity of these corrections  remains to be evaluated in detail. The
majority of the relevant nuclear masses has been measured as well.
The impact of remaining mass uncertainties has been discussed
elsewhere \citep{Schatz2006,Parikh2009,Kankainen2012}. 
However, the rates for the vast
majority of nuclear reactions occurring in X-ray bursts have not been
determined experimentally and have large uncertainties. 
We therefore focus here on the sensitivity of X-ray burst 
models to nuclear reaction rates to provide guidance for future experimental efforts
aimed at reducing the 
remaining nuclear physics uncertainties in X-ray burst models. 

Comprehensive studies of the sensitivity of X-ray burst models to
nuclear reaction rates have so far been limited to single-zone
post-processing studies.  \citet{Parikh2008} varied 3500 nuclear
processes individually, and concurrently in a Monte Carlo approach,
using fixed temperature and density profiles from various previously
published X-ray burst models. An important result was the lack of
additional sensitivities due to correlations among reaction rates in
the Monte Carlo approach, justifying the single reaction rate
variation approach to nuclear sensitivity studies for X-ray bursts
that we also adopt here.

Post-processing studies with fixed temperature and density profiles,
however, are not adequate to study nuclear sensitivities in X-ray
bursts \citep{Thielemann2001} because the entire reaction sequence
contributes to the energy release driving the burst. Changes in
reaction rates inevitably lead to changes in energy production, thereby
to changes in the temperature and density evolution of the burst
sequence.  It is therefore essential to ensure consistency between the
nuclear physics input and the temperature and density evolution.
Owing to computational limitations, however, sensitivity studies in
fully self-consistent dynamic X-ray burst models have so far been
limited to variations of a few individual reactions or a few groups of
reactions.  \citet{Thielemann2001} varied the proton capture rates on
\iso{Si}{27}, \iso{S}{31}, \iso{Ar}{35}, and \iso{Ca}{38} together, rather than 
individually,  in a 1D
multi-zone X-ray burst model and demonstrated their strong influence
on the light curve, in contrast to a post processing study that found
these reactions to be unimportant
\citep{Iliadis1999}. \citet{Fisker2004a} varied the
\iso{S}{30}($\alpha$,p) and \iso{Ar}{34}($\alpha$,p) reaction rates
and demonstrated their impact on doubly peaked burst
profiles. \citet{Woosley2004a} varied groups of $\beta$-decay rates to
simulate the potential impact of proton capture rate uncertainties on
the burst light curve and found a very strong sensitivity. However, 
this only provides a crude estimate of how the overall processing 
speed of the rp-process may affect observables. It does not 
provide any insights into which proton capture rates may be important, 
nor whether proton capture rates are important at all. 
\citet{Fisker2006} determined a new lower limit of the \iso{O}{15}($\alpha$,$\gamma$) reaction
rate and showed that bursts disappear altogether when that reaction
rate is at that lower limit. \citet{Davids11}, however, did not find
such an effect using a different X-ray burst model. More recently
\citet{Keek2014} explored the impact of variations of the 
3$\alpha$, $^{15}$O($\alpha$,$\gamma$) and $^{18}$Ne($\alpha$,$\gamma$)
reaction rates on the transition from unstable to stable burning in 
a multizone X-ray burst model. 

Here we present a comprehensive study of the sensitivity of X-ray
burst models to 1931 nuclear reaction rates using for the first time
self-consistent X-ray burst models that account for the coupling
between nuclear energy generation and the astrophysical conditions
that determine the reaction sequences. Our approach is enabled by the
use of two models.  A calibrated, self-consistent one-zone model
\citep{Schatz2001} is used to explore variations of all reaction
rates. A subset of relevant rates is then investigated further using a
state-of-the-art 1D multi-zone burst model based on the
\textsc{Kepler} code \citep{Woosley2004a}.

\section{Method}

 In nature, the
characteristics of X-ray bursts and the relevant nuclear reaction
sequences vary between sources depending on accreted
composition and neutron star properties, and even vary, for a given source, with time
depending on the mass accretion rate.  
See \citet{Galloway2008} for a comprehensive overview of the range of observed burst 
properties and their dependence on source state, accretion rate, and accreted fuel composition.
\citet{Lampe2015} provide a broad exploration of the impact of variations in astrophysical
 parameters on burst characteristics for the KEPLER burst model also employed here.
We focus
here on systems that accrete a mix of hydrogen and helium at high
accretion rates, resulting in helium ignition in a mixed hydrogen and
helium environment (burning regime 3 in \citealt{Strohmayer2006}).
Bursts in this regime are powered by the $\alpha$\textsl{p}-process
and the \textsl{rp}-process and exhibit extended burst tails.  GS
1826-24 is the most studied source in this regime, and has served as a
benchmark for X-ray burst models \citep{Heger2007}. Other sources in this 
regime are the low $\alpha$ high $\tau$ sources in Fig 14. of \citet{Galloway2008} with
$\alpha$ being the ratio of energy released as persistent flux to energy released in
bursts, and $\tau$ being the burst time scale. 

To examine the effect of nuclear reactions we use the multi-zone
hydrodynamics code {\Kepler}~\citep{Woosley2004a}. The full 1-D
multi-zone model divides the envelope of the star into zones which
have independent isotopic abundances.  This allows the model to
simulate burning processes and energy transport in full 1-D.  In
addition extended sequences of bursts are calculated where ashes from
earlier bursts affect later ones, an effect known as compositional
``inertia''~\citep{Taam1980}. {\Kepler} has been shown to reproduce the
GS 1826-24 light curve reasonably well \citep{Heger2007}.

Full 1-D models require a large amount of computing time. We therefore
use a single-zone model of ~\citet{Schatz2001} to first identify the
most sensitive reactions and then determine the multi-zone model
sensitivity to this subset of reactions.  We have found that our
single zone model approximates well the light curve and final composition
calculated by {\Kepler} if the ignition conditions are chosen
properly. Single-zone models that calculate changes of temperature and
density conditions induced by changes in reaction rates
self-consistently are more physical than a post-processing
calculations, given that many reactions along the \textsl{rp}- and
$\alpha$\textsl{p}-processes affect the energy generation in X-ray
bursts.

Our approach consists of the following steps: (1) We select ignition
conditions for our one-zone model (pressure and composition) from
among the conditions in the various zones of our {\Kepler} model just
prior to the ignition of a typical burst (see below), such that the light curve and
final composition agree as closely as possible with the {\Kepler}
results. (2) We select a set of 1931 (p,$\gamma$), ($\alpha$,$\gamma$), and ($\alpha$,p) 
reactions along and near the
reaction sequence in the one-zone model that spans the nuclear
mass range of $A=1$--$106$. (3) We vary each reaction,
together with its inverse as determined by detailed balance, by a
factor of 100 up and down and we rank reactions by their impact on
the light curve and, in a separate ranking, on the composition. (4) We then
vary the most important reactions in {\Kepler}, together with other
reactions that have been identified as being important in X-ray bursts
in the literature.

The factor 100 for the single zone variations was chosen as a worst
case uncertainty to ensure no sensitivity is missed. This ensures we
err on the inclusive side for the selection of reactions varied in
subsequent multi-zone calculations.  Such large uncertainties can
occur for reactions that are dominated by a small number of resonances
\citep{Clement2004} and were also found, as extreme cases, in 
comparisons of rates calculated with the theoretical Hauser-Feshbach 
approach \citep{NON-SMOKER} (the majority of reactions investigated here) with experimental
data \citep{Parikh2013,Deibel2011}.
 Even with these large variations, the number of
reactions that cause significant light curve changes was still
manageable. For the reaction rate variations in {\Kepler} we reduce the
factor of variation in cases where the reaction rate error is likely
to be much smaller. We emphasize that the goal here is not to provide
a realistic uncertainty estimate but to identify important
reactions. Variations are therefore chosen to overestimate rather than
underestimate uncertainties to ensure identification of important
reactions. Modifications to the light curve and composition for different variation factors may be roughly estimated using our results, if, for example
a detailed analysis of reaction rate uncertainty is carried out in a
particular case. 

As baseline nuclear physics input, we adopt thermonuclear and ground
state weak rates and the corresponding nuclide properties (e.g. masses
and partition functions) from the snapshot library ReaclibV1.0 from
the JINA REACLIB Database~\citep{Cyburt2010}, the most up-to-date
library at the beginning of this study. We also adopt temperature and
density dependent weak rates from \citet{FFN1982} and \citet{Pruet2003}. 
Test calculations carried out with the rates from \citet{Oda1994} did not 
show any significant differences. 

\subsection{Multi-zone Model}

The 1D multi-zone X-ray burst model used in this study is based on the
hydrodynamics code {\Kepler} \citep{WZW78}, and is discussed in detail
in~\citet{Woosley2004a}. This model couples the energy generation of a
complete, adaptive nuclear reaction network of over 1300 isotopes to a
one-dimensional hydrodynamic simulation of the accretion on an
adaptive Lagrangian grid, nuclear burning, radiative and convective
energy transport, and mixing of the isotopic composition both during and between
bursts. A ``smooth'' accretion scheme as well as an updated implementation of electron
conduction are employed \citep{Keek2011}. The radial dimension of the accreted layer is resolved into
typically around 200 zones.  The model follows a sequence of bursts
taking into account steady state nuclear burning in-between bursts and
compositional inertia effects, where ashes from preceding bursts
partially mix with freshly accreted material and affect the nuclear
processes in subsequent bursts.

The specific accretion model used for these studies is model ZM
from~\citet{Woosley2004a}, with solar metallicity material (hydrogen
mass fraction $X=0.7048$, helium mass fraction $Y=0.2752$, and
$^{14}$N mass fraction 0.0200) accreted at a rate of
$1.75\,\times10^{-9}\,\mathrm{M}_{\odot}/\textrm{yr}$, roughly
$10\,\%$ of the Eddington mass accretion rate. This model represents a
typical burster powered by mixed hydrogen and helium burning producing
longer bursts with timescales well beyond $10\,$s, up to
minutes~\citep{Galloway2004}. A version of this model at slightly 
lower accretion rate produced the best agreement with observations to date \citep{Heger2007}.

The neutron star is taken to be 10~km in radius and to have a mass of
$1.4\,\mathrm{M}_\odot$.  Since this study involves no comparison to
observational data, the general relativistic correction from the
neutron star frame to an observer at infinity will be neglected \citep{Woosley2004a}. All
times are therefore given in the reference frame of the neutron star
surface.

The multi-zone X-ray burst model was first run for a long sequence of
bursts ($\sim50$) using the unmodified baseline reaction rate
database, Reaclib v1.0. The long sequence of bursts serves as a
baseline and ensures enough burst statistics to characterize steady
state burst properties, including burst-to-burst variations that
affect the identification of sensitivities to reaction rate changes.
The models with varied reaction rates were run for shorter times for a
set number of simulation timesteps.  This limits the computational
time needed for each run.  Because the time step size in {\Kepler} is
dynamically adjusted to optimally resolve the time-dependent behavior, 
the total simulation time varies among the
multi-zone simulations.  The multi-zone model calculations with
individual reaction rate variations generally produced about 14
bursts, but in all cases, a continuous sequence of at least 12 bursts
was simulated. This is sufficient to ensure convergence into a steady
state bursting behavior.

\begin{figure}
\plotone{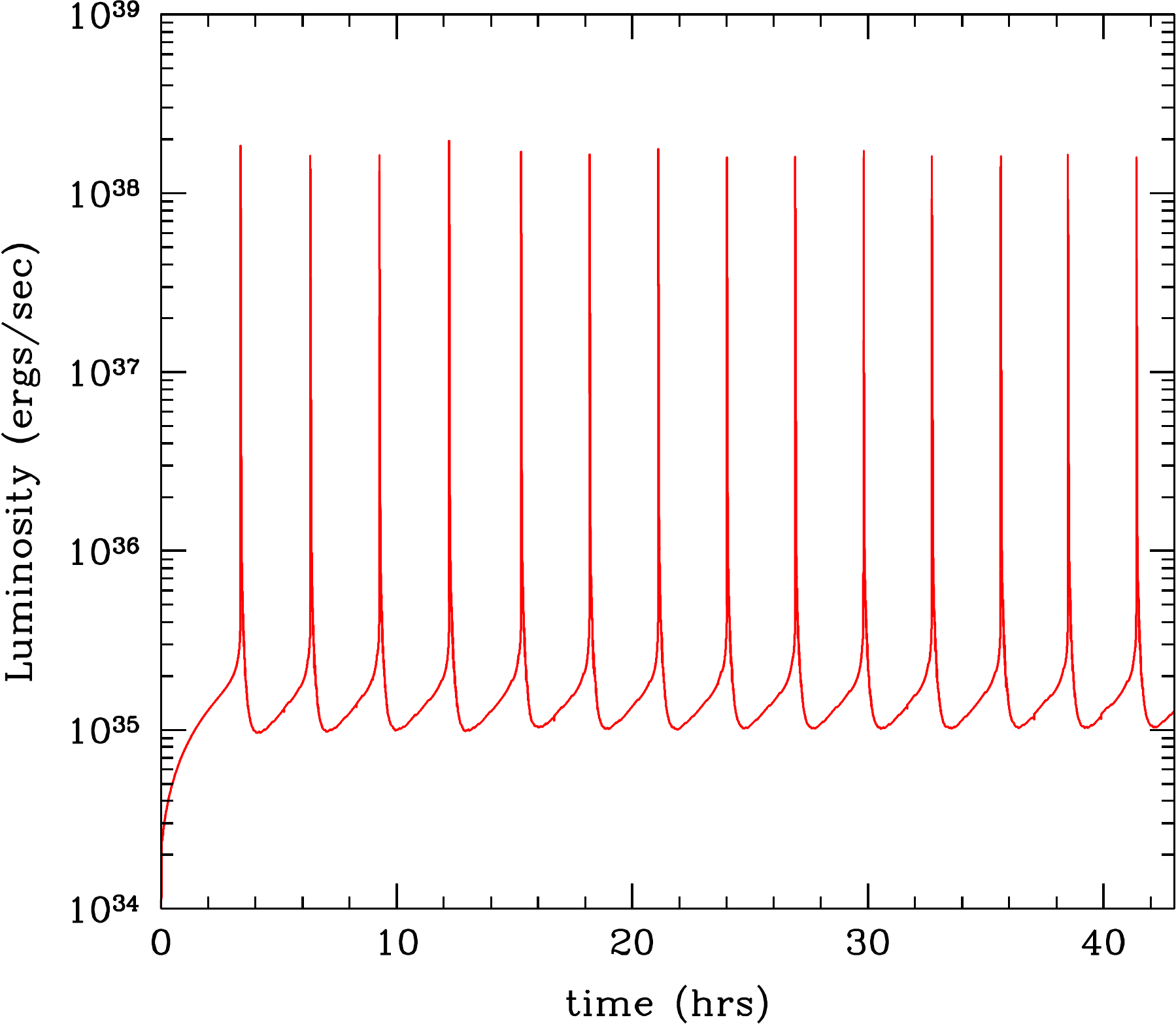}
\caption{\label{FigMZ_LCsequence} Luminosity as a function of time for
  the first 15 bursts of a burst sequence calculated with the multi-zone model.
  Note that this is only the luminosity from the neutron star itself,
  after accretion of the material; an observed XRB light curve would
  have the accretion luminosity of 2$\times 10^{37}$ erg/s added.}
\end{figure}

The burst lightcurves are shown in Fig.~\ref{FigMZ_LCsequence} and
Fig.~\ref{FigMZSZ_LC}.  The first burst is special, because it occurs
atop an inert neutron star substrate (assumed as iron for the purpose
of heat conduction) though in this model it turns out to be quite 
similar to the following bursts.  Already the second burst is very similar to the
remaining bursts and there is no evidence for systematic variation
beyond the second burst indicating that reasonable steady state
equilibrium is achieved beginning with the third burst.  At this point
bursts occur rather regularly, with a recurrence time of
$t_\mathrm{rec}=175\pm3$ minutes, peak luminosity of
$L_\mathrm{peak}=(1.7\pm0.1)\times10^{38}\,$ergs$\,$s$^{-1}$, and a
total energy released during the burst,
$E_\mathrm{tot}=(6.5\pm0.1)\times 10^{39}\,$ergs, (all numbers given
in the local neutron star surface reference frame).  The quoted
uncertainties are standard deviations and stem from the remaining
intrinsic burst-to-burst variations.  These variations do not decrease
with the number of bursts and are a feature of the steady state
behavior of the model. In steady state, bursts have a rise time of $5.3\pm0.1\,$s
from $10\,\%$ to
$90\,\%$  of $L_\mathrm{peak}$, and a duration of
$\tau=E_\mathrm{tot}/L_\mathrm{peak}=39\pm3\,$s.  The $10\,\%$ (rise) to
$10\,\%$ (decline) luminosity burst duration is $73\pm5\,$s.

In Fig.~\ref{FigMZSZ_LC} the burst luminosity profiles are shown,
shifted in time to align the peak luminosities of all bursts.
Burst-to-burst variations not only affect recurrence time and peak
luminosity, but also the shape of the light curve.

This study requires direct comparison of two light curve sequences
calculated with different nuclear physics. The burst-to-burst
variations need to be taken into account for this comparison.  In
order to define a light curve for comparison, we take the ensemble of
bursts starting with the third burst, shift each burst to align the
peak luminosity in time, and resample all bursts on a common time grid
using linear interpolation. The surface resolution of our model is finer 
than what was used by \citet{Heger2007}, with the mass of the outer zone
being up to 100 times lower. At times when the burst is brightest, 
an individual zone near the surface may get a ``kick'', and its behavior
differs briefly from the rest of the outer zones. To prevent this from influencing
the simulated light curve, we average the luminosity of
the outer-most 9 zones and
smooth over a scale of one second to remove any numerical noise.  At
each point in time we consider the variations in luminosity from all
remaining bursts, and determine the average luminosity and its error.
This average light curve with error band is used for our sensitivity
study.  Using an average is justified because observationally
precise burst
light curves can also be obtained by averaging long sequences of bursts
\citep{Galloway2004}.

\begin{figure}
\plotone{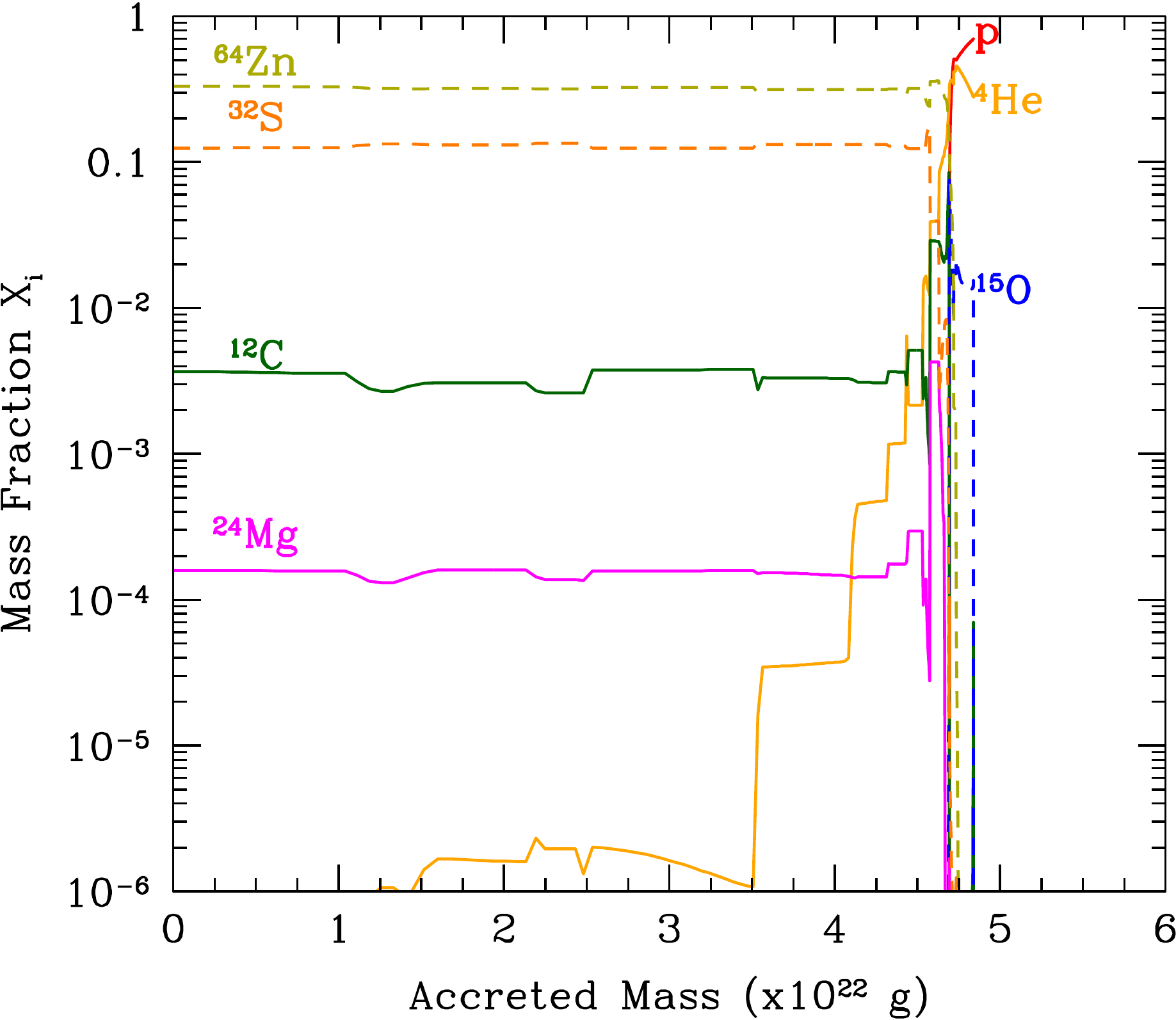}
\caption{\label{FigMZ_CompDepth}Average mass fraction of various
  nuclides as a function of accreted mass in the multi-zone
  model. Zero accreted mass marks the base of the accreted
  material. $4.8\times10^{22}\,$g corresponds to the surface and
  reflects the total accreted mass at the point in time shown.}
\end{figure}

We also investigate the sensitivity of the steady-state burst ashes to
nuclear reaction rate variations. This steady-state burst ashes of a
burst sequence is determined in the following way: The deepest zones
at the end of our multi-burst simulation sequence are special as they
represent the ashes of the first bursts before a steady-state burst
behavior was achieved.  Similarly to the procedure adopted for the light curves, we therefore remove
ashes from the first two bursts from our analysis to prevent anomalous
zones from contributing.  Too-shallow zones also have to be excluded
as they will continue to be modified by subsequent bursts and
therefore do not represent the steady-state ashes. Clearly this
includes zones that still contain hydrogen.  Deep unburned helium,
however, can also be burned during heating from bursts occurring in
shallower regions \citep{Woosley2004a}. This can be seen in
Fig.~\ref{FigMZ_CompDepth} where the helium mass fraction successively
decreases with increasing depth. As the helium mass fraction becomes
lower, the modifications in the ashes induced by burning of residual
helium become less important.  We therefore exclude shallow zones with
helium mass fractions above $\sim3.0\times 10^{-5}$ from the
calculation of our final composition.  For a typical burst model used
with the reaction rate variations, these constraints result in an
averaging of the final composition over a depth range of
$M_\mathrm{acc}\sim3-6\times 10^{21}\,$g (see
Fig~\ref{FigMZ_CompDepth}). We chose the time (rounded to nearest
100$^{th}$ time-step, as one in every 100 steps is stored) at which the light curve is minimum after the
$12^{th}$ burst as our point for determining this composition.

\begin{figure}
\plotone{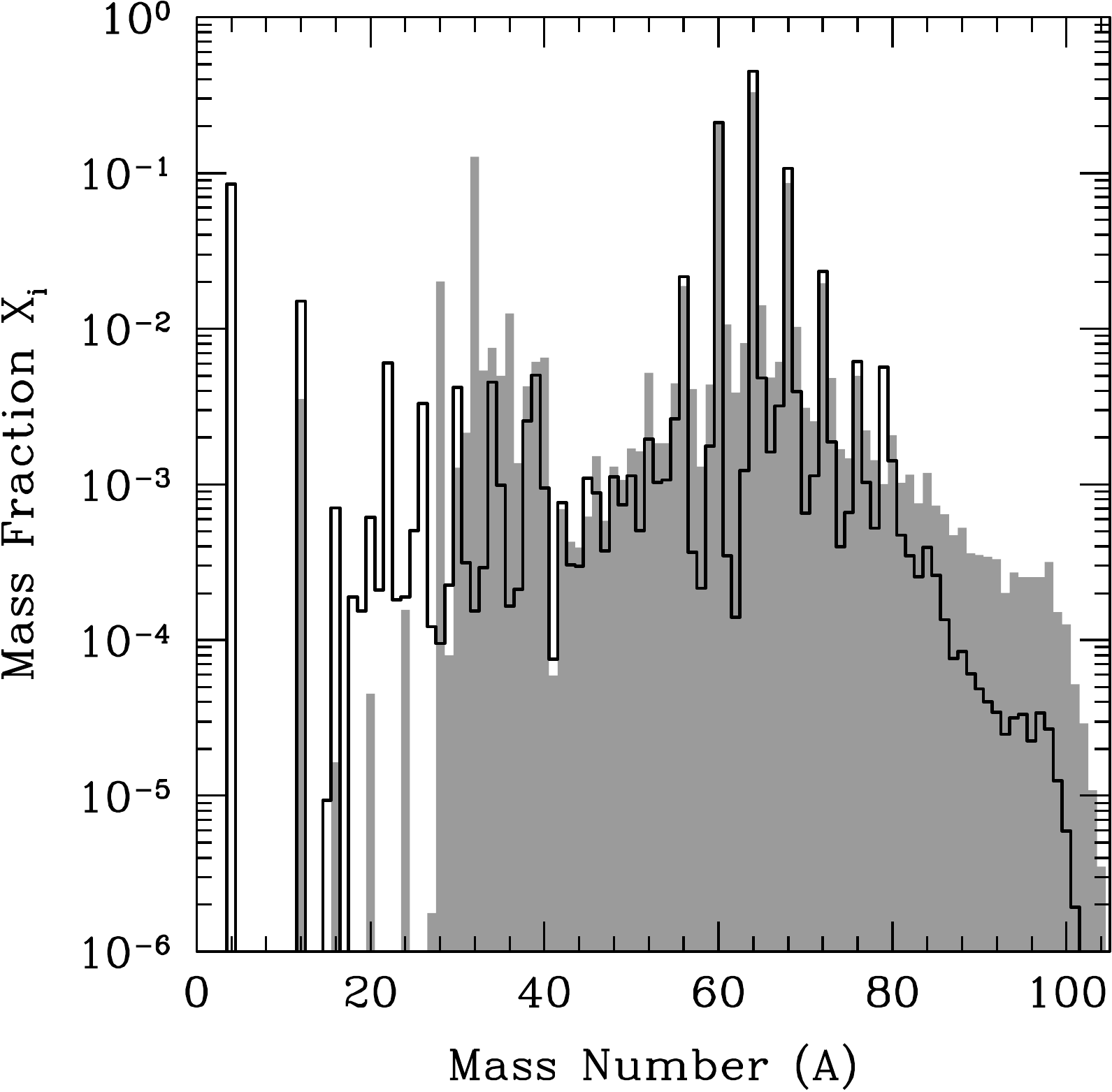}
\caption{\label{FigMZSZ_CompA}Steady state composition summed by mass
  number produced by a burst sequence in the multi-zone burst model
  (grey shaded histogram) and by the single-zone burst model (black
  line).}
\end{figure}

Fig.~\ref{FigMZSZ_CompA} shows the resulting isobaric ash composition.
The abundance peak around $A=40$ stems from the impedance of the
nuclear reaction sequence caused by the $Z=20$ shell closure at the
\iso{Ca}{40} waiting point, whereas the abundance peaks around
$A=60-64$ are produced by the \iso{Zn}{60}, \iso{Ge}{64}, and
\iso{Se}{68} \textsl{rp}-process waiting points, which have
particularly long half-lives despite being located at the proton drip
line \citep{Schatz1998}.  The NiCu and ZnGa cycles further enhance
abundances in this mass region \citep{VanWormer1994}. The $\alpha$-chain isotopes
($A=12,16,20,24,28,32$) are further enhanced by late helium burning
once hydrogen is consumed. The time and zone integrated reaction flow
during a typical X-ray burst is shown in Fig.~\ref{FigMZ_path} to
illustrate the type of reactions that occur.

\subsection{Single Zone model}

The one-zone model ONEZONE is similar to the model of
\citet{Koike1999} and has been used previously to estimate the
reaction sequence of the \textsl{rp}-process \citep{Schatz2001}. It
evolves thermodynamic conditions and composition in a single zone,
neglecting gradients in temperature, density, and composition, as well
as radiative transport and convection.  ONEZONE assumes nuclear
burning at constant pressure $P$. This is justified, as mass accretion
during the burst is negligible and only little material is ejected. As
the burning happens in a thin layer on the surface, the influence of
the neutron star is fully characterized through the surface gravity
$g$ (neglecting magnetic fields and rotation). The evolution of the
temperature starting at burst ignition is calculated for a time step
$dt$ using
 $dT=c_P^{-1} (\epsilon_{\rm nuc} + \epsilon_{\rm cool} + \epsilon_\nu ) dt$ with the specific heat
capacity at constant pressure $c_{\rm P}$, the positive specific nuclear energy
generation rate $\epsilon_{\rm nuc}$, the negative specific surface
cooling rate $\epsilon_{\rm cool}$, and the negative specific neutrino
loss rate $\epsilon_{\nu}$.
\begin{equation}
\epsilon_{\rm nuc}=\frac{\sum_i dY_i \Delta_i }{dt}
\end{equation}is calculated from the  abundance changes
$dY$ using atomic mass excesses $\Delta$, assuming positrons emitted
in $\beta^+$ decays are annihilated instantly.  For X-ray burst
conditions, $\epsilon_{\nu}$ is entirely due to neutrinos
emitted in $\beta^+$ decays. The respective energy losses are taken
from \citet{FFN1982,Pruet2003}. $\epsilon_{\rm cool}$ is
approximated as \citep{Fujimoto1981}
\begin{equation}
\epsilon_{\rm cool} =\frac{acT^4}{3\kappa y^2}
\end{equation}
where the opacity $\kappa$ is calculated according to
\citet{Schatz1999}. $a$ is the radiation constant, $c$ the speed of
light, and $y=P/g$ the column density.  The mass density change $d
\rho$ during a time step $dt$ is calculated from $P$ and $dT$ using an
equation of state that includes the pressure of the radiation and of a
partially degenerate electron gas \citep{Paczynski1983}.  The model is
coupled to an implicitly solved nuclear reaction network including 688
nuclei from hydrogen to tellurium, which for a given $T$ and $\rho$
calculates $dY$. We used the same reaction rates as in the multi-zone model. 

ONEZONE needs to be coupled with a model predicting the ignition
conditions. In the past, a full 1D steady-state hydrogen and helium
atmosphere model has been used that evolves the composition as a
function of depth during the accretion process until a thermal
instability criterion is fulfilled \citep{Cumming2000}. Pressure and
composition at that location are then used as initial conditions for
ONEZONE. The ignition temperature is not a critical parameter due to
the steep temperature rise at the beginning of an X-ray burst. The
initial temperature is chosen to be just high enough to initiate a
thermonuclear runaway in ONEZONE on a timescale that is sufficiently
short to not modify the composition further by any steady state
burning.  The results for burst timescale, major reaction sequence, and
final composition have been shown to agree reasonably well with the
first burst calculated with {\Kepler} \citep{Woosley2004a} and other
multi-zone models \citep{Fisker2008, Jose2010}. The details of the
light curve cannot be predicted accurately, as expected for a one-zone
model without radiation transport. As the energy generation and the reaction
network are coupled self-consistently, however, ONEZONE can be used to
explore the sensitivities of energy generation and the X-ray burst light curve
to nuclear reaction rate changes, provided realistic ignition
conditions are chosen.

Our goal here is for the single-zone model to most closely resemble a
typical burst of the multi-zone model calculation, not the first
burst. We therefore extract the ignition conditions for ONEZONE from
the baseline calculation of our multi-zone burst model shortly before
a typical burst is ignited. The ignition time is defined by
significant breakout from the CNO cycle via the
$^{15}$O($\alpha$,$\gamma$) reaction, and is identified from a peak in
the \iso{O}{15} abundance. The ONEZONE model ignition conditions are
taken at that time from the zone for which the ONEZONE results for
light curve and final composition most closely resemble the multi-zone
model results for total luminosity and steady state composition.  The chosen ignition conditions were a temperature of
$0.386\,$GK, a pressure of $1.73\times 10^{22}\,$erg$\,$cm$^{-3}$, and
hydrogen and helium mass fractions of $0.51$ and $0.39$, respectively.

\begin{figure}
\plotone{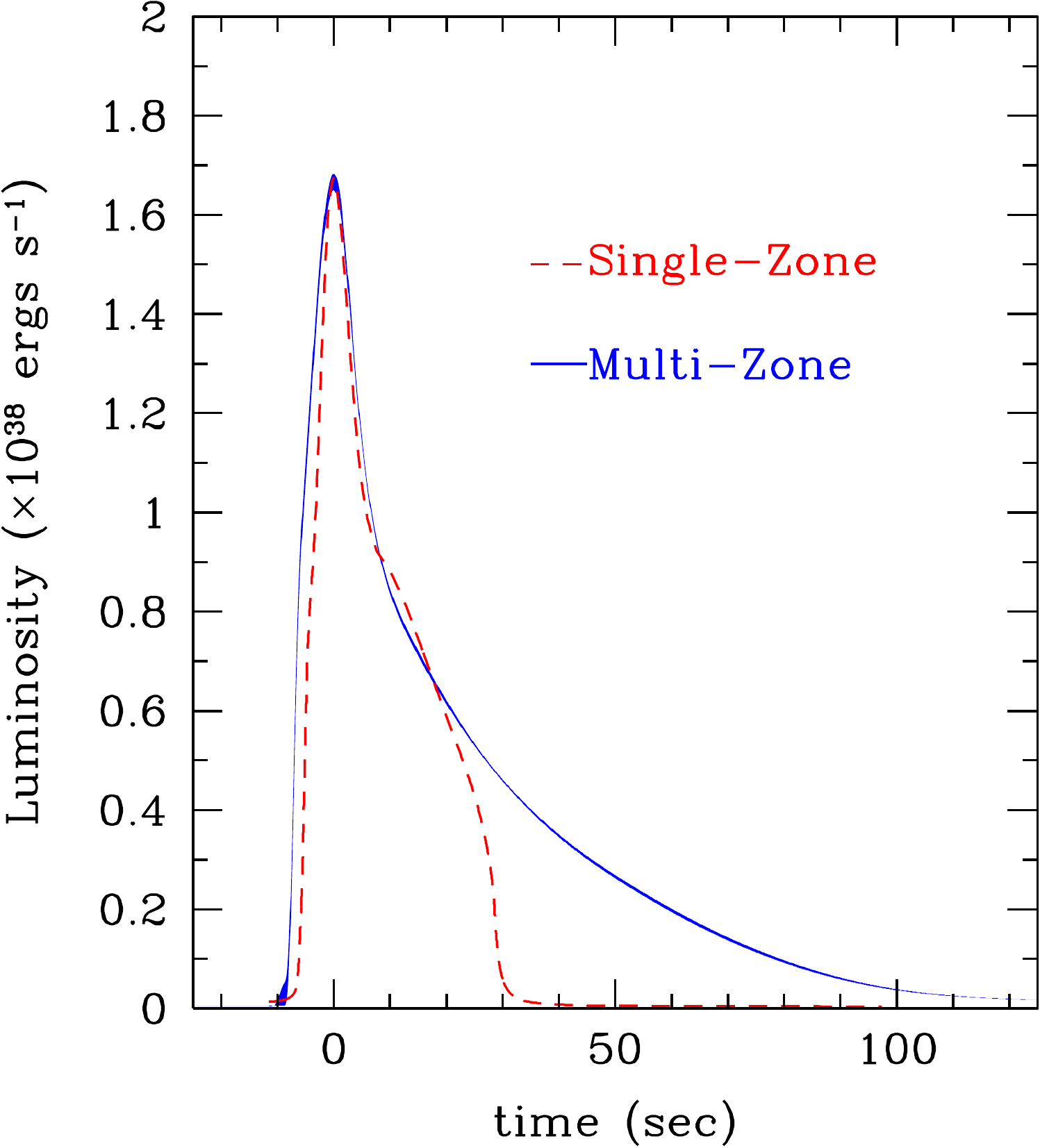}
\caption{\label{FigMZSZ_LC}X-ray burst light curves predicted by
  single-zone (red dashed) and multi-zone (blue solid) models. The line width for the multi-zone model light curve indicates the 1$\sigma$
  error of the average light curve.}
\end{figure}

With these ignition conditions, the peak temperature in ONEZONE
agrees with the peak temperature in the hottest zone of the multi-zone model (1.2~GK). 
Figs.~\ref{FigMZSZ_CompA} and \ref{FigMZSZ_LC} show that ONEZONE
does predict the burst light curve and ashes composition reasonably well. 
The main difference in the light curve is the sudden drop
in luminosity defining the end of the burst in the single-zone
model.  This is mainly the result of the absence of radiation transport
modeling in the one-zone model.  As far as the burst ashes are
concerned, ONEZONE predicts quite well the main features of the
composition, including the $A \approx 40$ peak and subsequent
abundance drop, and the main components of the ashes at $A=56$, $60$,
$64$, $68$, $72$, and $76$.  The main difference is the additional
helium burning in the multi-zone model, which is induced by heating of
the ashes in subsequent bursts and is not included in the one-zone
approach. Compared to the single-zone results, this helium burning
leads to the depletion of helium and other $A<24$ nuclei, which serve
as seeds for $\alpha$-capture reactions, and the build up of heavier
$\alpha$-chain nuclides at $A=28$ and $32$. A less important
difference is the enhancement in the multi-zone calculation of the
only weakly produced mass chains with $A>60$ owing to the broader
range of \textsl{rp}-process freeze out conditions in multiple zones.
The ONEZONE reaction flow shown in Fig.~\ref{FigSZ_path} is indeed
similar to the reaction flow in the multi-zone model during an X-ray
bursts (see Fig.~\ref{FigMZ_path}).

Based on the similarities between ONEZONE and the multi-zone model 
we expect that reaction rate sensitivities in the $\alpha$p- and rp-process
during a burst can be reasonably well approximated in ONEZONE, with the 
caveat that lower temperature and density zones contribute to some 
extent in the multi-zone model. On the other hand, reaction rate 
sensitivities related to inter-burst burning, for example in the CNO region, 
and reaction rate sensitivities related to deep helium burning triggered by subsequent bursts, such as 
sensitivity to $\alpha$-capture reactions, are not expected to be present in ONEZONE
as those burning regimes are not included in the single-zone simulation.  

\section{Results}

In order to quantify the impact of a particular reaction rate variation
on X-ray burst model light curve predictions, we define \beq
\label{eqn:LCrank}
M_{LC}^{(i)}=\int | \langle L_{i}(t) \rangle - \langle L_{0}(t)
\rangle |dt \eeq Here, $\langle L_i(t) \rangle $ is the lightcurve of
each variation $i$, and $\langle
L_0(t) \rangle $ is the luminosity of the baseline model. For the multi-zone model
$\langle L_i(t) \rangle $ and $\langle
L_0(t) \rangle $ are averaged over multiple bursts in the burst sequence.  In
addition we provide a qualitative classification of the degree of
light curve variation caused by the change of a particular reaction
rate.  Category 1 changes are the largest changes that likely would by themselves 
affect the interpretation of observational
data. Category 2 changes are smaller changes, that may nevertheless
be observable given, for example, observational error bars that can be 
achieved by averaging light curves \citep{Galloway2004}. Category 3 changes are insignificant, but confirm that
the varied rate is part of the reaction path and may become
significant for larger variations.

The composition of the burst ashes is summed by mass number as
electron captures will change the isotopic composition with increasing
depth within a constant mass number chain. Summing by mass number
allows averaging over a greater depth range as it removes the effects
of changes in the composition due to the decay of long-lived radioactive
isotopes and weak interactions in the ocean and crust of the neutron
star, which are not of interest here.

\subsection{Single-zone model results}
\label{sect:SZsens}

A total of 1,931 (p,$\gamma$), ($\alpha$,p), and ($\alpha,\gamma$)
reactions, together with their respective inverse reactions, were
varied individually in the single-zone burst model by factors of 100
up and down.  The reactions selected were those at or near the time
integrated reaction flow sequence of the baseline single zone model
(see Fig.~\ref{FigSZ_path}).

\begin{figure*}
\plottwo{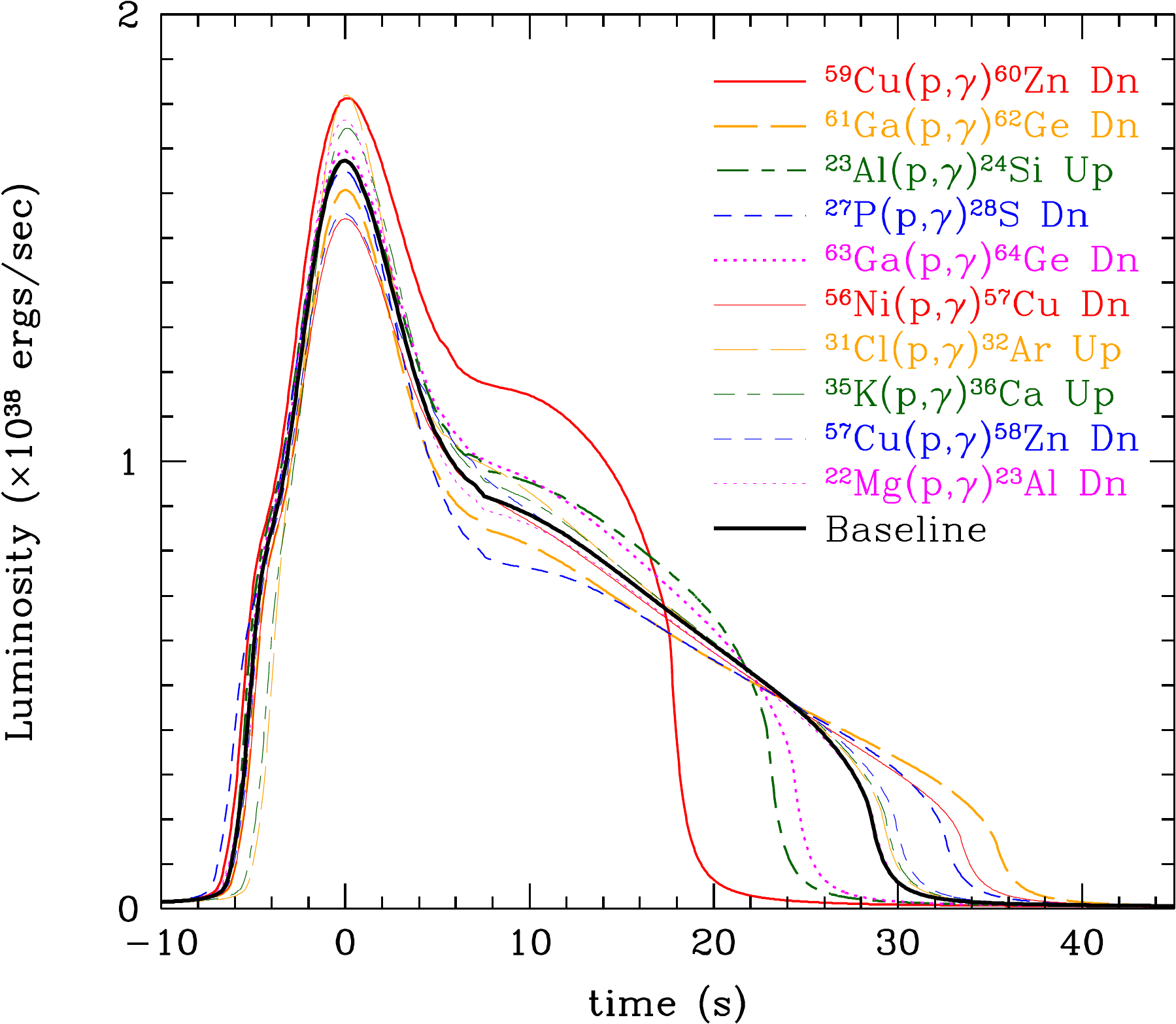}{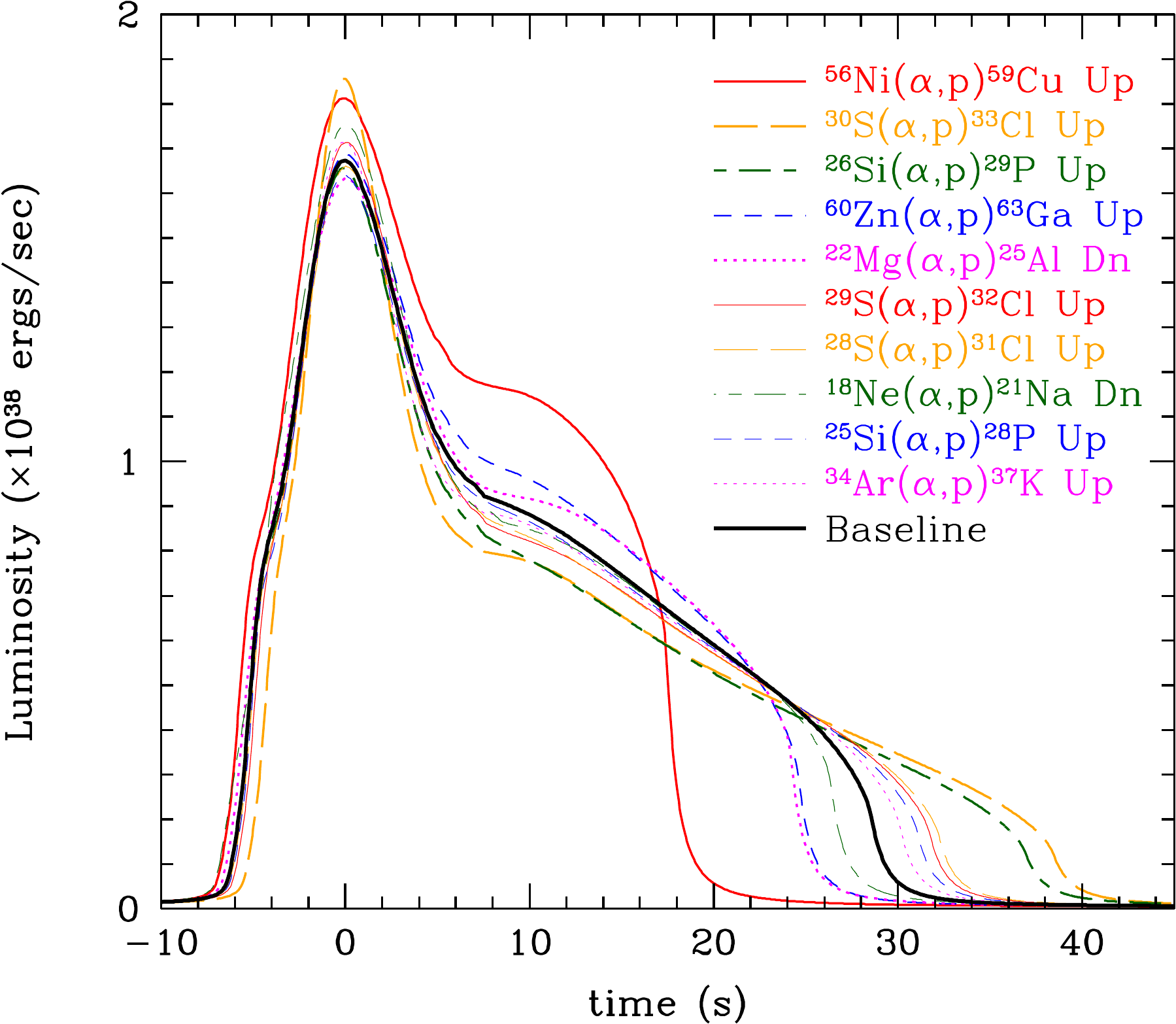}
\caption{\label{FigSZ_LC} The largest changes in single-zone model
  X-ray burst light curves induced by variations in (p,$\gamma$)
  reaction rates (left panel) and ($\alpha$,p) reaction rates (right
  panel). Up denotes a rate increase, Dn a rate decrease. For each
  rate, only the change with the larger impact is shown.}
\end{figure*}

\begin{figure}
\plotone{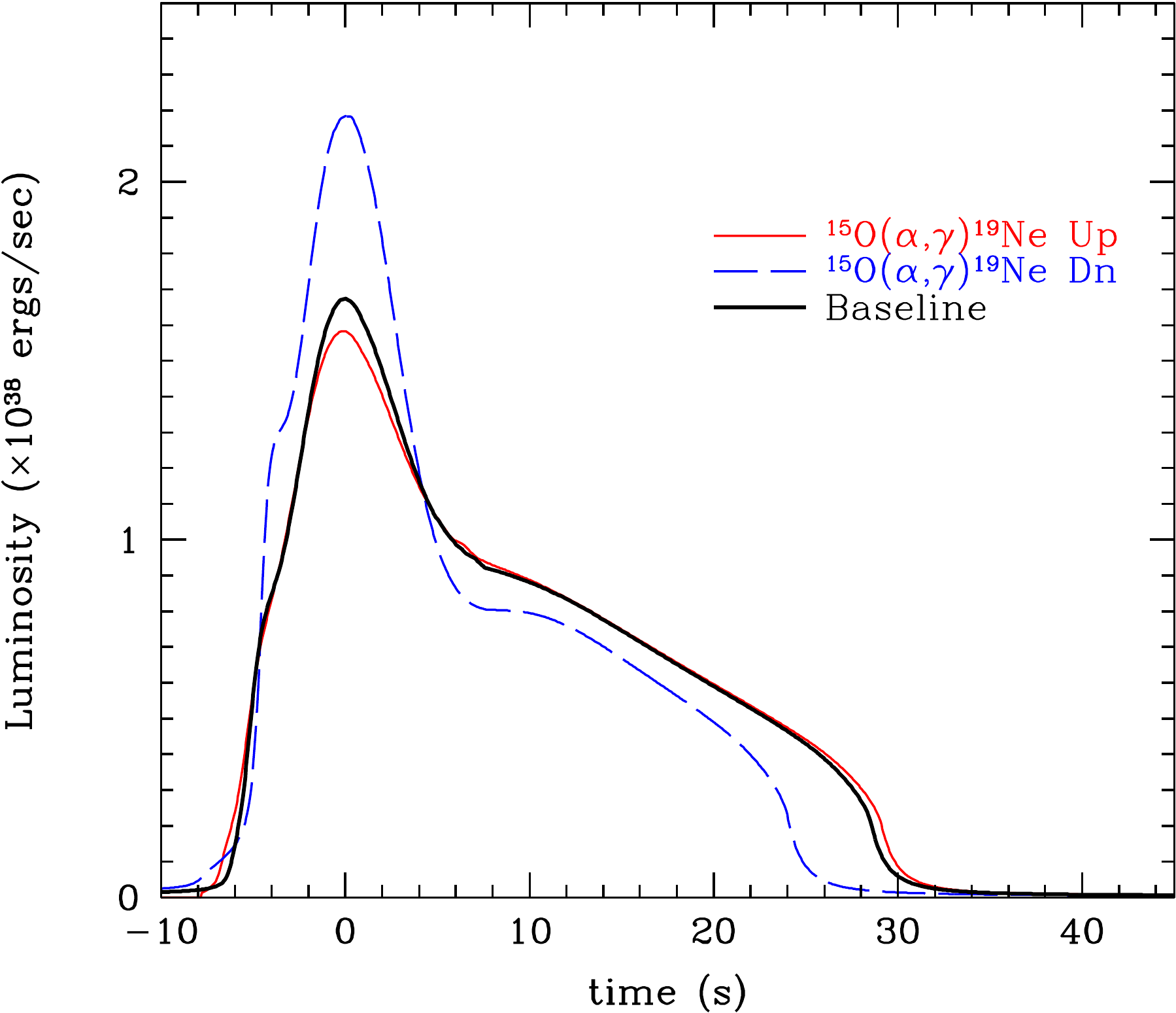}
\caption{\label{FigSZ_LC_ag} Change in single-zone model
  X-ray burst light curves induced by variation of the $^{15}$O($\alpha$,$\gamma$) 
  reaction up (Up) and down (Dn).}
\end{figure}

As expected, the number of individual reaction rate variations that
affect the burst light curve significantly (Category 1) is rather
small. Fig.~\ref{FigSZ_LC} and Fig.~\ref{FigSZ_LC_ag} show the largest resulting variations in
the light curve. A large light curve change is produced by variations
in 7 (p,$\gamma$) reactions, 7 ($\alpha$,p) reactions, and the
$^{15}$O($\alpha$,$\gamma$)$^{19}$Ne reaction (see also
Tab.~\ref{tbl:SZ_LCrank} and Fig.~\ref{FigSZ_path}). By far the
largest change is produced by varying the
$^{59}$Cu(p,$\gamma$)$^{60}$Zn and $^{59}$Cu(p,$\alpha$)$^{56}$Ni
rates, because a low $^{59}$Cu(p,$\gamma$) rate or a high
$^{59}$Cu(p,$\alpha$)$^{56}$Ni rate lead to the formation of a
stronger NiCu cycle \citep{VanWormer1994} that strongly limits
synthesis of heavier nuclei. The critical quantity determining the
strength of the NiCu cycle is the ratio of the (p,$\alpha$) to
(p,$\gamma$) reaction rates at $^{59}$Cu. The
$^{15}$O($\alpha$,$\gamma$)$^{19}$Ne reaction rate has a strong impact
on the total luminosity and leads to a strongly increased peak energy
release when lowered.  Variations of 4 additional Category 2 reactions
listed in Tab.~\ref{tbl:SZ_LCrank} cause smaller, but still
significant changes in the light curve. An additional 9 reactions do
have some noticeable impact on the light curve, but rate variations of
much more than a factor of 100 will be needed for a significant
change. We included the top 28 reactions in the multi-zone variations.

\begin{deluxetable}{ccccc}
  \tablecaption{\label{tbl:SZ_LCrank} Reactions that impact the burst
    light curve in the single-zone x-ray burst model.}
\tablewidth{0pt}
\tablehead{
\colhead{Rank} & \colhead{Reaction} & \colhead{Type\tablenotemark{a}} & \colhead{Sensitivity\tablenotemark{b}} & \colhead{Category}
}
\startdata
  1 & \iso{Ni}{56}($\alpha$,p)\iso{Cu}{59} & U & 12.5 &  1\\
  2 & \iso{Cu}{59}(p,$\gamma$)\iso{Zn}{60} & D & 12.1 &  1\\
  3 & \iso{O}{15}($\alpha$,$\gamma$)\iso{Ne}{19} & D &  7.9 &  1\\
  4 & \iso{S}{30}($\alpha$,p)\iso{Cl}{33} & U &  7.8 &  1\\
  5 & \iso{Si}{26}($\alpha$,p)\iso{P}{29} & U &  5.3 &  1\\
  6 & \iso{Ga}{61}(p,$\gamma$)\iso{Ge}{62} & D &  5.0 &  1\\
  7 & \iso{Al}{23}(p,$\gamma$)\iso{Si}{24} & U &  4.8 &  1\\
  8 & \iso{P}{27}(p,$\gamma$)\iso{S}{28} & D &  4.4 &  1\\
  9 & \iso{Ga}{63}(p,$\gamma$)\iso{Ge}{64} & D &  3.8 &  1\\
 10 & \iso{Zn}{60}($\alpha$,p)\iso{Ga}{63} & U &  3.6 &  1\\
 11 & \iso{Mg}{22}($\alpha$,p)\iso{Al}{25} & D &  3.5 &  1\\
 12 & \iso{Ni}{56}(p,$\gamma$)\iso{Cu}{57} & D &  3.4 &  1\\
 13 & \iso{S}{29}($\alpha$,p)\iso{Cl}{32} & U &  2.8 &  1\\
 14 & \iso{S}{28}($\alpha$,p)\iso{Cl}{31} & U &  2.7 &  1\\
 15 & \iso{Cl}{31}(p,$\gamma$)\iso{Ar}{32} & U &  2.7 &  1\\
 16 & \iso{K}{35}(p,$\gamma$)\iso{Ca}{36} & U &  2.5 &  2\\
 17 & \iso{Ne}{18}($\alpha$,p)\iso{Na}{21} & D &  2.3 &  2\\
 18 & \iso{Si}{25}($\alpha$,p)\iso{P}{28} & U &  1.9 &  2\\
 19 & \iso{Cu}{57}(p,$\gamma$)\iso{Zn}{58} & D &  1.7 &  2\\
 20 & \iso{Ar}{34}($\alpha$,p)\iso{K}{37} & U &  1.6 &  3\\
 21 & \iso{Si}{24}($\alpha$,p)\iso{P}{27} & U &  1.4 &  3\\
 22 & \iso{Mg}{22}(p,$\gamma$)\iso{Al}{23} & D &  1.1 &  3\\
 23 & \iso{As}{65}(p,$\gamma$)\iso{Se}{66} & U &  1.0 &  3\\
 24 & \iso{O}{14}($\alpha$,p)\iso{F}{17} & U &  1.0 &  3\\
 25 & \iso{Sc}{40}(p,$\gamma$)\iso{Ti}{41} & D &  0.9 &  3\\
 26 & \iso{Ar}{34}(p,$\gamma$)\iso{K}{35} & D &  0.8 &  3\\
 27 & \iso{Mn}{47}(p,$\gamma$)\iso{Fe}{48} & D &  0.8 &  3\\
 28 & \iso{Ca}{39}(p,$\gamma$)\iso{Sc}{40} & D &  0.8 &  3\\
\enddata
\tablenotetext{a}{Up (U) or down (D) variation that has the largest impact}
\tablenotetext{b}{$M_{LC}^{(i)}$ in units of 10$^{17}$ergs/g/s }
\end{deluxetable}

The composition of the burst ashes is affected by a much larger number
of reactions.  Tab.~\ref{tbl:SZ_ABrank} and Fig.~\ref{FigSZ_path} list
reactions for which a factor of 100 change (either up or down) of the
rate leads to at least a factor of 2 change in the mass fraction of a
mass chain with significant ($> 10^{-4}$) mass fraction.  The maximum
ratio listed in Tab.~\ref{tbl:SZ_ABrank} gives the largest change in
the mass fraction of a mass chain, calculated for each mass chain
as $\max(X_{\rm
  initial},10^{-4})/\max(X_{\rm final}, 10^{-4})$ or, if less than 1,
its inverse. Also listed are the mass chains with changes of a factor
of 2 or more and 10 or more, respectively. For 21 reactions, changes
in the rate by a factor of 100 lead to mass fraction changes in the
composition of the burst ashes of a factor of 10 or more (as defined
above). For an additional 75 reactions abundance changes range between
a factor of 2 and 9.  Whereas most reactions affect only a small number
of final mass chains in a significant way, typically mass numbers
close to the nuclei involved in the reaction, there are a few
reactions that affect the final composition broadly. As expected,
these tend to be the same reactions that have a significant impact on
the burst light curve (see Tab.~\ref{tbl:SZ_LCrank}). In some cases,
such as $^{15}$O($\alpha$,$\gamma$)$^{19}$Ne, the broad compositional
changes induced by a reaction that strongly affects the light curve
are less than a factor of 2 and therefore do not appear in
Tab.~\ref{tbl:SZ_ABrank}.

%

In total 84 reactions were selected to be explored further in the
multi-zone model analysis. This includes all reactions that were found
to affect the burst light curve significantly, but only a subset of
the reactions that affect the final composition (see reactions marked
in Tab.~\ref{tbl:SZ_ABrank} and in Fig.~\ref{FigMZ_path}).

\subsection{Multi-zone model results}

The selected 84 reactions were varied individually in the multi-zone
model {\Kepler}, for a total of 168 calculations, each with a full
sequence of typically 13-15 bursts. The rate variation factors are
listed in Tab.~\ref{tbl:MZ_Variation}. 
A factor of less than 100 was
chosen in cases where the reaction rate uncertainty is likely much
smaller.  
These are mostly reaction rates calculated with the Hauser-Feshbach approach that are closer to stability, where statistical methods should be well applicable (reduction to a factor of 10), ($\alpha$,p) reactions where limited experimental data hint at 
a total uncertainty span of a factor of 100 (for example \citep{Parikh2013,Deibel2011}, and, 
in a few cases, reaction rates where some experimental information is available. We do note that with a few exceptions none of the reactions are solely based on direct experimental measurements. 
The goal was not to determine realistic uncertainties but to
err on the side of a larger variation while still minimizing cases where a
reaction rate is flagged as important even though it is sufficiently
well known. In cases where reaction rates are calculated using shell
model information and are dominated by contributions from a few
resonances, we calculated upper and lower limits by varying resonance
energies by the experimental uncertainty or, when not known
experimentally, by 200~keV (see Tab.~\ref{tbl:MZ_Variation}). During
this study a compilation of reaction rate uncertainties was published
for a small subset of the reactions of interest here
\citep{Iliadis2010}. In cases where the new rate uncertainties did not
agree within an order of magnitude and, if available, the larger
variation showed a significant impact on light curve and composition,
we reran the multi-zone model calculation with a new variation factor estimate based on the data in
\citet{Iliadis2010} for a $99.7\,\%$ confidence range (see
Tab.~\ref{tbl:MZ_Variation}). This 3$\sigma$ confidence range was
chosen to be conservative as we want to make sure we are not missing
an important sensitivity.

\begin{figure*}
\begin{center}
\includegraphics*[clip=true,width=5.3cm]{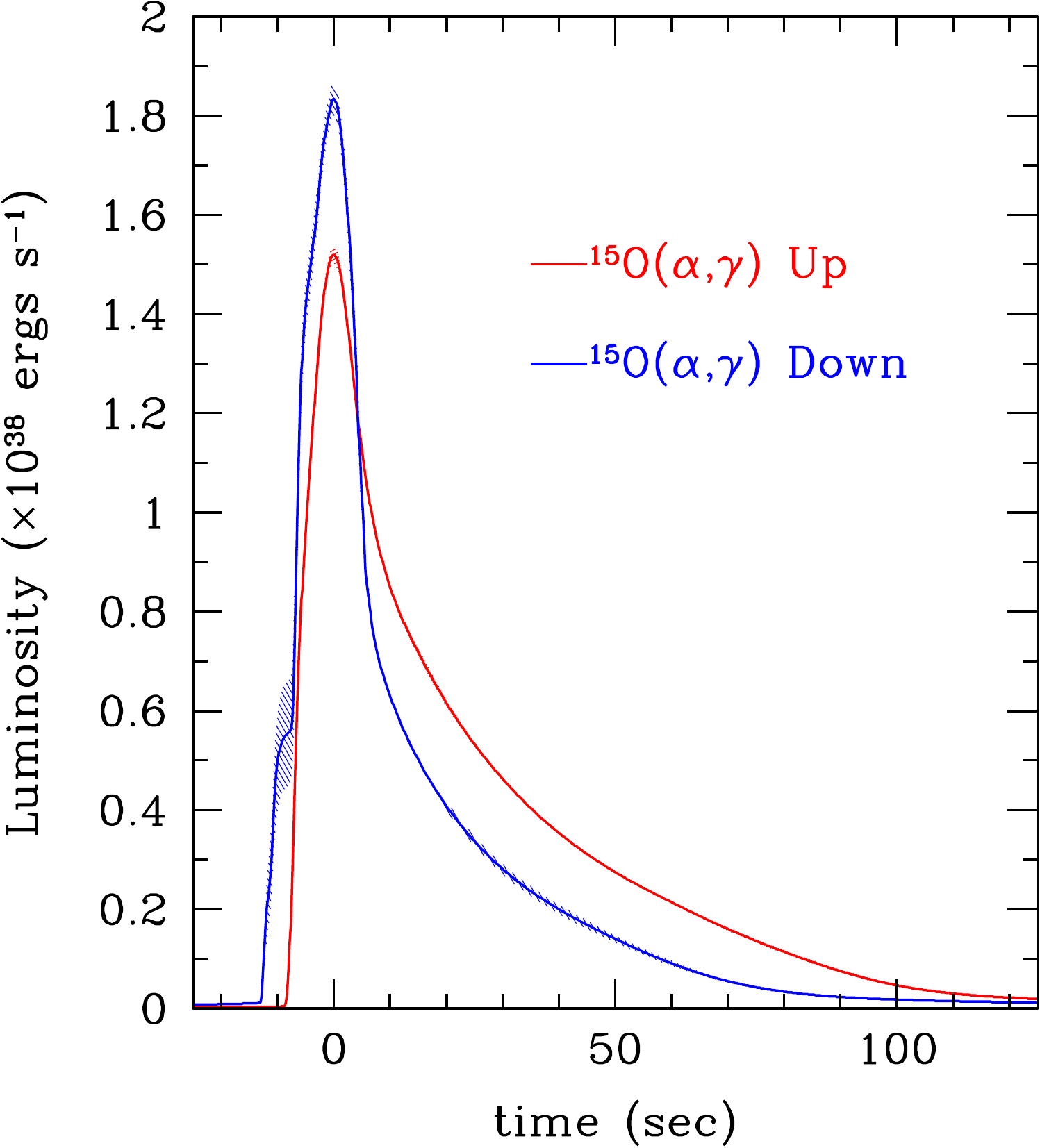}%
\hfill
\includegraphics*[clip=true,width=5.3cm]{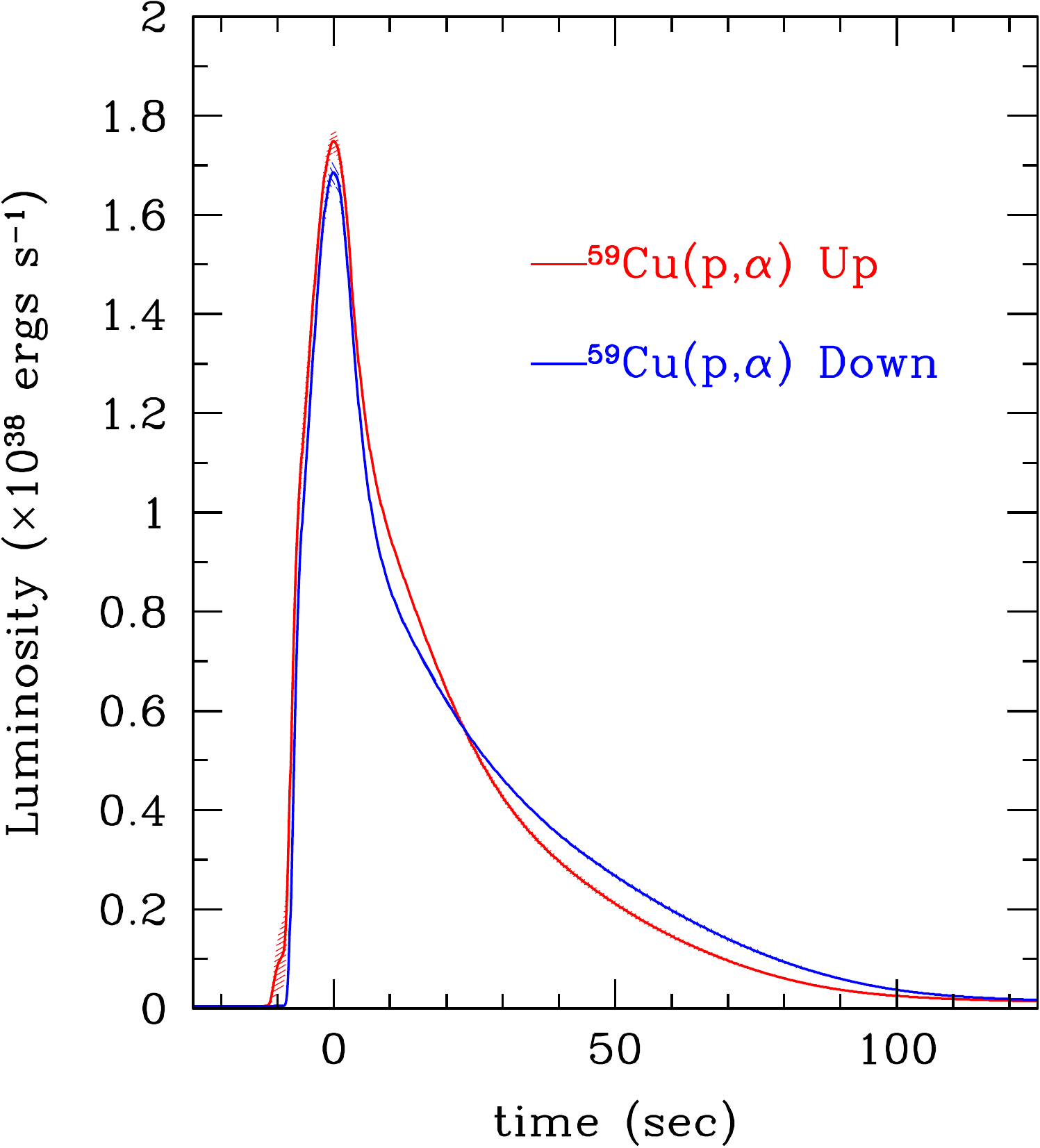}%
\hfill
\includegraphics*[clip=true,width=5.3cm]{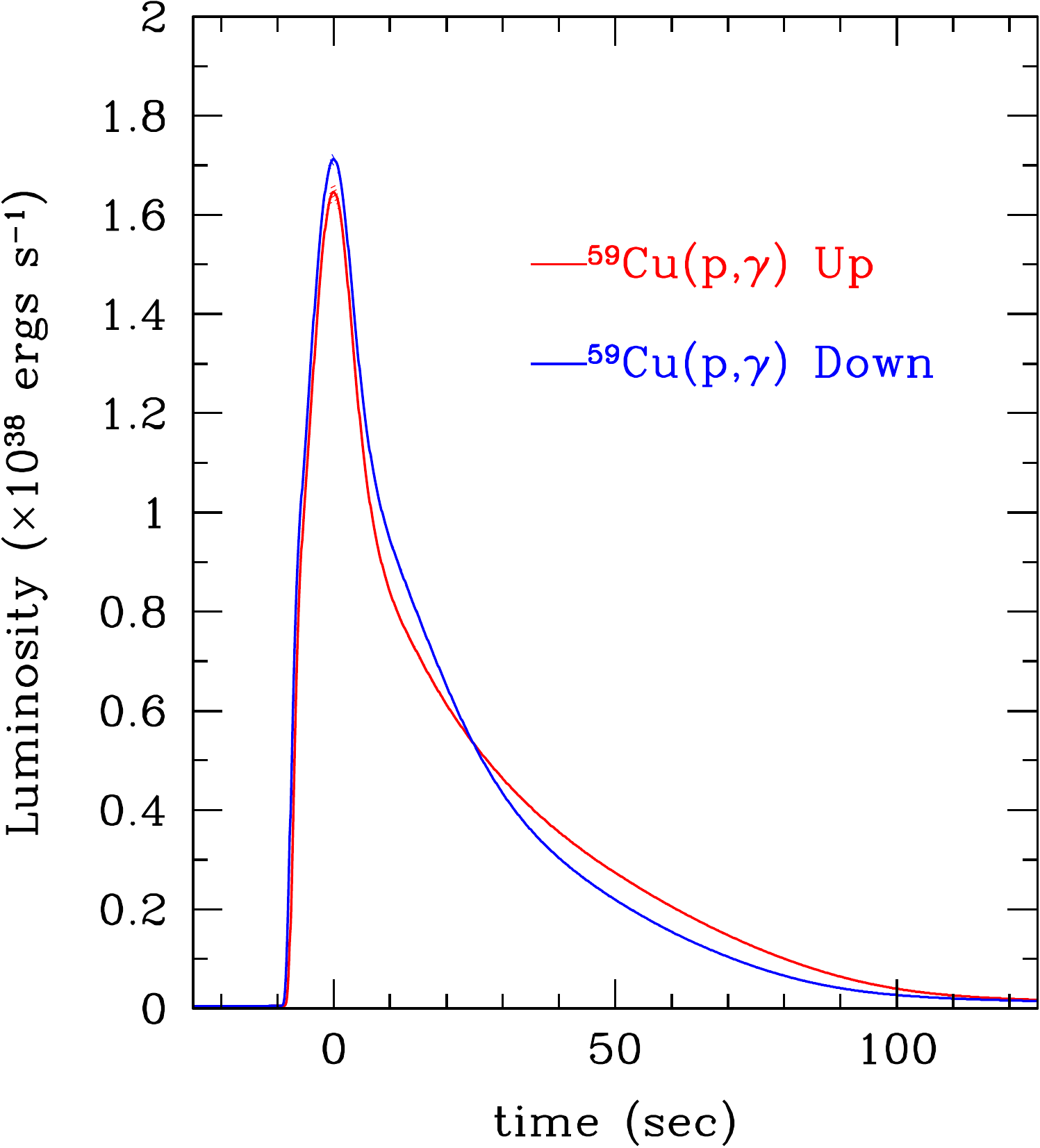}\\
\hfill
\includegraphics*[clip=true,width=5.3cm]{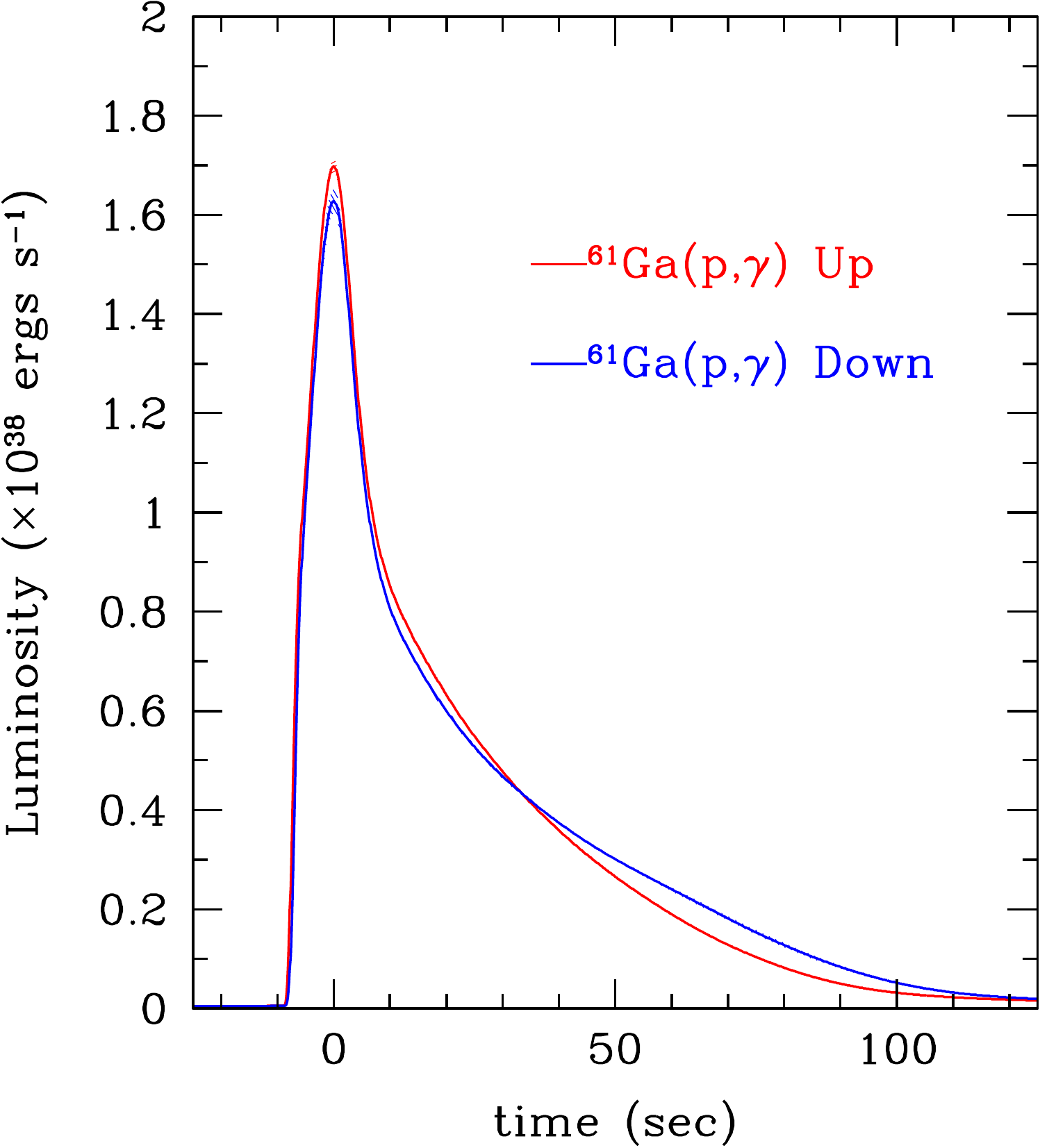}%
\hfill
\includegraphics*[clip=true,width=5.3cm]{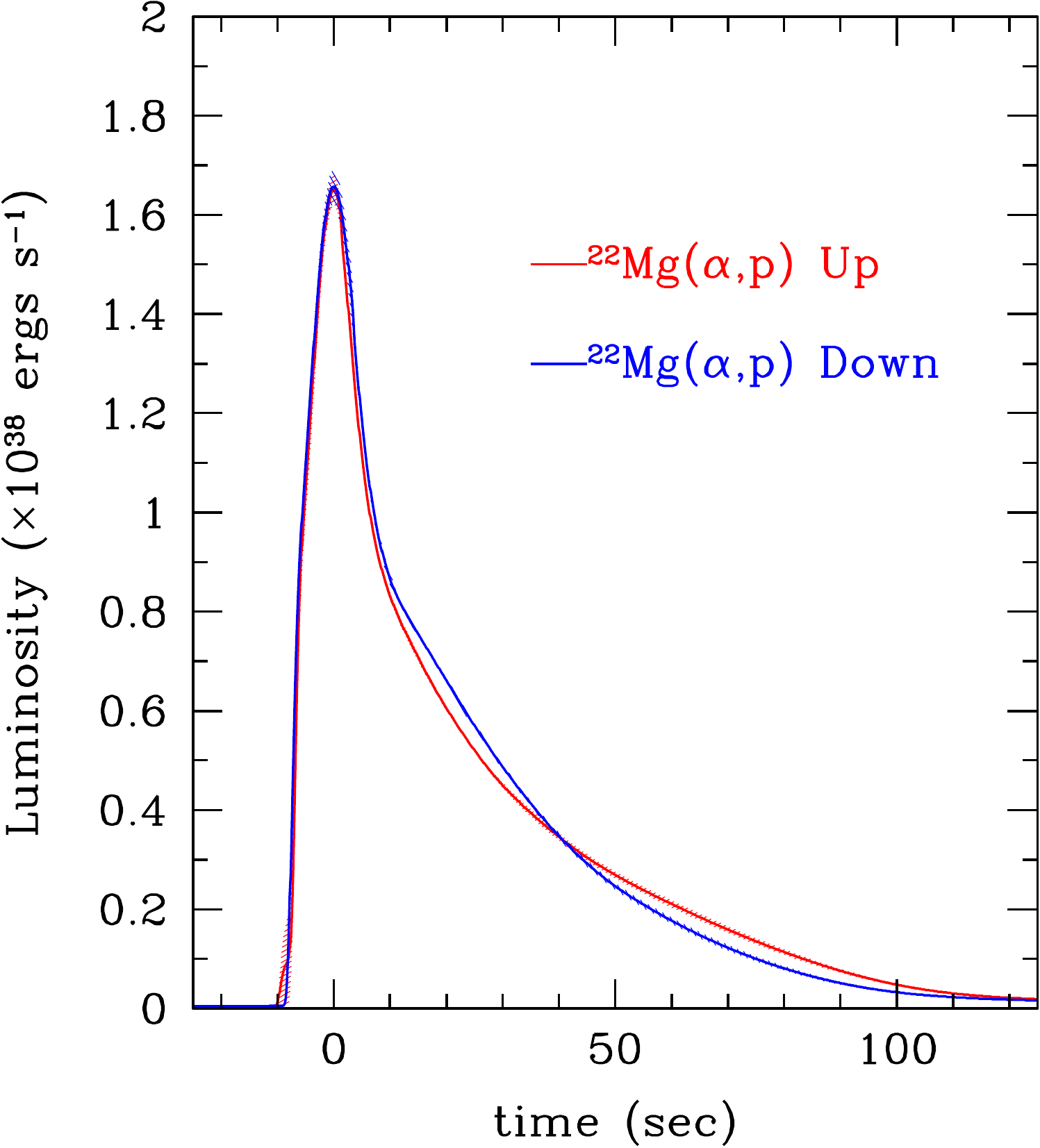}%
\hfill
\includegraphics*[clip=true,width=5.3cm]{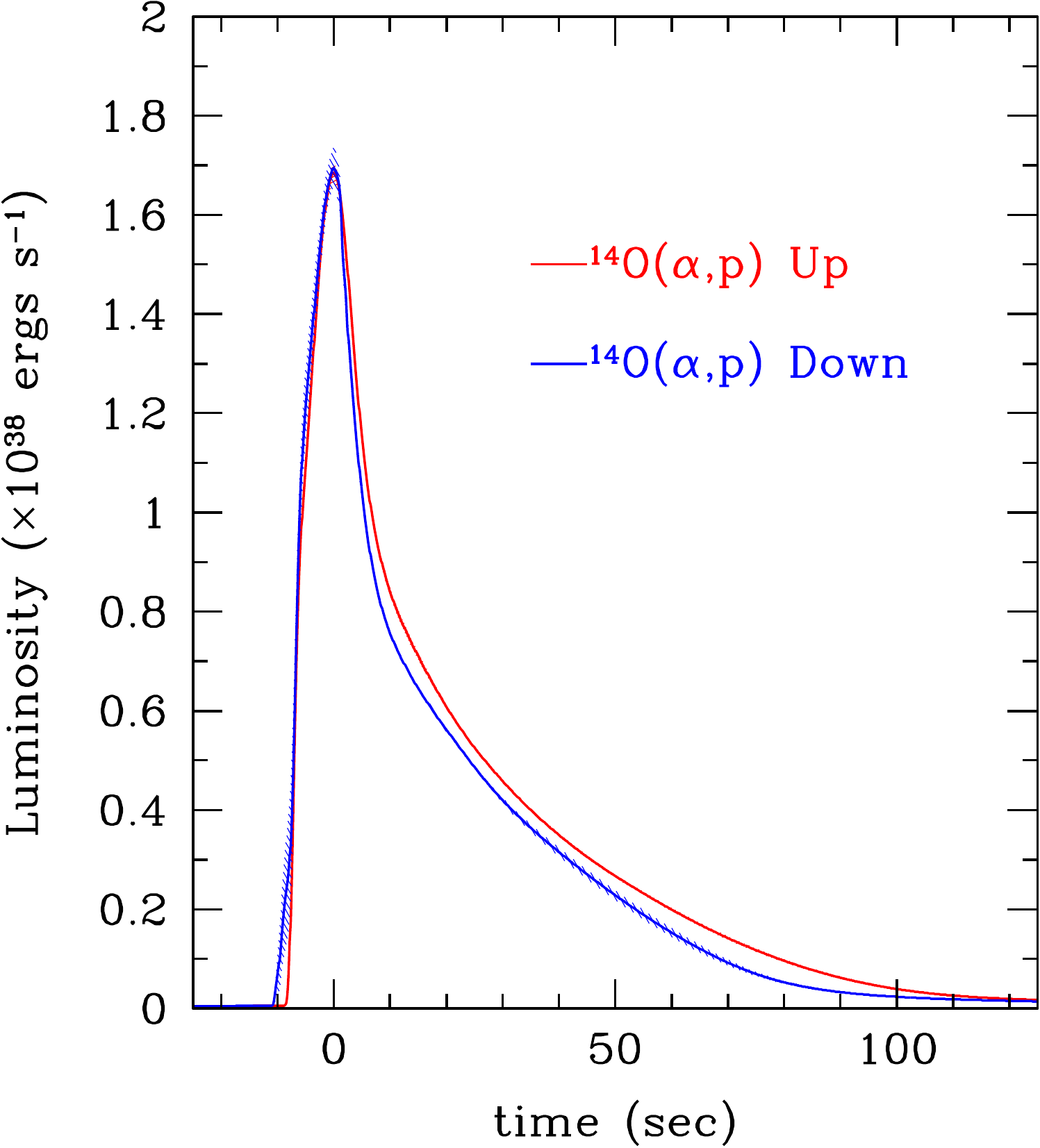}\\
\hfill
\includegraphics*[clip=true,width=5.3cm]{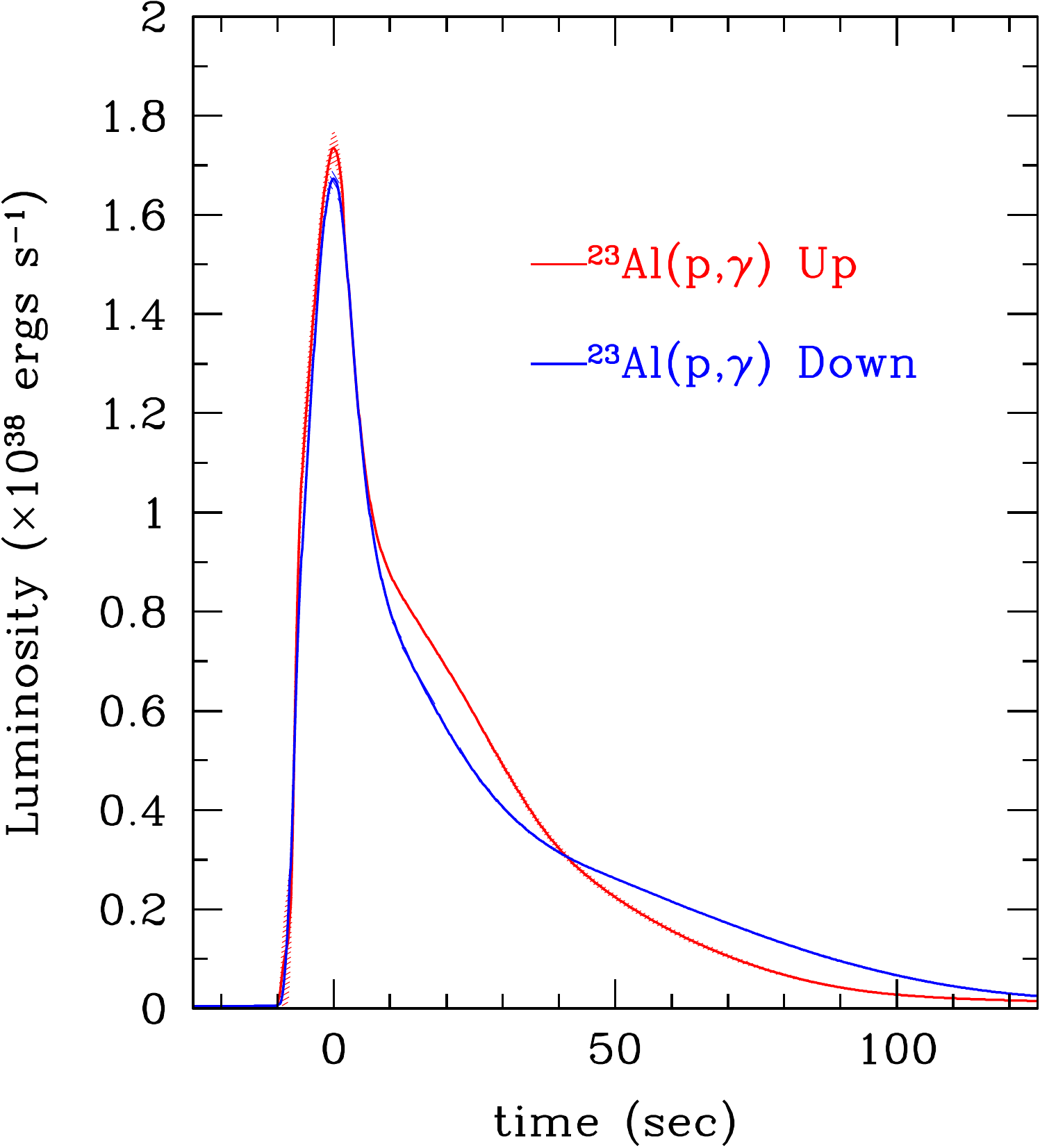}%
\hfill
\includegraphics*[clip=true,width=5.3cm]{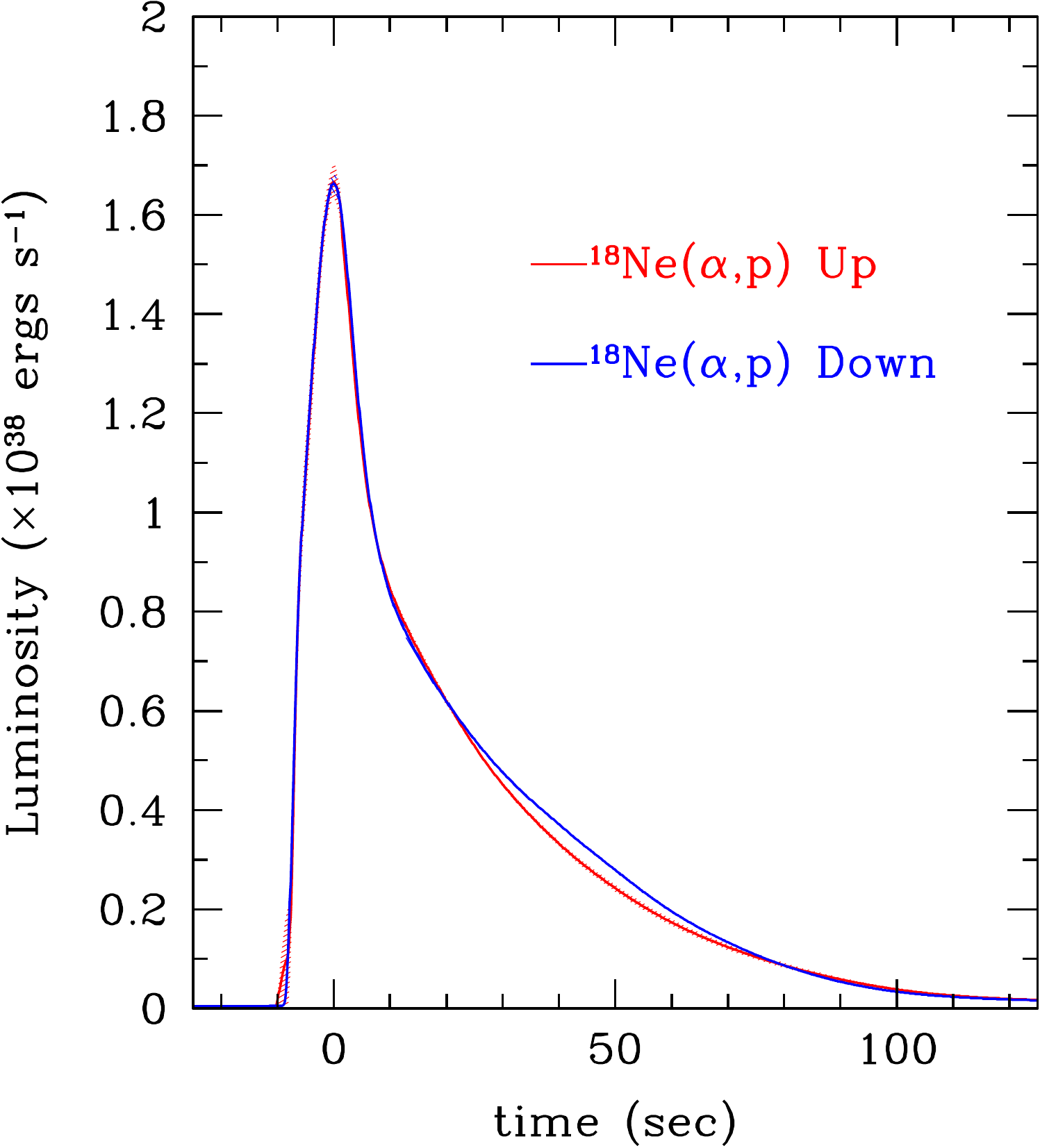}%
\hfill
\includegraphics*[clip=true,width=5.3cm]{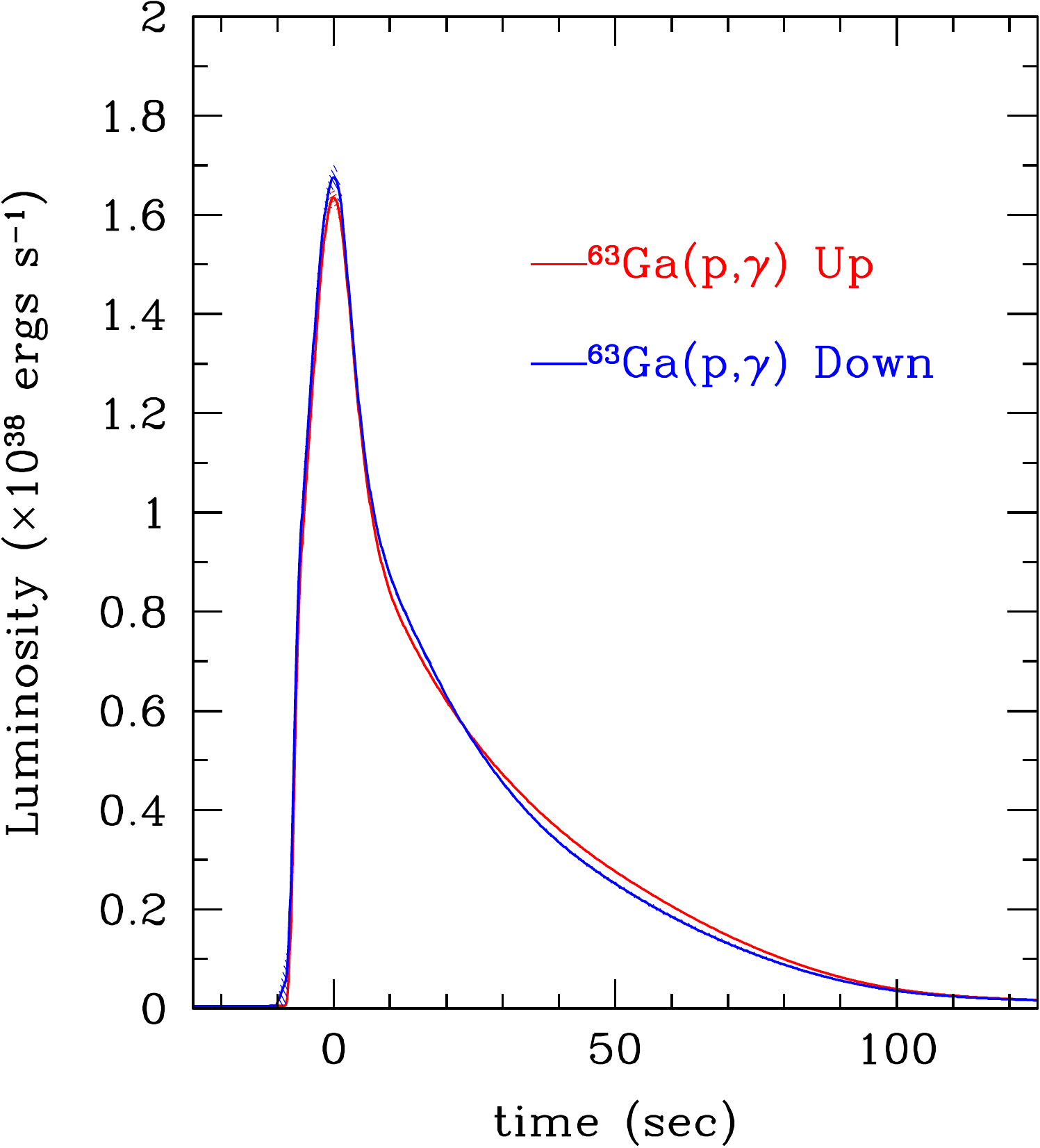}%
\end{center}
\caption{\label{FigMZ_LCcat1} The most significant (Category 1, except
  for $^{63}$Ga(p,$\gamma$)) X-ray burst light curve variations in the
  multi-zone model. Shown is the light curve average over all bursts
  of the calculated sequence. The line width indicates the 1$\sigma$
  error of the average light curve. }
\end{figure*}

\begin{figure*}
\begin{center}
\includegraphics*[clip=true,width=5.3cm]{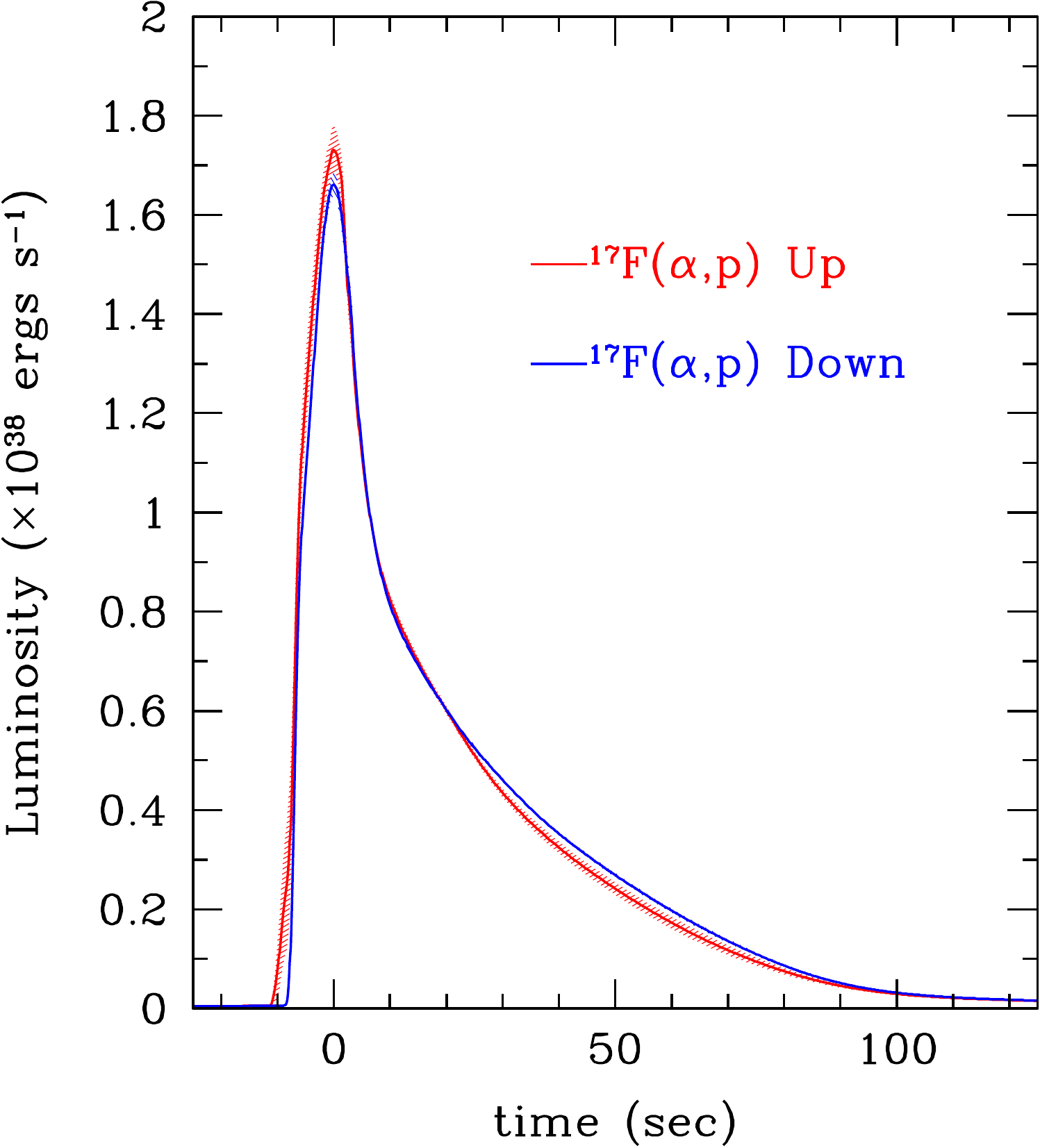}%
\hfill
\includegraphics*[clip=true,width=5.3cm]{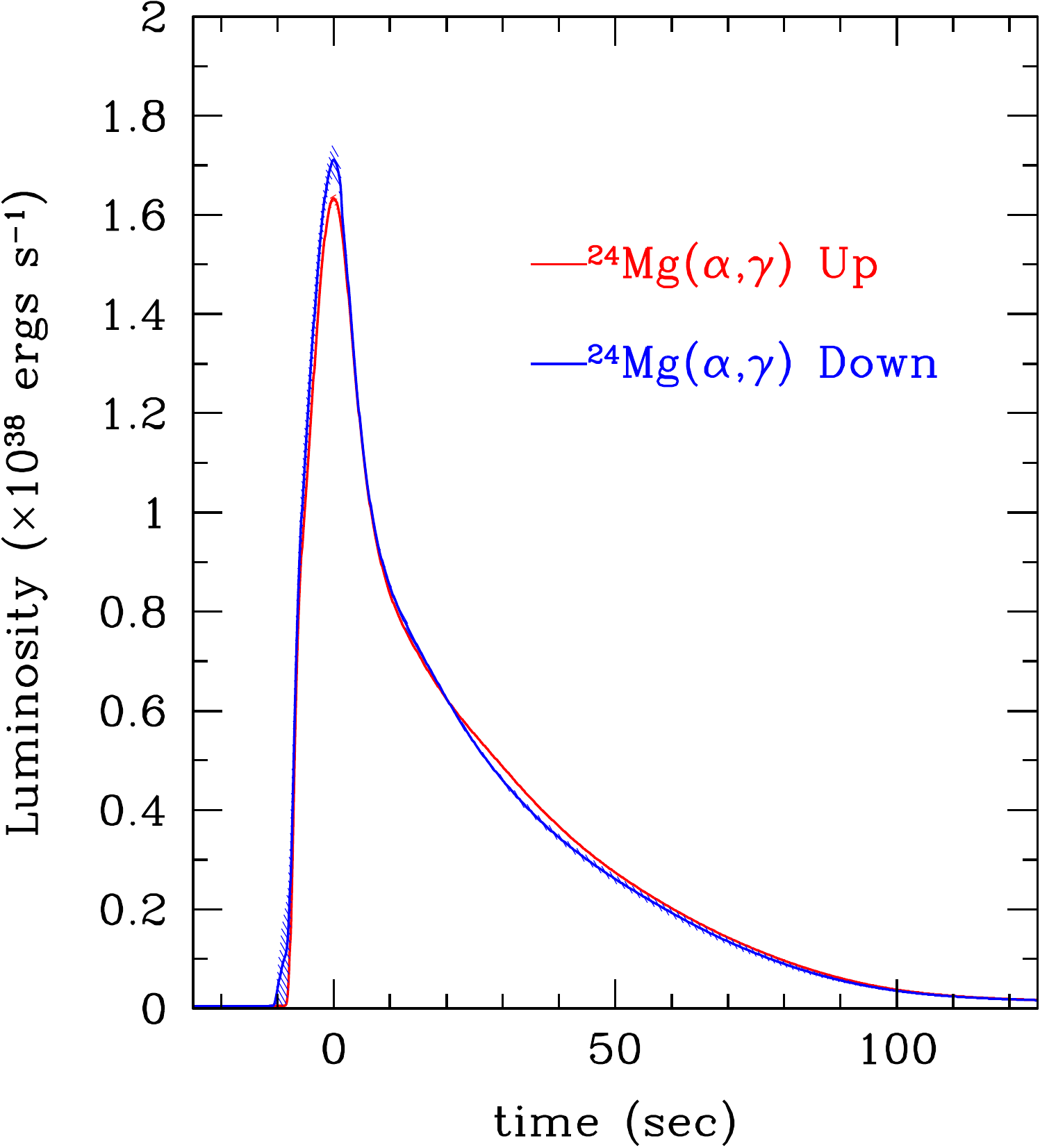}%
\hfill
\includegraphics*[clip=true,width=5.3cm]{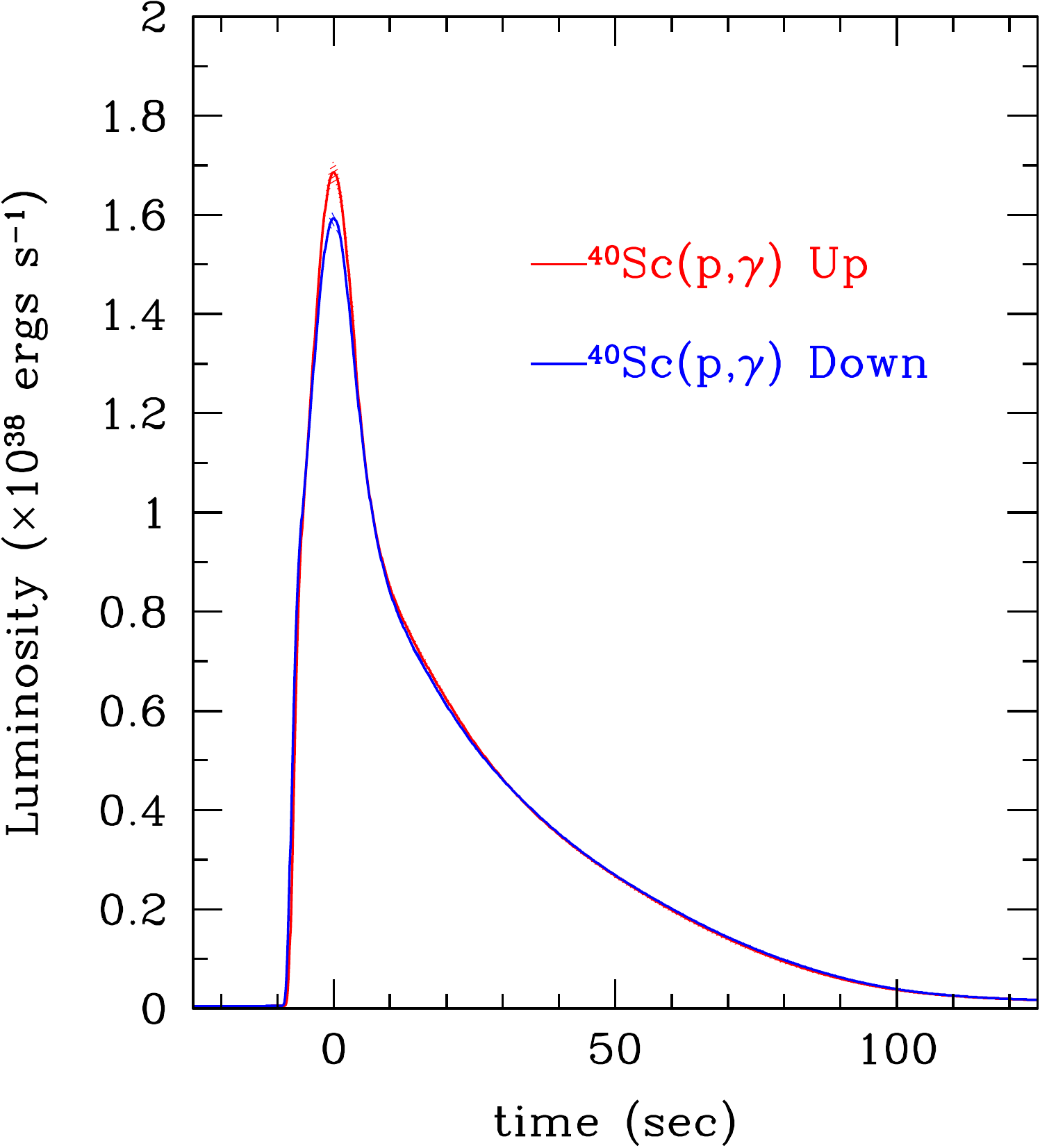}%
\end{center}
\caption{\label{FigMZ_LCcat2} Examples for less significant (Category
  2) X-ray burst light curve variations in the multi-zone model (see
  Fig.~\ref{FigMZ_LCcat1} for details).}
\end{figure*}

The reaction rate variations that were found to have a significant
impact on the burst light curve are listed in order of significance in
Tab.~\ref{tbl:MZ_LCrank} and Fig.~\ref{FigMZ_path}. There are 8
reaction rate variations that affect the light curve
strongly (Category 1) (Fig. \ref{FigMZ_LCcat1}). An additional 11
reaction rate variations lead to smaller but still significant changes
(Category 2, see Fig.~\ref{FigMZ_LCcat2} for
examples). We also provide the value for $M_{LC}^{(i)}$ as defined in 
Eq.~\ref{eqn:LCrank}. Note that while $M_{LC}^{(i)}$ tends to be large 
for reaction rate variations with a strong impact on the light curve, it 
does not always provide a reliable quantitative measure for ranking 
the significance of the light curve impact. $M_{LC}^{(i)}$ weighs
changes near peak luminosity more strongly (which may be relevant for 
some model applications but not for others), and it can become 
artificially large for small shifts in burst rise or decline, as is for example
the case for the  \iso{F}{17}($\alpha$,p)\iso{Ne}{20} rate variation. 

\begin{figure*}
\plotfour{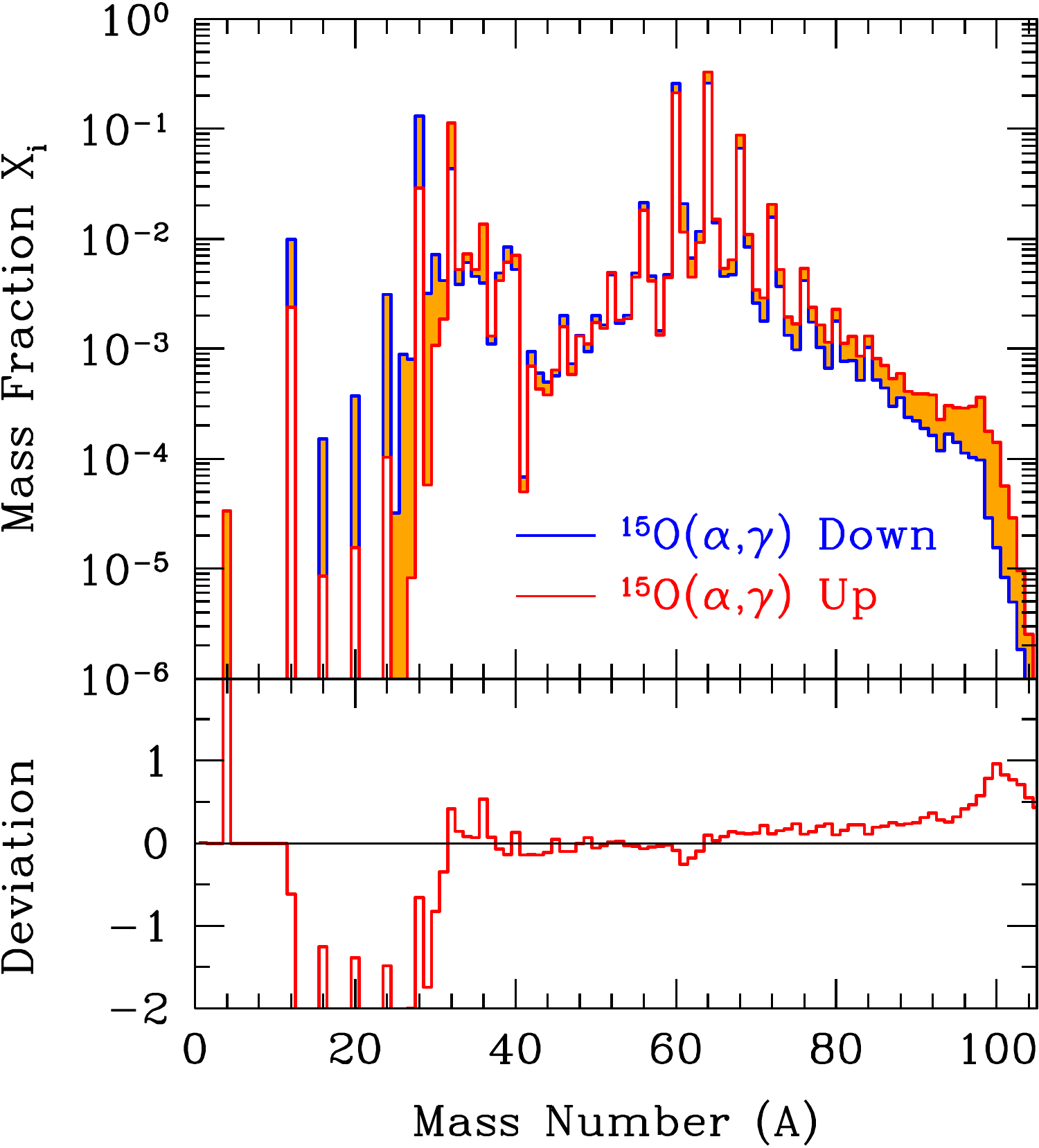}{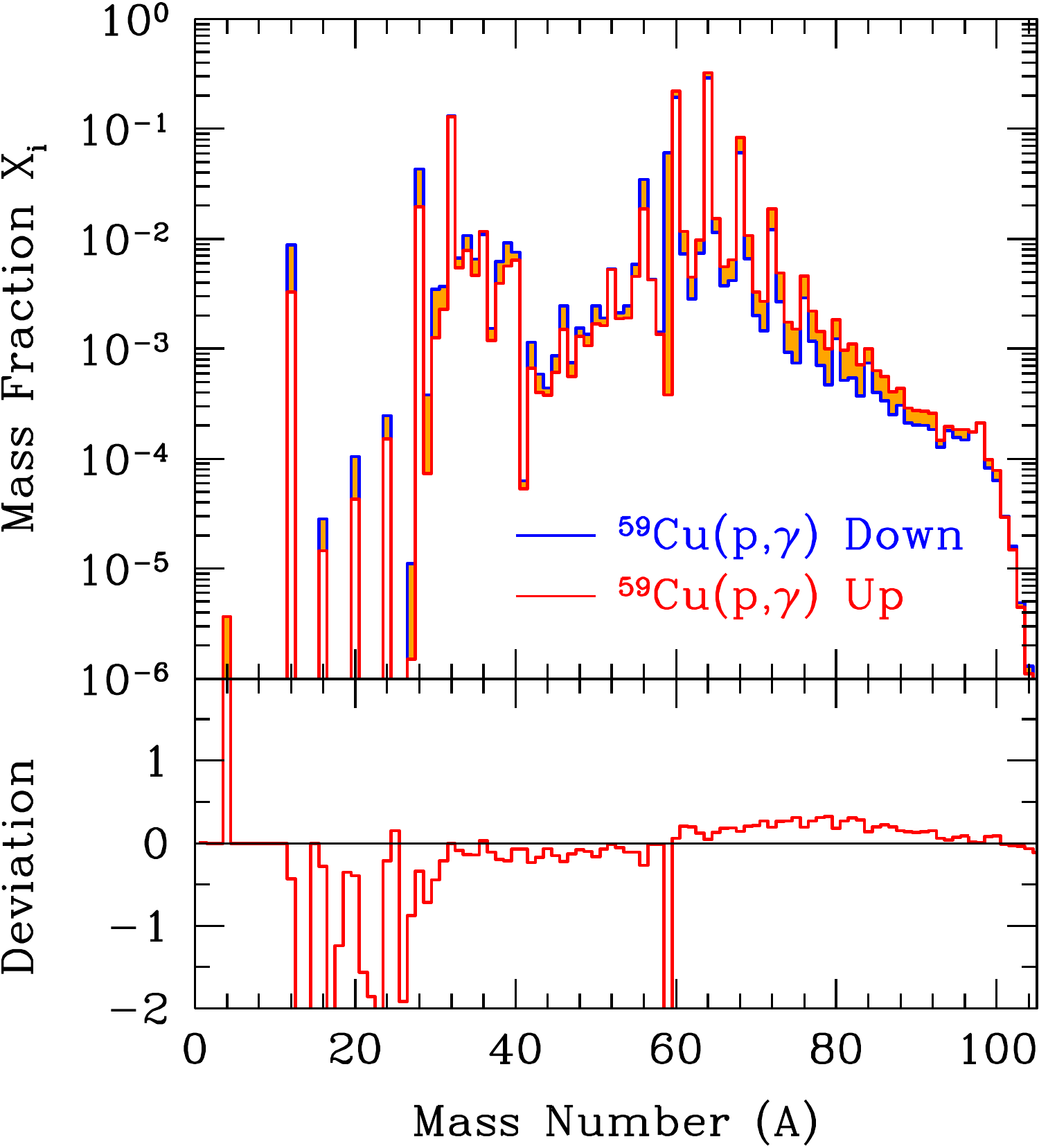}{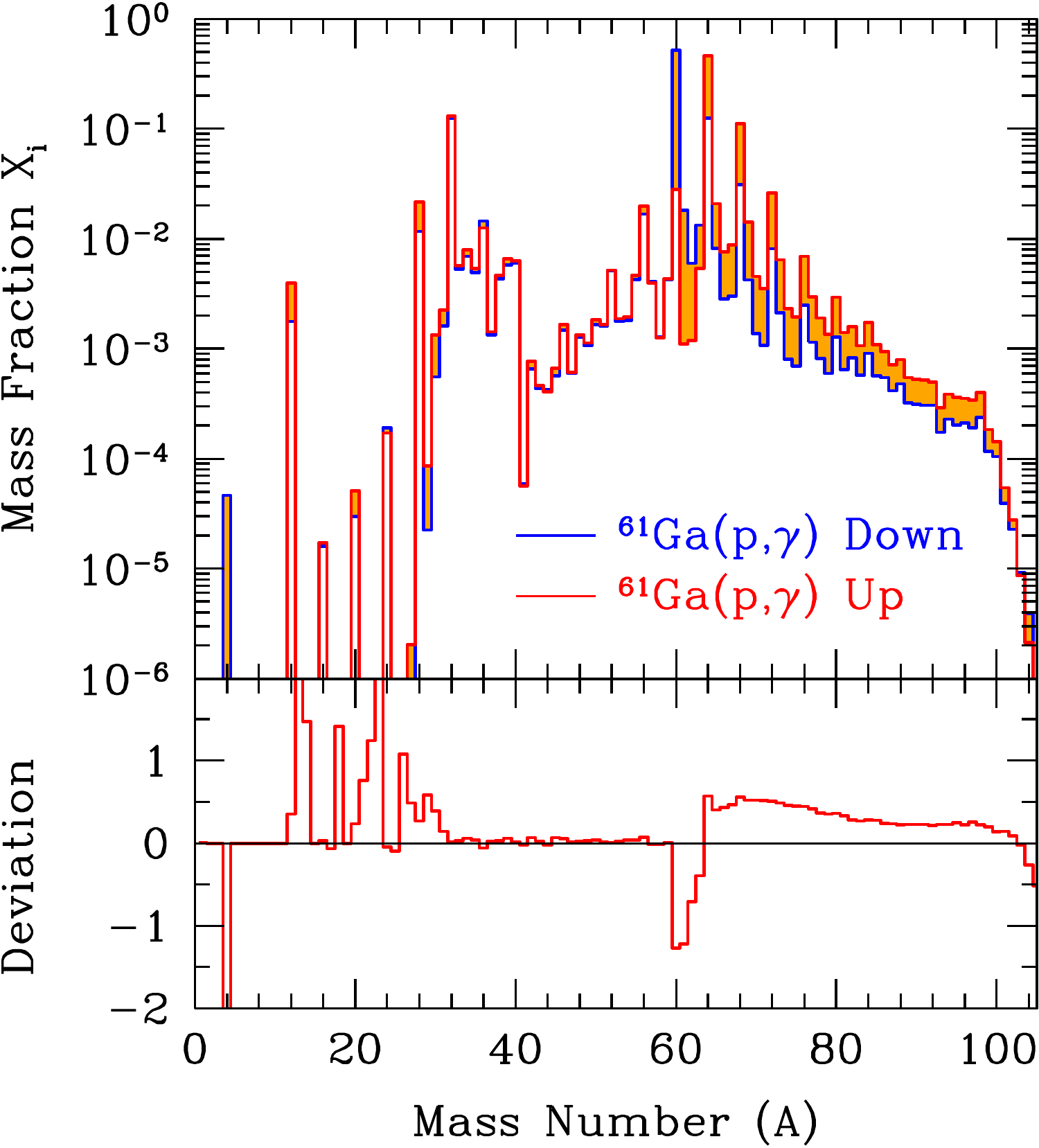}{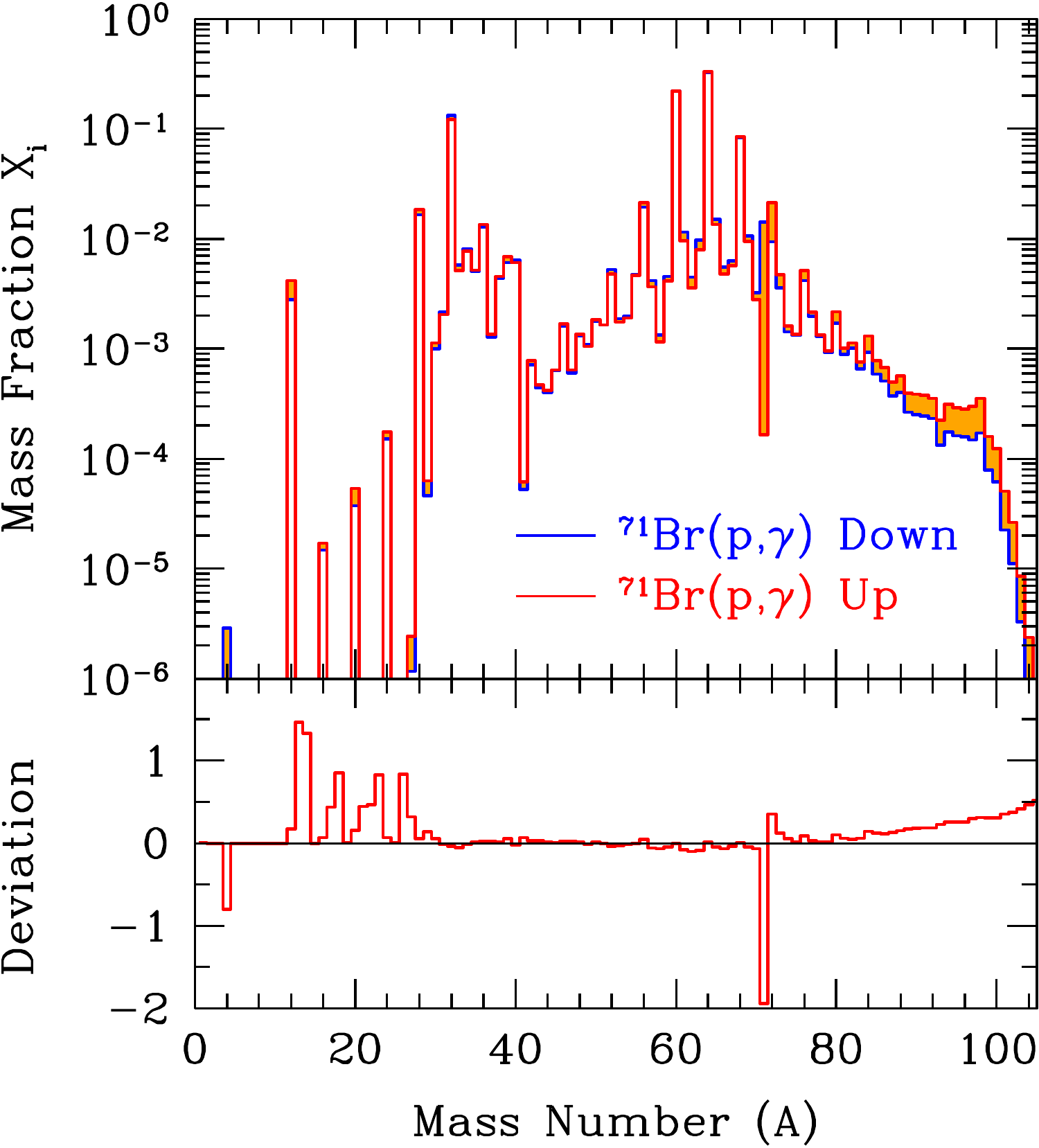}
\caption{\label{FigMZ_AB1} Examples for reaction rate variations that affect the 
composition of the multi-zone model burst ashes in a broad range of mass chains. The lower panel displays the base 10 logarithm
of the mass
  fraction ratio of the up to the down variation.}
\end{figure*}

\begin{figure*}
\plotfour{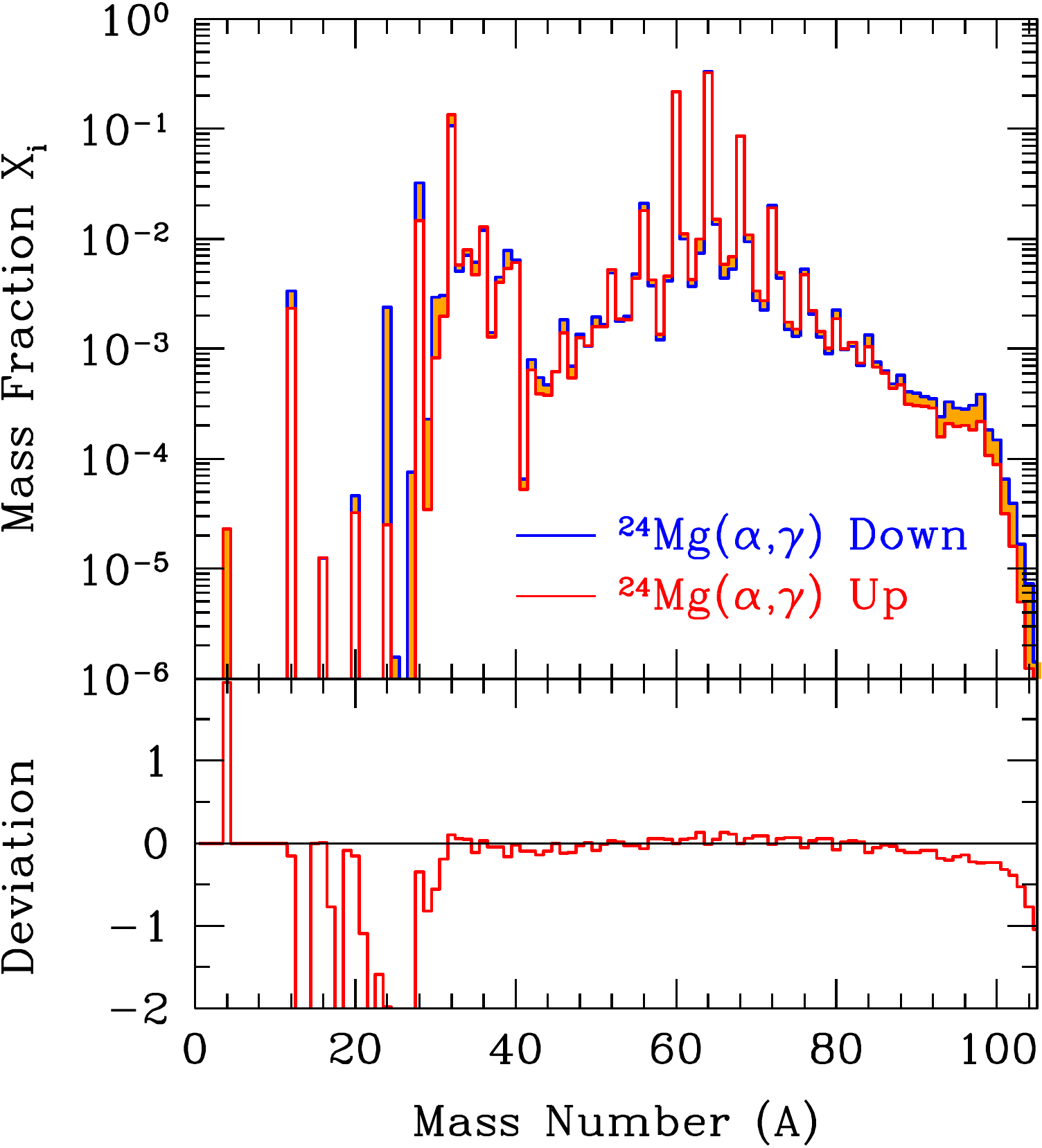}{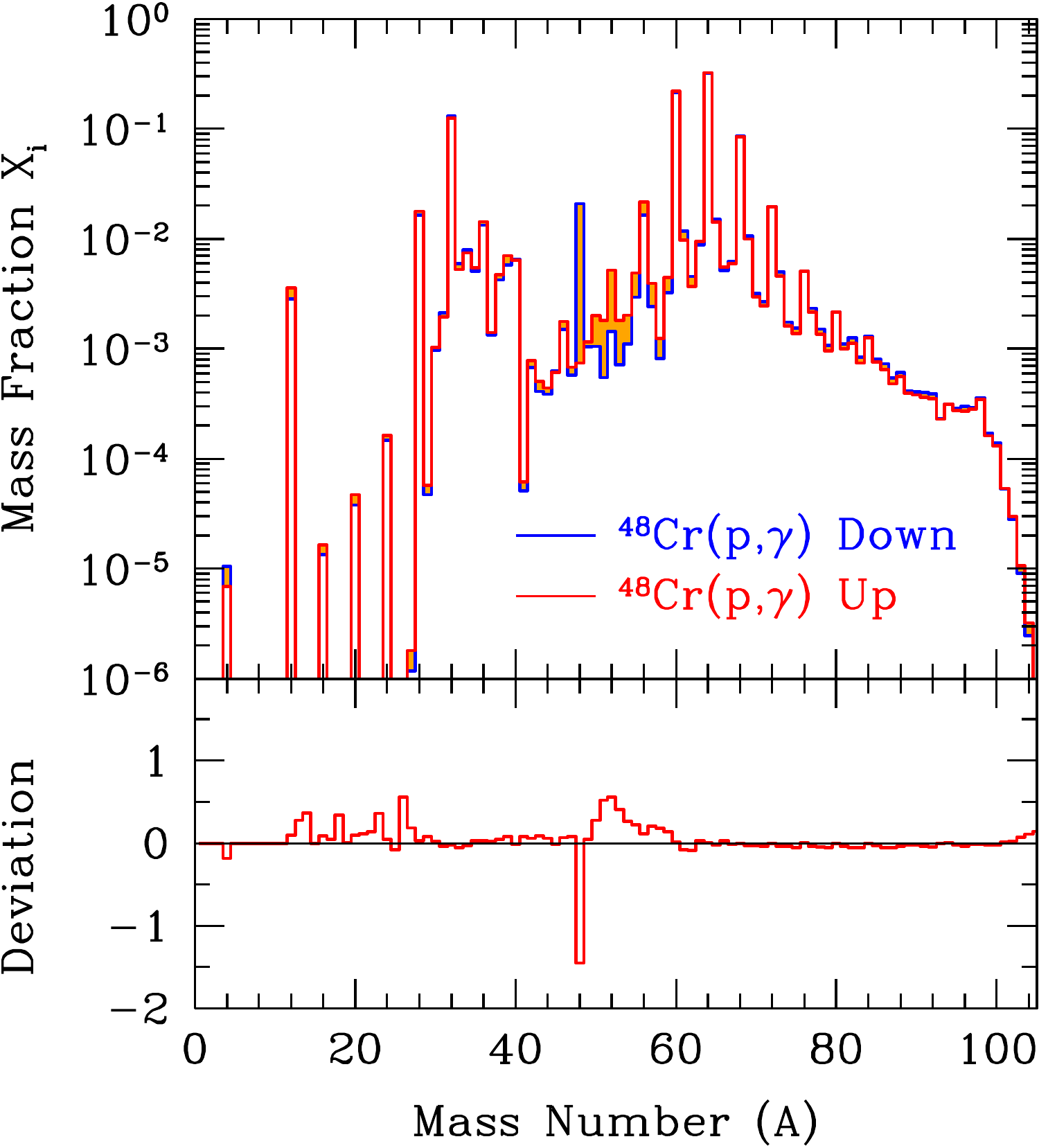}{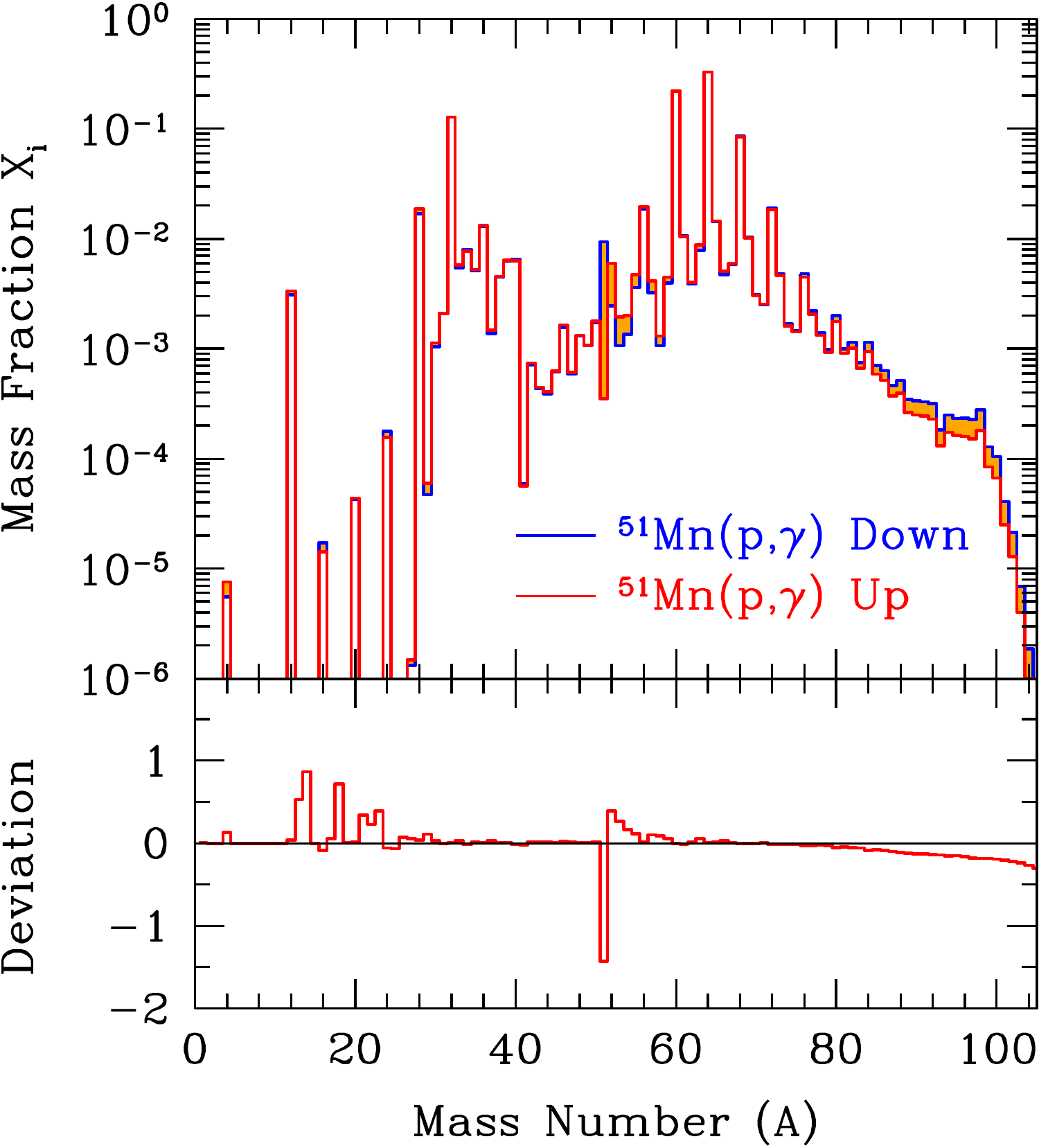}{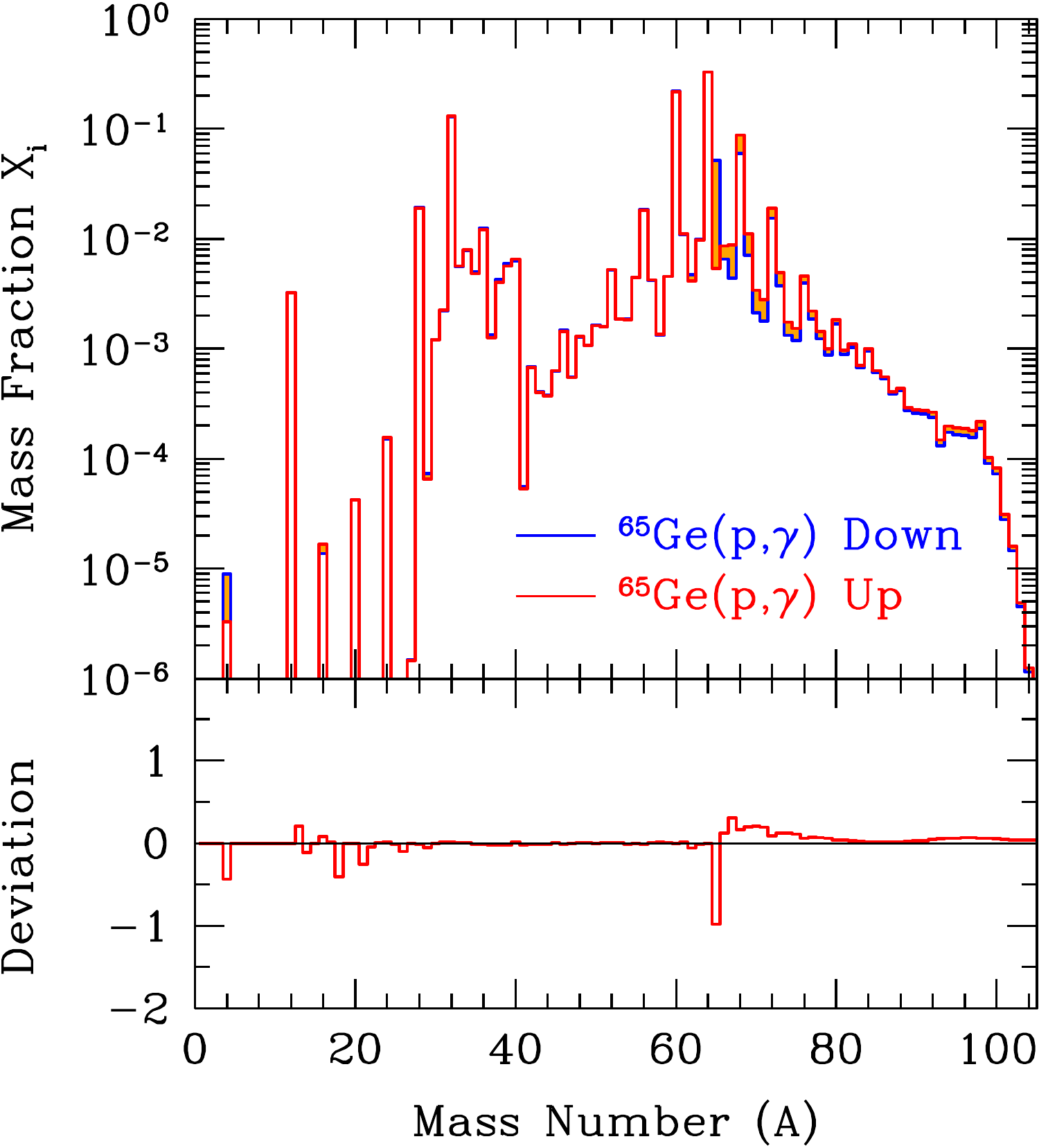}
\caption{\label{FigMZ_AB2} Examples for reaction rate variations that affect the 
composition of the multi-zone model burst ashes in a small number
  of mass chains. See Fig.~\ref{FigMZ_AB1} for details.}
\end{figure*}

Tab.~\ref{tbl:MZ_ABrank} and Fig.~\ref{FigMZ_path}
summarize the significant composition changes (at least a factor of 2
change for a mass chain with mass fraction $>10^{-4}$, see above)
induced by the reaction rate variations explored in the multi-zone
model. A total of 47 rate variations produce changes of more than a
factor of 2 in at least one mass chain, whereas 14 result in changes
of a factor of 10 or more. For the multi-zone calculations, changes are
defined as the ratio between the up and the down
variation. Note that this definition differs from what was used for the 
single-zone model analysis, where changes were defined relative to 
a baseline calculation. This addresses ambiguities in the definition of a 
baseline in the multi-zone model 
 that are a result of our limited burst statistics, 
combined
with rate dependent changes of recurrence time and burst to burst variations. 
Figs.~\ref{FigMZ_AB1} and \ref{FigMZ_AB2} show a few
examples for composition changes.

The multi-zone model calculations also offer the opportunity to
explore the impact of rate changes on the burst recurrence time. Only
two reaction variations produce significant changes beyond average burst to burst 
interval variations, the
$^{15}$O($\alpha$,$\gamma$) reaction rate variation leads to a
$11\,\%$ change, and the $^{14}$O($\alpha$,p) reaction rate variation
leads to a $7\,\%$ change. In both cases, only a decrease of the rate
has a significant impact and leads to a shortening of the recurrence
time (see discussion below).

\begin{deluxetable}{ccccc}
\tablewidth{0pt}
\tablecaption{\label{tbl:MZ_LCrank} Reactions that impact the burst
  light curve\\ in the multi zone x-ray burst model.}
\tablehead{
\colhead{Rank} & \colhead{Reaction} & \colhead{Type\tablenotemark{a}} & \colhead{Sensitivity\tablenotemark{b}} & \colhead{Category}
}
\startdata
\startdata
1 & \iso{O}{15}($\alpha$,$\gamma$)\iso{Ne}{19}  & D & 16 & 1\\
2 & \iso{Ni}{56}($\alpha$,p)\iso{Cu}{59}                     & U & 6.4 & 1\\
3 & \iso{Cu}{59}(p,$\gamma$)\iso{Zn}{60}                  & D & 5.1 & 1\\
4 & \iso{Ga}{61}(p,$\gamma$)\iso{Ge}{62}                 & D & 3.7 & 1\\
5 & \iso{Mg}{22}($\alpha$,p)\iso{Al}{25}                    & D & 2.3 & 1\\
6 & \iso{O}{14}($\alpha$,p)\iso{F}{17}                        & D & 5.8 & 1\\
7 & \iso{Al}{23}(p,$\gamma$)\iso{Si}{24}                   & D & 4.6 & 1\\
8 & \iso{Ne}{18}($\alpha$,p)\iso{Na}{21}                    & U & 1.8 & 1\\
9 & \iso{Ga}{63}(p,$\gamma$)\iso{Ge}{64}                 & D & 1.4 & 2\\
10 & \iso{F}{19}(p,$\alpha$)\iso{O}{16}                       &  U  & 1.3 & 2\\
11 & \iso{C}{12}($\alpha$,$\gamma$)\iso{O}{16}     &  U & 2.1 & 2\\
12 & \iso{Si}{26}($\alpha$,p)\iso{P}{29}                        & U & 1.8 & 2\\
13 & \iso{F}{17}($\alpha$,p)\iso{Ne}{20}                      & U & 3.5 & 2\\
14 & \iso{Mg}{24}($\alpha$,$\gamma$)\iso{Si}{28}   & U & 1.2 & 2\\
15 & \iso{Cu}{57}(p,$\gamma$)\iso{Zn}{58}                  & D & 1.3 & 2\\
16 & \iso{Zn}{60}($\alpha$,p)\iso{Ga}{63}                      & U & 1.1 & 2\\
17 & \iso{F}{17}(p,$\gamma$)\iso{Ne}{18}                     & U & 1.7 & 2\\
18 & \iso{Sc}{40}(p,$\gamma$)\iso{Ti}{41}                     & D & 1.1 & 2\\
19 & \iso{Cr}{48}(p,$\gamma$)\iso{Mn}{49}                  & D & 1.2 & 2\\
\enddata
\tablenotetext{a}{Up (U) or down (D) variation that has the largest impact}
\tablenotetext{b}{$M_{LC}^{(i)}$ in units of 10$^{38}$ergs/s }
\end{deluxetable}

%



\section{Discussion}

The objective of this study is to identify for the one-zone X-ray
burst model ONEZONE the important reaction rates that need to be known
to predict the burst light curve and composition, and in addition to use that
information to identify as many of such reactions as possible in the
full 1D multizone X-ray burst model {\Kepler}.

\subsection{X-ray Light Curves}

The reactions that we find to have a significant influence on the
burst light curve of the one-zone model are summarized in
Fig.~\ref{FigSZ_path}. The breakout reactions of the hot CNO cycle,
$^{15}$O($\alpha$,$\gamma$) and $^{18}$Ne($\alpha$,p) affect the burst light curve
onset, which sets the initial conditions for the later
stages. $^{14}$O($\alpha$,p) does not lead to CNO cycle breakout but
opens in the early stages of the burst a rapid pathway from $^{14}$O
to $^{15}$O that accelerates the hot CNO cycle.

A second set of reactions that are important for the light curve are the ($\alpha$,p) reactions in
the $\alpha$p process. In this model they occur on target nuclei
$^{22}$Mg, $^{24-26}$Si, $^{28-30}$S, and $^{34}$Ar. As the
temperature during the burst rise increases, these nuclei serve as
waiting points until it is hot enough for the ($\alpha$,p) reactions to
turn on. In addition, for $^{22}$Mg, $^{26}$Si, $^{30}$S, and $^{34}$Ar, a proton
capture pathway can compete with the ($\alpha$,p) reaction. Therefore,
proton capture rates in this pathway become also important. Typically
the waiting point is in equilibrium with the following isotone because
of the low proton capture Q-value and the strong inverse ($\gamma$,p)
reaction that enables the much slower ($\alpha$,p) reaction to compete
with the proton capture sequence. In this case, proton capture on the
$Z+1$ isotone determines the strength of the proton capture branch -
for $^{22}$Mg, $^{26}$Si, $^{30}$S, and $^{34}$Ar these are the proton
captures on $^{23}$Al, $^{27}$P, $^{31}$Cl, and $^{35}$K,
respectively. In the case of $^{22}$Mg and $^{34}$Ar the proton
capture on the waiting point itself also plays a role, indicating that
(p,$\gamma$)-($\gamma$,p) equilibrium is not always established.

Also important for the light curve are reactions related to the Ni-Cu and Zn-Ga cycles
\citep{VanWormer1994}.  Of key importance here is the branching into
the cycle at $^{59}$Cu and $^{63}$Ga, which is determined by the
competition of the proton capture rate and the (p,$\alpha$) rate. Note
the (p,$\alpha$) reactions are listed as ($\alpha$,p) reactions.

The proton capture rates in the \textsl{rp}-process are mostly
unimportant for determining the burst light curve. This is not
surprising, as the process is characterized by proton capture being
more rapid than $\beta^+$ decays, leaving the $\beta^+$ decay rates,
which we did not consider in this study, as the critical parameter
that determines energy generation. Nevertheless there are exceptions
at particular bottle necks. These are either associated with shell
structure - $^{39}$Ca and $^{56}$Ni are isotopes where the
\textsl{rp}-process crosses the $Z=20$ and $Z=28$ proton shells - or
with the \textsl{rp}-process waiting points $^{60}$Zn and $^{64}$Ge,
where half-lives are unusually long and proton capture Q-values are
low enough to slow down proton captures. As proton capture Q-values
tend to be low, (p,$\gamma$)-($\gamma$,p) equilibrium tends to play an
important role at these waiting points making the proton capture rate
on the $Z+1$ isotones $^{40}$Sc, $^{57}$Cu, $^{61}$Ga, and $^{65}$As
the important rate. In the case of $^{39}$Ca and $^{56}$Ni proton
capture on the waiting point is also important.

Fig.~\ref{FigMZ_path} and Tab.~\ref{tbl:MZ_LCrank} summarize the
reactions that have been found to influence the light curve in the
multi-zone model. The number of important reactions is somewhat
smaller than in the single-zone model, largely because we
did use smaller variations in several cases. Additional multi-zone
calculations demonstrate that the following reactions would appear in
Tab.~\ref{tbl:MZ_LCrank} when varied by the same factor of 100 as the
single-zone calculations: $^{22}$Mg(p,$\gamma$)$^{23}$Si,
$^{29}$S($\alpha$,p)$^{32}$Cl, $^{30}$S($\alpha$,p)$^{33}$Cl,
$^{31}$Cl($\alpha$,p)$^{34}$Ar (though with a very small effect),
$^{34}$Ar($\alpha$,p)$^{37}$K, $^{39}$Ca(p,$\gamma$)$^{40}$Sc,
$^{35}$K(p,$\gamma$)$^{36}$Ca, and $^{56}$Ni(p,$\gamma$)$^{57}$Cu.
Taking this into account, the qualitative agreement between the
single-zone model and the multi-zone model is quite reasonable, validating
the overall approach. Nevertheless, the quantitative impact of the
various rate variations can be quite different owing to the different
conditions in different zones. Only 6 out of the 28 reactions that
impact the light curve in the single-zone model have no impact at all
in the multi-zone model. 4 of these are ($\alpha$,p) reactions on the
very neutron deficient nuclei $^{24}$Si, $^{25}$Si, $^{28}$S. This may
be due to contributions from zones in the multi-zone model,
where lower peak temperatures
lead to a somewhat less pronounced and less extended
$\alpha$\textsl{p}-process. This may also explain the unimportance of
$^{34}$Ar(p,$\gamma$)$^{35}$K in the multi-zone model and the fact
that in general the ($\alpha$,p) reactions on heavier nuclei are less
important in the multi-zone model than they are in the single-zone
model. A detailed analysis of the origin of the differences between
single zone and multi-zone model sensitivities is beyond the scope of this work
and would require considerable additional computational effort.  

There are also a number of reaction variations that do not impact the
light curve in the single-zone model, but have a significant impact on
the light curve prediction of the multi-zone model. These reactions
fall mostly into two categories: reactions on $^{17}$F and $^{19}$F
that affect the CNO cycles and the helium burning reactions
$^{12}$C($\alpha$,$\gamma$)$^{16}$O and
$^{24}$Mg($\alpha$,$\gamma$)$^{28}$Si. The former group likely affects
burning between bursts and in shallower zones where the
\textsl{rp}-process is mostly absent or less dominant. The latter
group affects the helium burning phase once hydrogen is exhausted and
in particular the additional helium burning that occurs when a layer
of burst ashes is reheated from the ignition of a subsequent burst
above it. Both burning regimes are neglected in the single-zone
approximation. We were able to identify these reactions in the
multi-zone model because we chose to vary them for other reasons than
light curve impact in the single-zone model.  We cannot exclude the
possibility that there are a few additional reactions not listed here
that do not appear to be important in the single zone model, but that
would affect the multi-zone model predictions, however.

\subsection{Composition}

Fig.~\ref{FigSZ_path} and Tab.~\ref{tbl:SZ_ABrank} summarize the
reactions that affect the composition of the burst ashes in the
single-zone model significantly (by more than a factor of 2 for mass
chains with a mass fraction of more than 10$^{-4}$). Most reactions along the reaction path affect the
composition. An exception are reactions between Ca and Ni, where the
\textsl{rp}-process splits into multiple parallel paths. There, only reactions
on the path closest to stability are important for the final
composition, as this is where the longest $\beta^+$ decay half-lives
will be located that define the bottle-necks that determine the
composition in this mass region.

The most abundant mass chains that dominate the composition of the
burst ashes as calculated by the single-zone model (mass fractions $>
10^{-2}$) are $A=12$, 56, 60, 64, 68, 72, and 76. Only a small number
of reactions affect these mass chains significantly - these are proton
captures on $^{56}$Ni, $^{61}$Ga, $^{65}$Ge, $^{67}$As, $^{57}$Cu,
$^{55}$Co, and $^{59}$Cu together with the $^{56}$Ni($\alpha$,p)
reaction. An additional 38 reactions affect mass chains with mass
fractions $> 10^{-3}$.

It was not feasible to vary all the important reactions identified in
the single-zone model in individual multi-zone model
runs. Nevertheless, by varying a subset of rates we do find 47
reactions that affect the burst ashes significantly
(Fig.~\ref{FigMZ_path} and Tab.~\ref{tbl:MZ_ABrank}). We also find 37
reactions that do not affect the composition of the ashes
significantly and that are also indicated in Fig.~\ref{FigMZ_path}.
The most important reaction rate uncertainties among the subset of 47
relevant reaction rates will be the ones that affect the most abundant
mass chains, and the ones that affect mass chains that are of
particular interest because of their impact on ocean and crust
physics. The most abundant mass chains in the multi-zone model
($>10^{-2}$) are $A=$28, 32, 56, 60, 64, 68, and 72. Overall these
abundances are quite robust, with no reaction causing a change by more
than a factor of 10. The set of reactions affecting the $A=56, 60, 64,
68$, and 72 mass chains by more than a factor of 2 is very
small. $^{56}$Ni($\alpha$,p)$^{59}$Cu affects $A=56$,
$^{61}$Ga(p,$\gamma$)$^{62}$Ge has a broad impact on $A=60, 64$, and
68, and both, $^{69}$Se(p,$\gamma$)$^{70}$Br and
$^{71}$Br(p,$\gamma$)$^{72}$Kr affect $A=72$.  The $A=28$ mass chain
is affected by $^{12}$C($\alpha$,$\gamma$)$^{16}$O,
$^{15}$O($\alpha$,$\gamma$)$^{19}$Ne, $^{16}$O($\alpha$,p)$^{19}$F,
$^{18}$Ne($\alpha$,p)$^{21}$Na, $^{22}$Mg($\alpha$,p)$^{25}$Al,
$^{23}$Mg($\alpha$,p)$^{26}$Al and
$^{24}$Mg($\alpha$,$\gamma$)$^{28}$Si, whereas $A=32$ is only affected by
$^{15}$O($\alpha$,$\gamma$)$^{19}$Ne. Reactions such as 
$^{12}$C($\alpha$,$\gamma$)$^{16}$O, $^{16}$O($\alpha$,p)$^{19}$F, 
and $^{24}$Mg($\alpha$,$\gamma$)$^{28}$S are not expected to occur in 
the presence of hydrogen, as proton capture rates on these target 
isotopes are much faster. This indicates late time helium
burning as the source of these abundance peaks, which also explains
their absence in the single-zone model.

Among the mass chains that affect ocean and crust physics, $A=12$
stands out. $^{12}$C produced by X-ray bursts and reignited deeper in
the neutron star ocean is the leading model to explain the
occasionally observed superbursts. The small mass fraction of $^{12}$C
produced in current X-ray burst models (0.4\% in our model) is a major issue in this
context because super burst models require $^{12}$C mass fractions in
the $20\,\%$ range \citep{Cumming2006} to achieve ignition and explain
burst energetics.  We find that only four reactions,
$^{15}$O($\alpha$,$\gamma$)$^{19}$Ne,
$^{56}$Ni($\alpha$,p)$^{59}$Cu, $^{59}$Cu(p,$\gamma$)$^{60}$Zn, and
$^{61}$Ga(p,$\gamma$)$^{62}$Ge affect $^{12}$C production by more than
a factor of 2, and none affect it by more than a factor of 10, as
would be required to explain the origin of superbursts. With the
caveat that we studied only a limited set of reactions, this may
indicate that the problem of the small $^{12}$C production in X-ray
bursts is not caused by nuclear physics uncertainties.

It has recently been found that the presence of nuclei with odd mass
numbers may lead to strong crust cooling via electron
capture-$\beta$-decay Urca cycles \citep{Schatz2014}. The cooling
rates are directly proportional to the abundance of the respective
mass chain. Reaction rate uncertainties that affect odd mass chains
will therefore be of particular importance. The most abundant odd mass
chains in the burst ashes are $A=61,63,65$, and 69, which have mass
fractions of close to $10^{-2}$ or larger. In our subset of reaction
rate variations we identify proton captures on $^{61}$Zn, $^{61}$Ga,
$^{63}$Ga, $^{65}$Ge, $^{66}$Ge, $^{67}$As, and $^{69}$Se as important
for determining the amount of these nuclides in the burst ashes.

\subsection{Recurrence time}
Only three reaction variations were found to significantly (beyond average burst to burst 
interval variations) affect
burst recurrence times: $^{15}$O($\alpha$,$\gamma$)$^{19}$Ne,
$^{14}$O($\alpha$,p)$^{17}$F, and the 3$\alpha$ reaction. Variation of
the 3$\alpha$ reaction rate has a small effect because the rate is
relatively well known. A change of the rate of $20\,\%$ up or down
results in a recurrence time change of about $4\,\%$.  As expected, an
increase of the 3$\alpha$-reaction rate leads to faster burst ignition
and a shorter recurrence time, indicating that indeed as predicted the
3$\alpha$-reaction is the main burst ignition mechanism
\citep{Strohmayer2006}. On the other hand, only a decrease of the
$^{14}$O($\alpha$,p)$^{17}$F and $^{15}$O($\alpha$,$\gamma$)$^{19}$Ne
reactions has a strong impact, and, somewhat counterintuitively, leads
to faster burst ignition and a decrease in recurrence time by $7\,\%$
and $11\,\%$, respectively.

\begin{figure}
\plotone{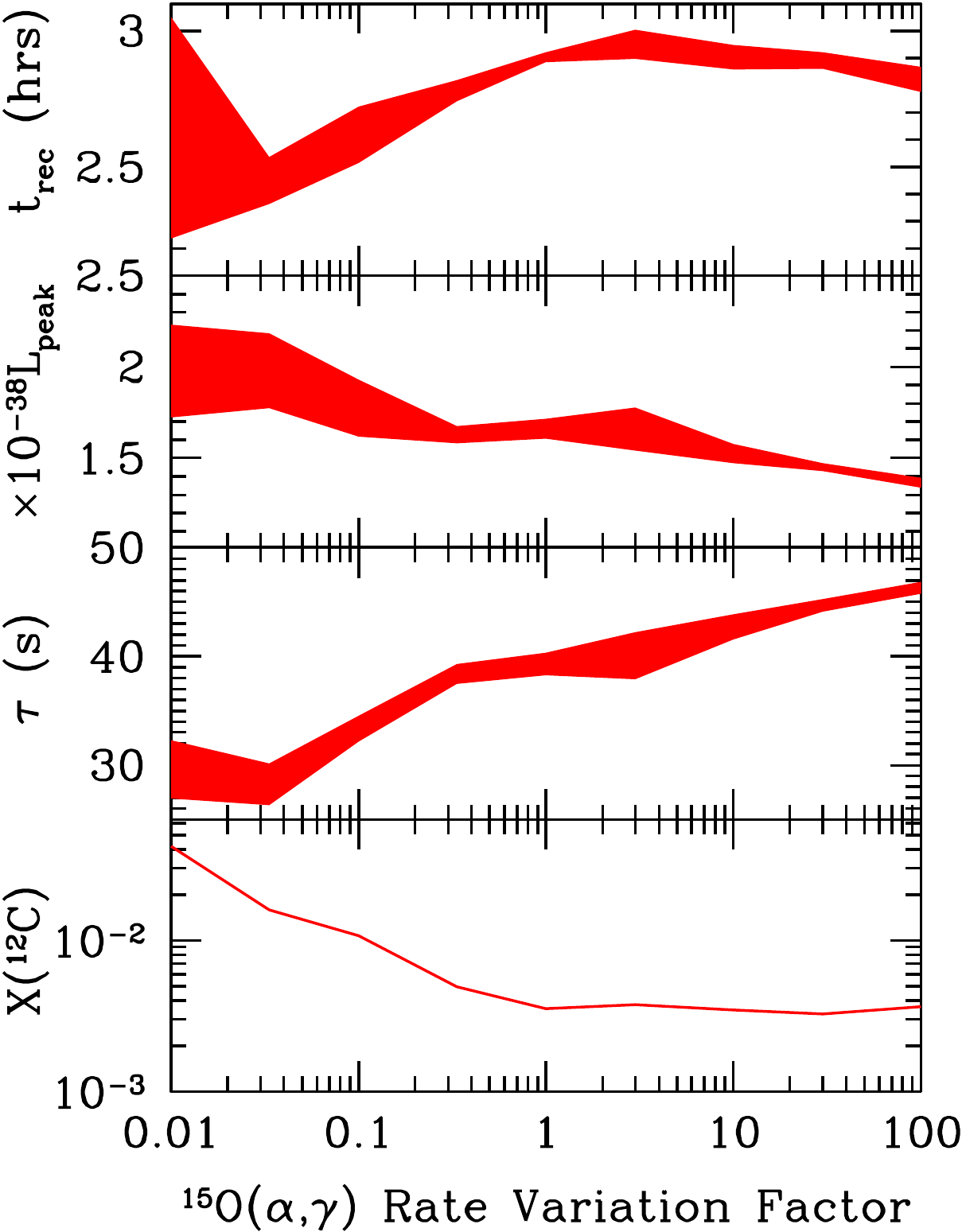}
\caption{\label{FigMZ_o15ag}Multi-zone model results for burst
  recurrence time, peak luminosity in erg/s, burst duration, and final
  $^{12}$C mass fraction as functions of the rate multiplier for the
  $^{15}$O($\alpha$,$\gamma$)$^{19}$Ne reaction. The band denotes 
  the standard deviation due to burst-to-burst variations. }
\end{figure}

\begin{figure*}
\plottwolarge{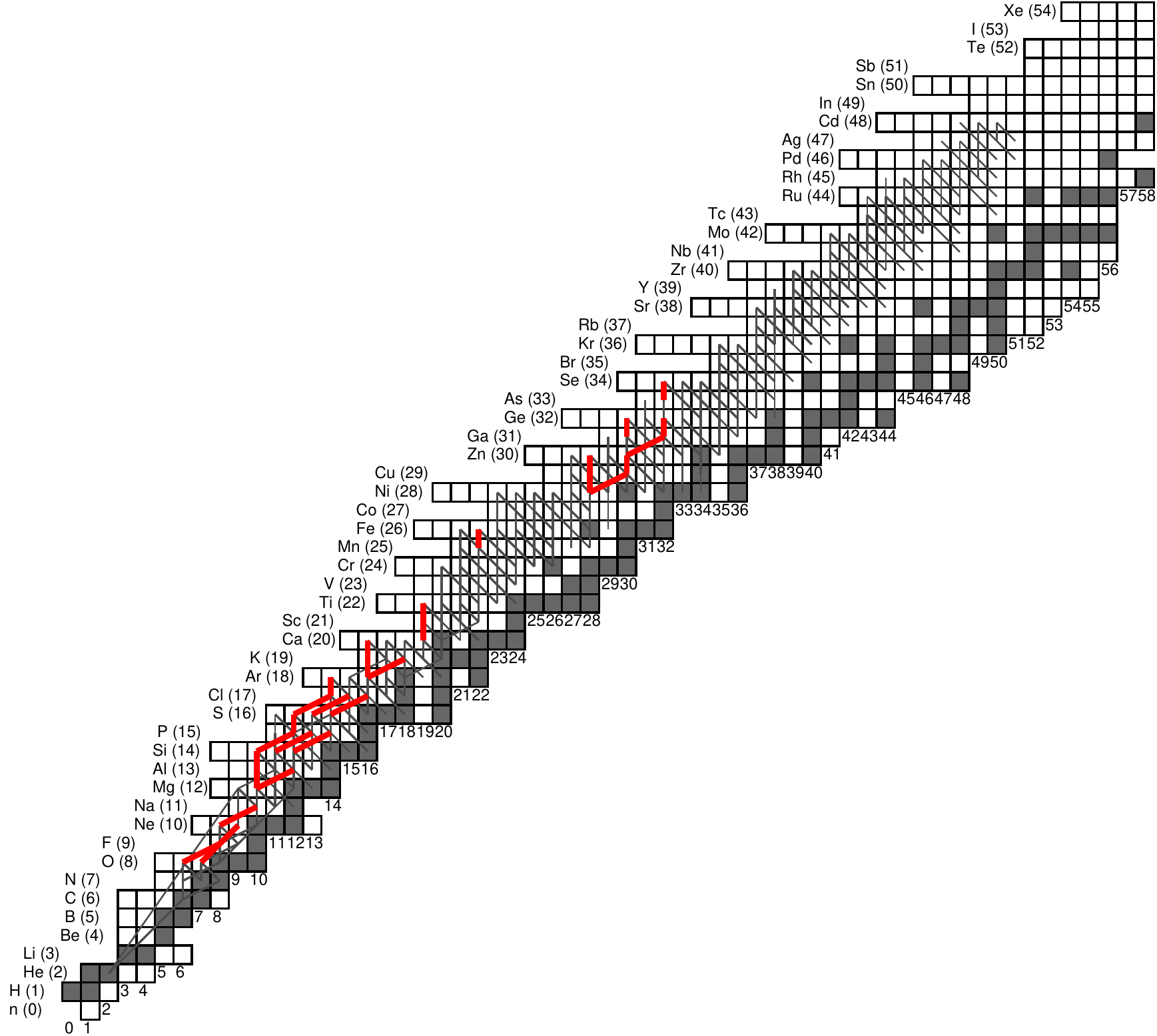}{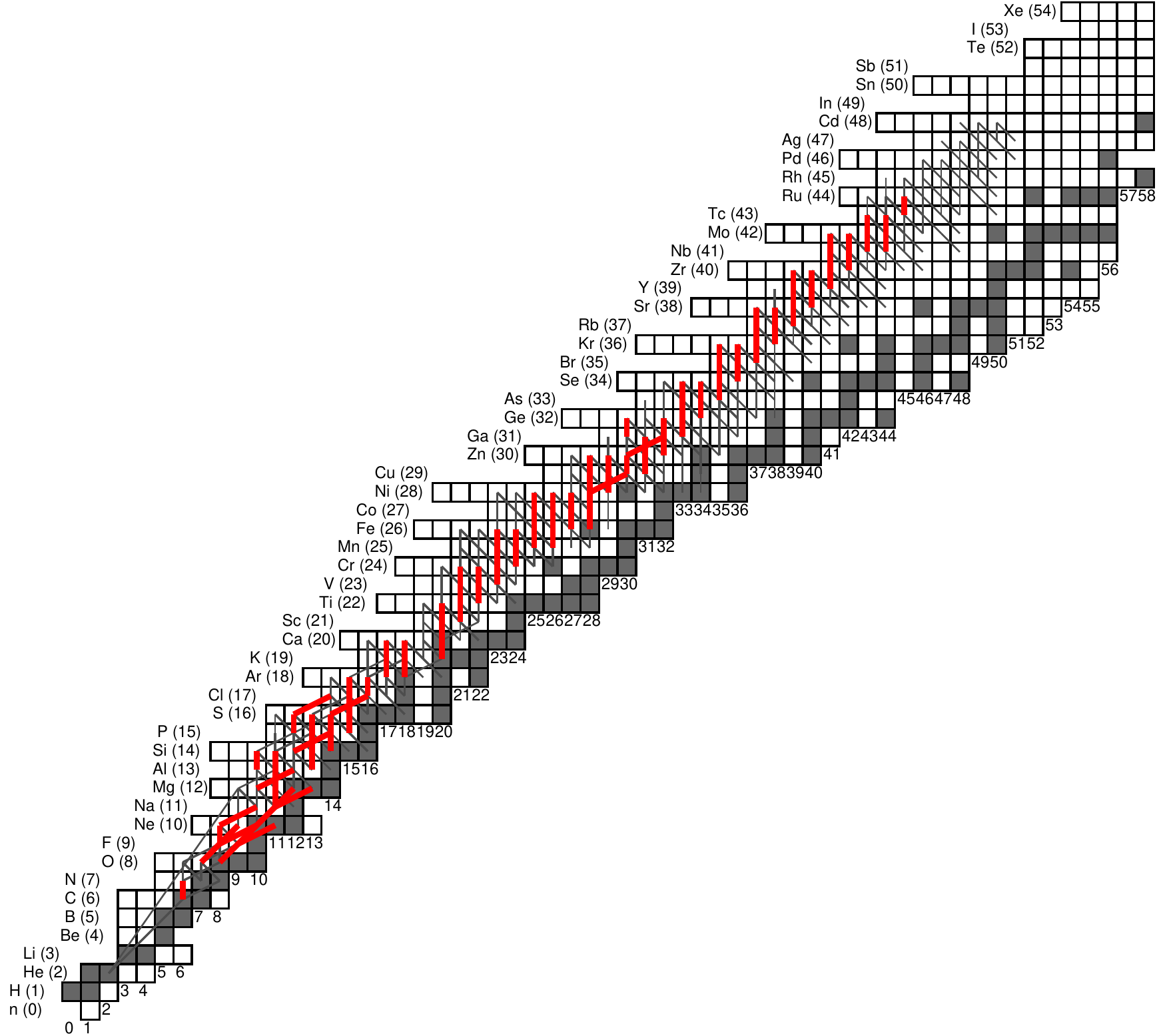}
\caption{\label{FigSZ_path}Time integrated net reaction flows in the single-zone model (black). Thick red lines indicate reactions that significantly affect the light curve (upper panel) and composition of the burst ashes (lower panel). Note that proton capture flows carry large uncertainties because they are determined from the difference between proton capture and the inverse photodisintegration flows, which both can be very large.}
\end{figure*}

\begin{figure*}
\plottwolarge{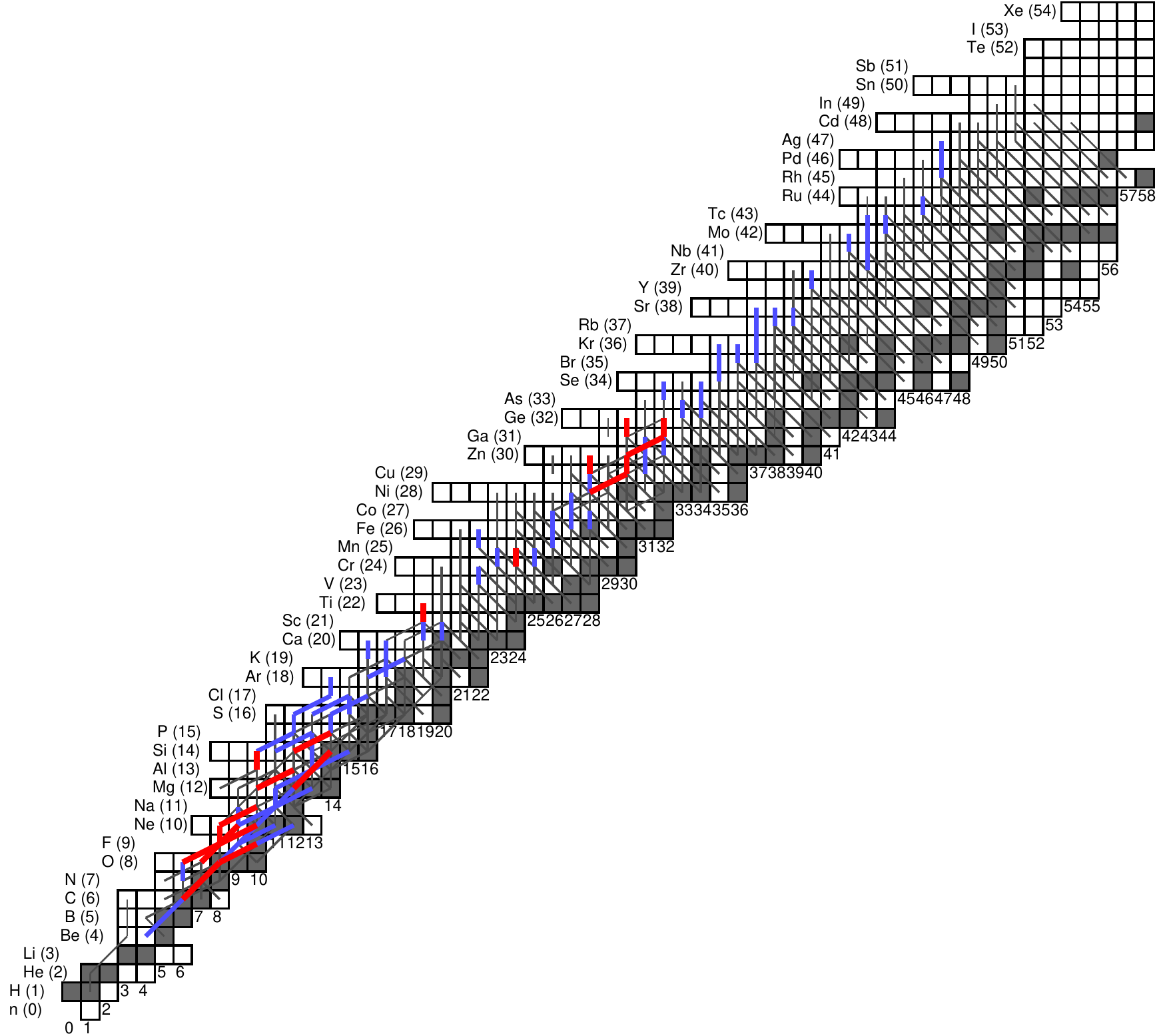}{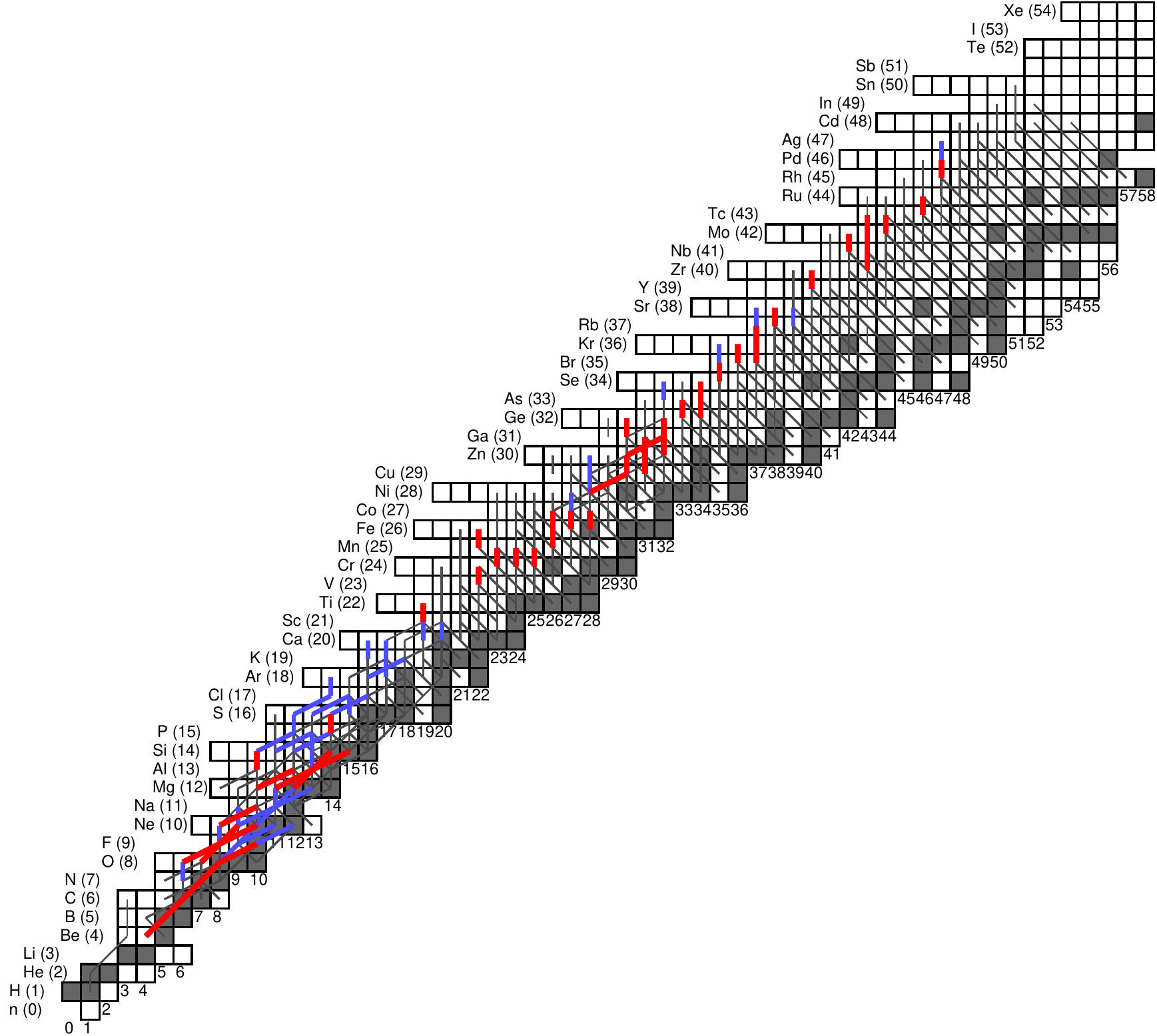}
\caption{\label{FigMZ_path}Time integrated net reaction flows in the multi-zone model (black). Flows are a weighted average over all zones. Thick red lines indicate multi-zone model reaction variations that significantly affect the light curve (upper panel) and composition of the burst ashes (lower panel). The thick blue lines indicate reaction rates that were varied in the multi-zone mode but did not lead to significant changes in the respective observable. Note that proton capture flows carry large uncertainties because they are determined from the difference between proton capture and the inverse photodisintegration flows, which both can be very large.}
\end{figure*}

A possible explanation is the effect of these reactions on the
operation of the hot CNO cycle prior to burst ignition (see also the
discussion in \citealt{Fisker2006}).  Energy generation by the hot CNO
cycle is given by $5.8 \times 10^{15} Z_{\rm CNO}$~ ergs g$^{-1}$
s$^{-1}$ with $Z_{\rm CNO}$ being the mass fraction of CNO nuclei
\citep{Strohmayer2006}. There exists a positive feedback loop between
the hot CNO cycle and the 3$\alpha$ reaction. As the hot CNO cycle
increases the $^4$He abundance, the 3$\alpha$ reaction rate, which
depends strongly on the $^4$He abundance, will increase. This will
increase the amount of CNO nuclei $Z$ in the hot CNO cycle further
increasing $^4$He production. As the energy generation in the hot CNO
cycle increases with $Z$ this may significantly affect burst
ignition. The $^{15}$O($\alpha$,$\gamma$)$^{19}$Ne reaction may then
act as a ``valve'' removing material from the CNO cycle, damping the
$^4$He abundance feedback loop, slowing down energy generation, and
delaying burst ignition. Thus, a smaller
$^{15}$O($\alpha$,$\gamma$)$^{19}$Ne reaction increases energy
generation between bursts via the CNO cycle, decreases recurrence
times, and for a given accretion rate pushes models closer to the
boundary to stable burning. This interpretation is supported by the
trends in peak luminosity and burst duration. Fig.~\ref{FigMZ_o15ag}
shows recurrence time, peak luminosity, and burst timescale $\tau$ for
additional multi-zone model variations of the
$^{15}$O($\alpha$,$\gamma$)$^{19}$Ne reaction rate. A smaller
$^{15}$O($\alpha$,$\gamma$)$^{19}$Ne reaction rate leads to shorter
recurrence times, higher peak brightness, and shorter bursts, just as
one would expect for a larger $^4$He/H ratio at burst ignition that
would be the result of a stronger CNO cycle in-between bursts. We also
confirm previous results \citep{Fisker2006} that a lower
$^{15}$O($\alpha$,$\gamma$)$^{19}$Ne reaction rate leads to an
increase in $^{12}$C production (Fig.~\ref{FigMZ_o15ag}). It has been
shown previously that stable burning of hydrogen and helium can lead
to the production of large amounts of $^{12}$C
\citep{Schatz2003NPA,Stevens14}. Increased $^{12}$C production would therefore
be expected for an increase in stable burning between bursts caused by
a reduced $^{15}$O($\alpha$,$\gamma$)$^{19}$Ne reaction rate.

The $^{14}$O($\alpha$,p)$^{17}$F reaction may further support this
mechanism by opening a pathway to bypass the $^{14}$O $\beta^+$ decay
in the hot CNO cycle via the
$^{14}$O($\alpha$,p)$^{17}$F(p,$\gamma$)$^{18}$Ne($\beta^+$)$^{18}$F(p,$\alpha$)$^{15}$O
reaction sequence.  This will increase the $^{15}$O abundance relative
to the $^{14}$O abundance in the hot CNO cycle. An increase of the
$^{14}$O($\alpha$,p)$^{17}$F reaction rate will therefore also lead to
more efficient breakout of the CNO cycle via
$^{15}$O($\alpha$,$\gamma$)$^{19}$Ne.

This effect can only occur if conditions for breakout via
$^{15}$O($\alpha$,$\gamma$) are reached prior to burst ignition for a
long enough time for the hot CNO cycle to operate so that breakout
occurs in parallel to the hot CNO cycle. If on the other hand, the
time between the onset of the $^{15}$O($\alpha$,$\gamma$)$^{19}$Ne
reaction and ignition is short, one would expect the opposite
behavior, with a larger $^{15}$O($\alpha$,$\gamma$)$^{19}$Ne reaction
leading to ignition sooner. This may explain why different burst
models show different sensitivities to
$^{15}$O($\alpha$,$\gamma$)$^{19}$Ne reaction rate variations (see
below).

\subsection{Comparison with Other Sensitivity Studies}

Our results can be compared with previous studies of the sensitivities
of X-ray burst models to reaction rate variations. We emphasize that
we do not expect the results to be the same for different models (see
discussion below). Nevertheless, such a comparison is useful as
experimentalists or nuclear theorists may prefer to focus on reaction
rates that serve multiple model needs.

Only very few and extremely limited fully self-consistent one-zone or
1D multi-zone model based sensitivity studies, that take into account
the changes in astrophysical conditions induced by rate changes, have
been carried out so far, however. \citet{Thielemann2001} varied proton
capture rates on \iso{Si}{27}, \iso{S}{31}, \iso{Ar}{35}, and
\iso{Ca}{38} and found a strong influence on the burst light curve in
their AGILE based 1D multi-zone model. We find that these reactions do
not play a significant role in either our single-zone or multi-zone
burst model. This agrees with post-processing studies
\citep{Iliadis1999} though the post-processing approach cannot be used
to draw firm conclusions (see Section 1). The same multi-zone X-ray burst model has
been used by \citet{Fisker2004a} to investigate the impact of
variations in the $^{30}$S($\alpha$,p)$^{33}$Cl and
$^{34}$Ar($\alpha$,p)$^{37}$K reaction rates on a double peak
structure in their burst light curve. We also find small changes
in the burst light curve when these reactions are varied, though the details of the burst shape
variations are different. In particular our chosen burst model does not exhibit  the double peak 
structure that \citet{Fisker2004a} obtain. 

The impact of rate variations of the
$^{15}$O($\alpha$,$\gamma$)$^{19}$Ne reaction has been investigated in
a number of multi-zone models \citep{Fisker2006,Fisker2007a,Davids11,Keek2014}.
\citet{Fisker2007a} used the AGILE model and carried out detailed
calculations for 8 accretion rates with 9 variations of the
$^{15}$O($\alpha$,$\gamma$)$^{19}$Ne rate each, ranging from a
reduction by a factor of 1000 to an increase by a factor of 10 of
their recommended rate. They found that for very low
$^{15}$O($\alpha$,$\gamma$)$^{19}$Ne reaction rates burst rise times
become longer and longer and eventually burst activity is replaced by
an oscillatory burning behavior that resembles a stable burning
regime. For an accretion rate and recurrence times that are similar to
our study ($10\,\%$ of the Eddington accretion rate and 200~min,
respectively) they find that this transition occurs for a rate
reduction of a factor of 30 or more.  To explore this effect we
carried out additional variations of the
$^{15}$O($\alpha$,$\gamma$)$^{19}$Ne rate with our multi-zone model
spanning a reaction rate change of /100 to x100. Even for our lowest
$^{15}$O($\alpha$,$\gamma$)$^{19}$Ne rate we do not find the
oscillatory behavior found by \citet{Fisker2006}.  We do find a
decrease in recurrence time (see Fig.~\ref{FigMZ_o15ag}) with
decreasing rate that confirms the effect of the
$^{15}$O($\alpha$,$\gamma$)$^{19}$Ne rate on the hot CNO cycle that
\citet{Fisker2006} propose to explain their transition to stable
burning, however.  Therefore, whereas we confirm that a lower
$^{15}$O($\alpha$,$\gamma$)$^{19}$Ne rate pushes the model towards
larger energy generation between bursts and towards stable burning,
the effect in our model is not strong enough to actually trigger the
transition to stable burning, in contrast to the model of
\citet{Fisker2006}.

 \citet{Davids11} used the multi-zone X-ray burst code SHIVA to carry
 out calculations with the same $^{15}$O($\alpha$,$\gamma$)$^{19}$Ne
 rate variation used in this work. They find the opposite behavior,
 with a decreasing $^{15}$O($\alpha$,$\gamma$)$^{19}$Ne rate leading
 to longer recurrence times. This may indicate that in their model
 $^{15}$O($\alpha$,$\gamma$)$^{19}$Ne does not play a significant role
 during the hot CNO cycle and rather contributes on short timescales
 to burst ignition with an additional energy boost.  The different
 behavior is not necessarily surprising as \citet{Davids11} explore a
 somewhat different burst regime with longer recurrence times of around
 300~min (compared to 175 min and 200 min in this work and
 \citet{Fisker2006}, respectively). In addition, they follow only four
  bursts and the results may be affected by
 additional uncertainties from intrinsic burst-to-burst variations.

The only previous large-scale sensitivity study for X-ray bursts was
carried out in the post-processing approximation by
\citet{Parikh2008}. In this approach, a fixed temperature and density
profile from a single-zone X-ray burst model is used to calculate
energy generation and final composition using a nucleosynthesis
network code. As temperature and density are fixed, changes in
temperature and density induced by reaction rate changes are
neglected. This is a major limitation for X-ray burst studies, where
the same reactions that create the burst ashes are the sole source of
energy and entirely determine temperature and density evolution. In
general, therefore changes in reaction rates in X-ray burst models
will lead to changes in temperature and density and associated changes
in reaction pathways. Nevertheless post-processing studies are fast
and efficient, and can be used to identify qualitatively likely
candidates for important reactions by identifying changes in energy
generation and composition.  It is not generally possible to make
reliable predictions of the quantitative impact of a rate change,
however.  Similarly to our study, \citet{Parikh2008} find that a large
number of reactions tend to impact the burst composition. Detailed
comparisons are not useful, as \citet{Parikh2008} post processed
models with very different temperature and density profiles. They also use a different importance criterion
(rates must affect more than three mass chains) than our
study. \citet{Parikh2008} do identify a set of 17 reactions that lead
to significant changes in energy generation and composition. Indeed 8
of the 17 reactions have an impact on the burst light curve in either
our single-zone or our multi-zone study. These are
\iso{Ni}{56}($\alpha$,p)\iso{Cu}{59},
\iso{Cu}{59}(p,$\gamma$)\iso{Zn}{60},
\iso{O}{15}($\alpha$,$\gamma$)\iso{Ne}{19},
\iso{Al}{23}(p,$\gamma$)\iso{Si}{24},
\iso{S}{30}($\alpha$,p)\iso{Cl}{33},
\iso{Mg}{22}($\alpha$,p)\iso{Al}{25},
\iso{Cl}{31}(p,$\gamma$)\iso{Ar}{32}, and
\iso{Ne}{18}($\alpha$,p)\iso{Na}{21}.  Many rates that we find
strongly impact our light curves, however, are not listed in
\citet{Parikh2008}. Again, this is not surprising as different models
are used. For example, we find a strong impact of the
$^{14}$O($\alpha$,p)$^{17}$F reaction in our multi-zone model, which
is not expected to appear prominently in a single-zone post processing
study as it mainly impacts the pre-burst phase and ignition
conditions (see \citet{Hu2014} for a post-processing analysis).
An example for a rate that we find to be unimportant
even though it appears prominently in \citet{Parikh2008} is
$^{103}$Sn($\alpha$,p). The burst model S01 used for post-processing
by \citet{Parikh2008} has different astrophysical conditions that lead
to a more extended main \textsl{rp}-process path that reaches
$^{103}$Sn, whereas in our model the main \textsl{rp}-process ends at
lower masses and only a very weak reaction flow reaches the $A=103$
mass region.

This comparison reinforces an important point. Sensitivity studies by
design are specific to a particular model and identify specifically
the important nuclear physics in that model. This is of particular
importance for X-ray bursts, where a broad range of astrophysical
parameters occurs in nature and leads to variations in bursts and
nuclear processes.  Various models reported in the literature
typically choose different astrophysical parameters, that may all be
reasonable and reflect different types of systems and sources.
Differences among sensitivity studies are therefore expected and
reflect the broad needs of nuclear physics data. This is also true
for models that use the same input parameters but different
astrophysical approximations, for example descriptions for convection
or diffusion. The goal of a sensitivity study is to enable the nuclear
physics uncertainties of a particular model to be addressed, so that
this particular model can be validated against observations. With
reliable nuclear physics this validation process can then be used to
adjust astrophysical parameters (to effectively extract these from
observations) or to guide improvements of the astrophysical
approximations.

\section{Conclusions}

We present results from the first large-scale investigation of the
influence of nuclear reaction rate uncertainties on X-ray burst light
curves and ashes that uses a self-consistent multi-zone 1D X-ray burst model. 
This approach 
fully accounts for the influence of reaction rate changes on
temperature and density conditions, radiation transport, and
compositional inertia along a sequence of bursts. The results serve 
as a road map for nuclear experimental and theoretical work 
towards reducing nuclear uncertainties in X-ray burst models 
that describe bursts in the common mixed hydrogen and helium 
burning regime. 

The most important reaction rate uncertainties
identified here are the ones that affect the burst light curve (listed
in Tab.~\ref{tbl:MZ_LCrank}), the major components of the ashes,
$^{12}$C production, and the production of odd mass isotopes
(Tab.~\ref{tbl:MZ_ABrank}). We note that the light curve variations 
we find from varying a number of single reaction rates with reasonable multiplication factors 
are comparable or larger than the effects of different surface gravities \cite{Zamfir2012}. 
Clearly reaction rate uncertainties have to be addressed before 
burst light curve tails can be used to constrain surface gravity and therefore 
neutron star compactness. We also identified a number of reactions
that are not important (Fig.~\ref{FigMZ_path}). 

We also used a one-zone X-ray burst model as closely matched to {\Kepler} ZM
as possible to maximize the number of identified critical reactions
with a limited number of multi-zone calculations. For this one-zone
model we identified the complete set of important reaction rate
uncertainties (Tab.~\ref{tbl:SZ_LCrank}, Tab.~\ref{tbl:SZ_ABrank} and
Fig.~\ref{FigSZ_path}). This is the first complete identification of
important reaction rates for mixed hydrogen and helium 
burning bursts in a fully self-consistent one-zone burst
model.

 We
emphasize that our goal was not to calculate realistic uncertainties
in a statistical sense, but to flag important reaction rates, and
provide one data point on the dependence of observables on this
reaction rate. Modifications to the light curve and composition for different variation factors may then be roughly estimated using our results. 
For a precise analysis calculations would have to be repeated for a particular reaction rate uncertainty, 
and uncertainty correlations with temperature also had to be included. 

More work is needed to quantify the uncertainties of the important reaction 
rates identified in this work, and to develop targeted approaches 
to reduce these uncertainties using experiments or nuclear theory. In addition,
more sensitivity studies along the lines of this work are needed to 
arrive at a complete picture of the nuclear physics needs for 
X-ray burst models. This includes variations of additional 
reaction rates in the multi-zone burst model investigated here, 
as the selection based on the single-zone model sensitivity may 
be incomplete. In addition, sensitivity studies of burst in other 
burning regimes should be performed. 

\section{Acknowledgments}

We thank R. Ferguson, M.\ Klein, S.\ Warren for help with the data
analysis, F.-K.\ Thielemann for providing the network solver, and
L.\ Bildsten for contributions to the one-zone model.This material is
based upon work supported by the National Science Foundation under
Grant Numbers PHY-02-016783, PHY-08-22648, and PHY-1430152 (JINA
Center for the Evolution of the Elements).  AH was supported by an ARC
Future Fellowship (FT120100363) and the US Department of Energy
(SC0005012). LK is supported by NASA under award number NNG06EO90A.

\bibliographystyle{apj}
\bibliography{hsref_v3}

\vspace{5cm}
\hspace{1cm}

\clearpage
\LongTables
\begin{deluxetable*}{clclcl}
\tablecaption{\label{tbl:MZ_Variation} Rate variations in the
  multi-zone model calculations.}  
\tablewidth{0pt} 
\tablehead{  \colhead{Reaction} & \colhead{Variation\tablenotemark{b}} & 
\colhead{Reaction} & \colhead{Variation\tablenotemark{b}} & 
\colhead{Reaction} & \colhead{Variation\tablenotemark{b}}
 }
\startdata
3$\alpha$ & x1.2 [2] &
\iso{C}{12}($\alpha$,$\gamma$)\iso{O}{16} & 2.0 [3]&
\iso{C}{12}(p,$\gamma$)\iso{N}{13} & 1.1 [1]\\
\iso{N}{13}(p,$\gamma$)\iso{O}{14} & 10 &
\iso{O}{14}($\alpha$,p)\iso{F}{17} & 10 &
\iso{O}{15}($\alpha$,$\gamma$)\iso{Ne}{19} & 10\\
\iso{O}{16}($\alpha$,$\gamma$)\iso{Ne}{20} & 1.80 [1]&
\iso{O}{16}($\alpha$,p)\iso{F}{19} & 10 &
\iso{F}{17}($\alpha$,p)\iso{Ne}{20} & 10\\
\iso{F}{17}(p,$\gamma$)\iso{Ne}{18} & 6.33 \tablenotemark{d}  &
\iso{F}{18}($\alpha$,p)\iso{Ne}{21} & 100 &
\iso{F}{19}($\alpha$,p)\iso{Ne}{22} & 10\\
\iso{Ne}{18}($\alpha$,p)\iso{Na}{21} & 30 [4]&
\iso{Ne}{19}($\alpha$,p)\iso{Na}{22} & 10 &
\iso{Ne}{19}(p,$\gamma$)\iso{Na}{20} & 100\\
\iso{Ne}{20}($\alpha$,$\gamma$)\iso{Mg}{24} & 1.40 [1]&
\iso{Na}{22}($\alpha$,p)\iso{Mg}{25} & 10 &
\iso{Na}{22}(p,$\gamma$)\iso{Mg}{23} & 2 [1]\\
\iso{Mg}{22}($\alpha$,p)\iso{Al}{25} & 10 &
\iso{Mg}{23}($\alpha$,p)\iso{Al}{26} & 10 &
\iso{Mg}{24}($\alpha$,$\gamma$)\iso{Si}{28} & 10\\
\iso{Al}{23}(p,$\gamma$)\iso{Si}{24} & 30-100\tablenotemark{a} &
\iso{Al}{26}($\alpha$,p)\iso{Si}{29} & 10 &
\iso{Al}{26}(p,$\gamma$)\iso{Si}{27} & 2\tablenotemark{d}\\
\iso{Si}{24}($\alpha$,p)\iso{P}{27} & 10 &
\iso{Si}{25}($\alpha$,p)\iso{P}{28} & 10 &
\iso{Si}{26}($\alpha$,p)\iso{P}{29} & 10\\
\iso{Si}{27}(p,$\gamma$)\iso{P}{28} & 3\tablenotemark{d} &
\iso{P}{27}(p,$\gamma$)\iso{S}{28} & 2-3\tablenotemark{a} &
\iso{P}{29}(p,$\gamma$)\iso{S}{30} & 10\\
\iso{S}{28}($\alpha$,p)\iso{Cl}{31} & 10 &
\iso{S}{29}($\alpha$,p)\iso{Cl}{32} & 10 &
\iso{S}{30}($\alpha$,p)\iso{Cl}{33} & 10\\
\iso{S}{31}(p,$\gamma$)\iso{Cl}{32} & 6\tablenotemark{d} &
\iso{Cl}{31}(p,$\gamma$)\iso{Ar}{32} & 2-3\tablenotemark{a} &
\iso{Ar}{34}($\alpha$,p)\iso{K}{37} & 10\\
\iso{Ar}{35}(p,$\gamma$)\iso{K}{36} & 100 &
\iso{K}{35}(p,$\gamma$)\iso{Ca}{36} & 3-10\tablenotemark{a} &
\iso{K}{36}(p,$\gamma$)\iso{Ca}{37} & 10\\
\iso{Ca}{39}(p,$\gamma$)\iso{Sc}{40} & 3\tablenotemark{d} &
\iso{Ca}{40}(p,$\gamma$)\iso{Sc}{41} & 1.40\tablenotemark{d}  &
\iso{Sc}{40}(p,$\gamma$)\iso{Ti}{41} & 100\\
\iso{V}{45}(p,$\gamma$)\iso{Cr}{46} & 100 &
\iso{Cr}{47}(p,$\gamma$)\iso{Mn}{48} & 10 &
\iso{Cr}{48}(p,$\gamma$)\iso{Mn}{49} & 100\\
\iso{Cr}{49}(p,$\gamma$)\iso{Mn}{50} & 10 &
\iso{Mn}{47}(p,$\gamma$)\iso{Fe}{48} & 100 &
\iso{Mn}{51}(p,$\gamma$)\iso{Fe}{52} & 100\\
\iso{Fe}{52}(p,$\gamma$)\iso{Co}{53} & 100 &
\iso{Fe}{53}(p,$\gamma$)\iso{Co}{54} & 100 &
\iso{Fe}{54}(p,$\gamma$)\iso{Co}{55} & 10\\
\iso{Co}{54}(p,$\gamma$)\iso{Ni}{55} & 10 &
\iso{Ni}{56}($\alpha$,p)\iso{Cu}{59} & 100 &
\iso{Ni}{56}(p,$\gamma$)\iso{Cu}{57} & 5\tablenotemark{a}\\
\iso{Cu}{57}(p,$\gamma$)\iso{Zn}{58} & 100\tablenotemark{c} &
\iso{Cu}{59}(p,$\gamma$)\iso{Zn}{60} & 100 &
\iso{Cu}{60}(p,$\gamma$)\iso{Zn}{61} & 10\\
\iso{Zn}{60}($\alpha$,p)\iso{Ga}{63} & 100 &
\iso{Zn}{61}(p,$\gamma$)\iso{Ga}{62} & 100 &
\iso{Zn}{62}(p,$\gamma$)\iso{Ga}{63} & 10\\
\iso{Ga}{61}(p,$\gamma$)\iso{Ge}{62} & 100 &
\iso{Ga}{63}(p,$\gamma$)\iso{Ge}{64} & 10 &
\iso{Ge}{61}(p,$\gamma$)\iso{As}{62} & 100\\
\iso{Ge}{65}(p,$\gamma$)\iso{As}{66} & 100 &
\iso{Ge}{66}(p,$\gamma$)\iso{As}{67} & 100 &
\iso{As}{65}(p,$\gamma$)\iso{Se}{66} & 100\\
\iso{As}{67}(p,$\gamma$)\iso{Se}{68} & 100 &
\iso{Se}{69}(p,$\gamma$)\iso{Br}{70} & 100 &
\iso{Br}{70}(p,$\gamma$)\iso{Kr}{71} & 100\\
\iso{Br}{71}(p,$\gamma$)\iso{Kr}{72} & 100 &
\iso{Br}{72}(p,$\gamma$)\iso{Kr}{73} & 10 &
\iso{Kr}{73}(p,$\gamma$)\iso{Rb}{74} & 10\\
\iso{Rb}{74}(p,$\gamma$)\iso{Sr}{75} & 10 &
\iso{Rb}{75}(p,$\gamma$)\iso{Sr}{76} & 100 &
\iso{Rb}{76}(p,$\gamma$)\iso{Sr}{77} & 10\\
\iso{Y}{79}(p,$\gamma$)\iso{Zr}{80} & 10 &
\iso{Zr}{83}(p,$\gamma$)\iso{Nb}{84} & 10 &
\iso{Nb}{83}(p,$\gamma$)\iso{Mo}{84} & 10\\
\iso{Nb}{84}(p,$\gamma$)\iso{Mo}{85} & 10 &
\iso{Mo}{85}(p,$\gamma$)\iso{Tc}{86} & 100 &
\iso{Mo}{86}(p,$\gamma$)\iso{Tc}{87} & 100\\
\iso{Tc}{89}(p,$\gamma$)\iso{Ru}{90} & 10 &
\iso{Rh}{92}(p,$\gamma$)\iso{Pd}{93} & 10 &
\iso{Pd}{93}(p,$\gamma$)\iso{Ag}{94} & 10\\
\tablenotetext{a}{Values approximate. Variation calculated using
  resonance uncertainty (see text) and depends on temperature.} 
   \tablenotetext{b}{Reaction rates were each
  multiplied and divided by the factor given. References refer to work that was used
  in the estimate of the variation factor.}
\tablenotetext{c}{Recent experiment reduced uncertainty significantly
  \citep{Langer2014} } \tablenotetext{d}{Adjusted with data from
  \citet{Iliadis2010} }
\tablerefs{[1] \citet{NACRE1}, [2] \citet{Herwig2006}, [3] \citet{Buchmann2006}, [4] \citet{Matic2009}}
\end{deluxetable*}

\clearpage

\LongTables
\begin{deluxetable*}{cccccc}
\tablecaption{\label{tbl:SZ_ABrank} Reactions that impact the
  composition in the single zone x-ray burst model.}  \tablewidth{0pt}
\tablehead{ \colhead{Count} & \colhead{Reaction} &
  \colhead{Max. Ratio\tablenotemark{a}} & \multicolumn{3}{c}{Affected Mass Numbers with
    Mass Fraction $> 10^{-4}$}\\ \colhead{} & \colhead{} & \colhead{}
  & \colhead{max} & \colhead{$> \times 10$ change} & \colhead{$\times 2 <$ change $< \times 10$} }
\startdata 
  1 & \iso{C}{12}(p,$\gamma$)\iso{N}{13}     &     6 &  16 &  & 16\\
  2 & \iso{O}{15}($\alpha$,$\gamma$)\iso{Ne}{19}$^*$ &     4 &  15 &  & 15\\
  3 & \iso{O}{16}($\alpha$,$\gamma$)\iso{Ne}{20}$^*$ &     7 &  16 &  & 16,20-21\\
  4 & \iso{F}{17}($\alpha$,p)\iso{Ne}{20}$^*$ &     2 &  21 &  & 21\\
  5 & \iso{F}{17}(p,$\gamma$)\iso{Ne}{18}$^*$ &     3 &  19 &  & 18-19,21\\
  6 & \iso{F}{18}($\alpha$,p)\iso{Ne}{21}$^*$ &     2 &  23 &  & 18,21,23\\
  7 & \iso{Ne}{18}($\alpha$,p)\iso{Na}{21}$^*$ &     7 &  19 &  & 18-19,21,24,57\\
  8 & \iso{Ne}{20}($\alpha$,$\gamma$)\iso{Mg}{24}$^*$ &     6 &  20 &  & 20,24\\
  9 & \iso{Na}{22}($\alpha$,p)\iso{Mg}{25}$^*$ &     4 &  27 &  & 16,21,25,27\\
 10 & \iso{Na}{22}(p,$\gamma$)\iso{Mg}{23}$^*$ &     4 &  23 &  & 23\\
 11 & \iso{Mg}{22}($\alpha$,p)\iso{Al}{25}$^*$ &    20 &  22 & 22 & 21,23-27,57\\
 12 & \iso{Mg}{23}(p,$\gamma$)\iso{Al}{24}     &     2 &  23 &  & 23\\
 13 & \iso{Al}{23}(p,$\gamma$)\iso{Si}{24}$^*$ &     4 &  57 &  & 16,19,21,24,26-28,46,57,70,74,81-82\\
 14 & \iso{Al}{24}(p,$\gamma$)\iso{Si}{25}     &     2 &  24 &  & 24\\
 15 & \iso{Al}{26}(p,$\gamma$)\iso{Si}{27}$^*$ &     6 &  27 &  & 27\\
 16 & \iso{Si}{26}($\alpha$,p)\iso{P}{29}$^*$ &     3 &  18 &  & 18-21,46,57,70,73-75,78,82,86\\
 17 & \iso{Si}{27}(p,$\gamma$)\iso{P}{28}$^*$ &     3 &  27 &  & 27\\
 18 & \iso{Si}{28}(p,$\gamma$)\iso{P}{29}     &     2 &  28 &  & 28\\
 19 & \iso{P}{27}(p,$\gamma$)\iso{S}{28}$^*$ &     2 &  26 &  & 26\\
 20 & \iso{P}{28}(p,$\gamma$)\iso{S}{29}     &     2 &  28 &  & 28\\
 21 & \iso{P}{29}(p,$\gamma$)\iso{S}{30}$^*$ &     4 &  29 &  & 29\\
 22 & \iso{P}{30}(p,$\gamma$)\iso{S}{31}     &     3 &  31 &  & 31\\
 23 & \iso{S}{28}($\alpha$,p)\iso{Cl}{31}$^*$ &     2 &  19 &  & 18-19\\
 24 & \iso{S}{30}($\alpha$,p)\iso{Cl}{33}$^*$ &     3 &  18 &  & 18-21,46,57,70-71,73-75,77-78,81-83,86\\
 25 & \iso{S}{31}(p,$\gamma$)\iso{Cl}{32}$^*$ &     6 &  31 &  & 31,33\\
 26 & \iso{Cl}{32}(p,$\gamma$)\iso{Ar}{33}     &     5 &  32 &  & 32-33\\
 27 & \iso{Cl}{33}(p,$\gamma$)\iso{Ar}{34}     &     4 &  33 &  & 33\\
 28 & \iso{Ar}{35}(p,$\gamma$)\iso{K}{36}$^*$ &     7 &  35 &  & 35-37\\
 29 & \iso{Ar}{36}(p,$\gamma$)\iso{K}{37}     &     4 &  36 &  & 36\\
 30 & \iso{K}{36}(p,$\gamma$)\iso{Ca}{37}$^*$ &     5 &  36 &  & 36\\
 31 & \iso{K}{37}(p,$\gamma$)\iso{Ca}{38}     &     6 &  37 &  & 37\\
 32 & \iso{K}{39}(p,$\gamma$)\iso{Ca}{40}     &     3 &  40 &  & 40\\
 33 & \iso{Ca}{40}(p,$\gamma$)\iso{Sc}{41}$^*$ &     6 &  40 &  & 40,42-46\\
 34 & \iso{Sc}{41}(p,$\gamma$)\iso{Ti}{42}     &    10 &  41 & 41 & 42\\
 35 & \iso{Sc}{42}(p,$\gamma$)\iso{Ti}{43}     &     3 &  42 &  & 42-43\\
 36 & \iso{Ti}{43}(p,$\gamma$)\iso{V}{44}     &     8 &  43 &  & 43-45\\
 37 & \iso{Ti}{44}(p,$\gamma$)\iso{V}{45}     &     2 &  44 &  & 44\\
 38 & \iso{V}{44}(p,$\gamma$)\iso{Cr}{45}     &     7 &  44 &  & 44-45\\
 39 & \iso{V}{45}(p,$\gamma$)\iso{Cr}{46}$^*$ &     6 &  45 &  & 45-46\\
 40 & \iso{V}{46}(p,$\gamma$)\iso{Cr}{47}     &     2 &  46 &  & 46\\
 41 & \iso{Cr}{47}(p,$\gamma$)\iso{Mn}{48}$^*$ &     8 &  47 &  & 47\\
 42 & \iso{Cr}{48}(p,$\gamma$)\iso{Mn}{49}$^*$ &     3 &  48 &  & 48\\
 43 & \iso{Mn}{48}(p,$\gamma$)\iso{Fe}{49}     &    10 &  48 & 48 & \\
 44 & \iso{Mn}{49}(p,$\gamma$)\iso{Fe}{50}     &     7 &  49 &  & 49\\
 45 & \iso{Mn}{50}(p,$\gamma$)\iso{Fe}{51}     &     3 &  50 &  & 50-51\\
 46 & \iso{Mn}{51}(p,$\gamma$)\iso{Fe}{52}$^*$ &     2 &  51 &  & 51\\
 47 & \iso{Fe}{51}(p,$\gamma$)\iso{Co}{52}     &     5 &  51 &  & 51\\
 48 & \iso{Fe}{52}(p,$\gamma$)\iso{Co}{53}$^*$ &     7 &  52 &  & 52\\
 49 & \iso{Fe}{53}(p,$\gamma$)\iso{Co}{54}$^*$ &     3 &  53 &  & 53\\
 50 & \iso{Fe}{54}(p,$\gamma$)\iso{Co}{55}$^*$ &     3 &  54 &  & 54\\
 51 & \iso{Co}{52}(p,$\gamma$)\iso{Ni}{53}     &     5 &  52 &  & 52\\
 52 & \iso{Co}{53}(p,$\gamma$)\iso{Ni}{54}     &    10 &  53 & 53 & \\
 53 & \iso{Co}{54}(p,$\gamma$)\iso{Ni}{55}$^*$ &    10 &  54 & 54 & \\
 54 & \iso{Co}{55}(p,$\gamma$)\iso{Ni}{56}$^*$ &    10 &  55 & 55 & 56\\
 55 & \iso{Ni}{56}($\alpha$,p)\iso{Cu}{59}$^*$ &     7 &  24 &  & 12,16,21-22,24-34,36-39,42,44,46,50,52,54-57,61,63,65-67,69-71,73-75,77-78,80-85\\
 56 & \iso{Ni}{56}(p,$\gamma$)\iso{Cu}{57}$^*$ &    10 &  56 & 56 & 18-19,21,57-59\\
 57 & \iso{Ni}{57}(p,$\gamma$)\iso{Cu}{58}     &     5 &  57 &  & 57\\
 58 & \iso{Cu}{57}(p,$\gamma$)\iso{Zn}{58}$^*$ &     6 &  57 &  & 56-58\\
 59 & \iso{Cu}{58}(p,$\gamma$)\iso{Zn}{59}     &    30 &  58 & 58 & \\
 60 & \iso{Cu}{59}(p,$\gamma$)\iso{Zn}{60}$^*$ &    10 &  59 & 59 & 12,16,21-22,24-34,36-39,42,44,46,50,52,54-57,61,65-67,69-71,73-75,77-78,80-85\\
 61 & \iso{Cu}{60}(p,$\gamma$)\iso{Zn}{61}$^*$ &     3 &  61 &  & 61,63\\
 62 & \iso{Zn}{60}($\alpha$,p)\iso{Ga}{63}$^*$ &     2 &  57 &  & 21,24,27,57\\
 63 & \iso{Zn}{61}(p,$\gamma$)\iso{Ga}{62}$^*$ &    10 &  61 & 61 & 63\\
 64 & \iso{Zn}{62}(p,$\gamma$)\iso{Ga}{63}$^*$ &     5 &  62 &  & 62\\
 65 & \iso{Ga}{61}(p,$\gamma$)\iso{Ge}{62}$^*$ &     8 &  60 &  & 12,18-19,60-68\\
 66 & \iso{Ga}{63}(p,$\gamma$)\iso{Ge}{64}$^*$ &    10 &  63 & 63 & 21,24,27,57\\
 67 & \iso{Ga}{64}(p,$\gamma$)\iso{Ge}{65}     &     6 &  65 &  & 65-67\\
 68 & \iso{Ge}{65}(p,$\gamma$)\iso{As}{66}$^*$ &    30 &  65 & 65 & 66-67\\
 69 & \iso{Ge}{66}(p,$\gamma$)\iso{As}{67}$^*$ &    20 &  66 & 66 & 67\\
 70 & \iso{As}{66}(p,$\gamma$)\iso{Se}{67}     &     5 &  66 &  & 66\\
 71 & \iso{As}{67}(p,$\gamma$)\iso{Se}{68}$^*$ &    30 &  67 & 67 & \\
 72 & \iso{As}{68}(p,$\gamma$)\iso{Se}{69}     &     5 &  69 &  & 69-71\\
 73 & \iso{Se}{69}(p,$\gamma$)\iso{Br}{70}$^*$ &    20 &  69 & 69 & 70-71\\
 74 & \iso{Se}{70}(p,$\gamma$)\iso{Br}{71}     &    10 &  70 & 70 & 71\\
 75 & \iso{Br}{70}(p,$\gamma$)\iso{Kr}{71}$^*$ &     3 &  70 &  & 70\\
 76 & \iso{Br}{71}(p,$\gamma$)\iso{Kr}{72}$^*$ &    10 &  71 & 71 & \\
 77 & \iso{Br}{72}(p,$\gamma$)\iso{Kr}{73}$^*$ &     5 &  73 &  & 73-75\\
 78 & \iso{Kr}{73}(p,$\gamma$)\iso{Rb}{74}$^*$ &    20 &  73 & 73 & 74-78\\
 79 & \iso{Kr}{74}(p,$\gamma$)\iso{Rb}{75}     &    10 &  74 & 74 & 75\\
 80 & \iso{Rb}{74}(p,$\gamma$)\iso{Sr}{75}$^*$ &     3 &  74 &  & 74\\
 81 & \iso{Rb}{75}(p,$\gamma$)\iso{Sr}{76}$^*$ &     9 &  75 &  & 75-76\\
 82 & \iso{Rb}{76}(p,$\gamma$)\iso{Sr}{77}$^*$ &     5 &  77 &  & 77-78\\
 83 & \iso{Sr}{77}(p,$\gamma$)\iso{Y}{78}     &    10 &  77 & 77 & \\
 84 & \iso{Sr}{78}(p,$\gamma$)\iso{Y}{79}     &     6 &  78 &  & 78,80-85\\
 85 & \iso{Y}{78}(p,$\gamma$)\iso{Zr}{79}     &    10 &  80 & 80 & 78-79,81-92\\
 86 & \iso{Y}{79}(p,$\gamma$)\iso{Zr}{80}$^*$ &     5 &  80 &  & 80\\
 87 & \iso{Y}{80}(p,$\gamma$)\iso{Zr}{81}     &     5 &  81 &  & 81-85\\
 88 & \iso{Zr}{81}(p,$\gamma$)\iso{Nb}{82}     &     5 &  81 &  & 81\\
 89 & \iso{Zr}{82}(p,$\gamma$)\iso{Nb}{83}     &     3 &  82 &  & 82-85\\
 90 & \iso{Nb}{82}(p,$\gamma$)\iso{Mo}{83}     &     2 &  82 &  & 82\\
 91 & \iso{Nb}{83}(p,$\gamma$)\iso{Mo}{84}$^*$ &     3 &  84 &  & 83-84\\
 92 & \iso{Nb}{84}(p,$\gamma$)\iso{Mo}{85}$^*$ &     3 &  85 &  & 84-85\\
 93 & \iso{Nb}{85}(p,$\gamma$)\iso{Mo}{86}     &     2 &  85 &  & 85\\
 94 & \iso{Mo}{85}(p,$\gamma$)\iso{Tc}{86}$^*$ &     3 &  85 &  & 85\\
 95 & \iso{Mo}{86}(p,$\gamma$)\iso{Tc}{87}$^*$ &     3 &  86 &  & 86\\
 96 & \iso{Tc}{88}(p,$\gamma$)\iso{Ru}{89}     &     4 &  88 &  & 88\\
\enddata
\tablenotetext{*}{Reaction also varied in multi-zone model.} 
\tablenotetext{a}{Abundance ratio relative to the baseline calculation}
\end{deluxetable*}

\clearpage

\LongTables
\begin{deluxetable*}{cccccc}
\tablecaption{\label{tbl:MZ_ABrank} Reactions that impact the composition in the multi-zone x-ray burst model.}
\tablewidth{0pt}
\tablehead{
\colhead{Count} & \colhead{Reaction} & \colhead{Max. Ratio\tablenotemark{a}} & \multicolumn{3}{c}{Affected Mass Numbers with Mass Fraction $> 10^{-4}$}\\
\colhead{} & \colhead{} & \colhead{} & \colhead{max} &  \colhead{$> \times 10$ change} & \colhead{$\times 2 <$ change $< \times 10$}
}
\startdata
  1 & \iso{Be}{ 8}($\alpha$,$\gamma$)\iso{C}{12} &     3 &  98 &  & 30,93-99\\
  2 & \iso{C}{12}($\alpha$,$\gamma$)\iso{O}{16} &     2 &  28 &  & 28\\
  3 & \iso{O}{14}($\alpha$,p)\iso{F}{17} &     4 &  29 &  & 29\\
  4 & \iso{O}{15}($\alpha$,$\gamma$)\iso{Ne}{19} &    30 &  29 & 24,29 & 12,20,26-28,30-32,36,91-92,95-98\\
  5 & \iso{O}{16}($\alpha$,p)\iso{F}{19} &     3 &  28 &  & 28\\
  6 & \iso{F}{17}($\alpha$,p)\iso{Ne}{20} &     3 &  98 &  & 90,93-99\\
  7 & \iso{Ne}{18}($\alpha$,p)\iso{Na}{21} &     4 &  30 &  & 28,30,98-99\\
  8 & \iso{Mg}{22}($\alpha$,p)\iso{Al}{25} &     4 &  30 &  & 28-30\\
  9 & \iso{Mg}{23}($\alpha$,p)\iso{Al}{26} &     2 &  28 &  & 28\\
 10 & \iso{Mg}{24}($\alpha$,$\gamma$)\iso{Si}{28} &    20 &  24 & 24 & 28-30\\
 11 & \iso{Al}{23}(p,$\gamma$)\iso{Si}{24} &     2 &  75 &  & 75,79\\
 12 & \iso{Al}{26}($\alpha$,p)\iso{Si}{29} &     4 &  26 &  & 26,29\\
 13 & \iso{P}{29}(p,$\gamma$)\iso{S}{30} &     3 &  98 &  & 90,92-99\\
 14 & \iso{Sc}{40}(p,$\gamma$)\iso{Ti}{41} &     3 &  29 &  & 29\\
 15 & \iso{V}{45}(p,$\gamma$)\iso{Cr}{46} &    10 &  45 & 45 & 46\\
 16 & \iso{Cr}{47}(p,$\gamma$)\iso{Mn}{48} &     2 &  47 &  & 47\\
 17 & \iso{Cr}{48}(p,$\gamma$)\iso{Mn}{49} &    30 &  48 & 48 & 51-53\\
 18 & \iso{Cr}{49}(p,$\gamma$)\iso{Mn}{50} &     5 &  49 &  & 49\\
 19 & \iso{Mn}{47}(p,$\gamma$)\iso{Fe}{48} &     2 &  98 &  & 97-99\\
 20 & \iso{Mn}{51}(p,$\gamma$)\iso{Fe}{52} &    30 &  51 & 51 & 52\\
 21 & \iso{Fe}{52}(p,$\gamma$)\iso{Co}{53} &    40 &  52 & 52 & 53-55,57\\
 22 & \iso{Fe}{53}(p,$\gamma$)\iso{Co}{54} &    40 &  53 & 53 & 57-58\\
 23 & \iso{Fe}{54}(p,$\gamma$)\iso{Co}{55} &     6 &  54 &  & 54\\
 24 & \iso{Ni}{56}($\alpha$,p)\iso{Cu}{59} &     5 &  29 &  & 12,29-30,56,75,78-79,82\\
 25 & \iso{Cu}{59}(p,$\gamma$)\iso{Zn}{60} &   200 &  59 & 59 & 12,29-30,75,78-79\\
 26 & \iso{Cu}{60}(p,$\gamma$)\iso{Zn}{61} &     2 &  98 &  & 93-98\\
 27 & \iso{Zn}{60}($\alpha$,p)\iso{Ga}{63} &     2 &  98 &  & 97-98\\
 28 & \iso{Zn}{61}(p,$\gamma$)\iso{Ga}{62} &     9 &  61 &  & 61-63\\
 29 & \iso{Zn}{62}(p,$\gamma$)\iso{Ga}{63} &     7 &  62 &  & 62,90-99\\
 30 & \iso{Ga}{61}(p,$\gamma$)\iso{Ge}{62} &    20 &  60 & 60-61 & 12,30,62-81\\
 31 & \iso{Ga}{63}(p,$\gamma$)\iso{Ge}{64} &     3 &  63 &  & 63\\
 32 & \iso{Ge}{65}(p,$\gamma$)\iso{As}{66} &    10 &  65 & 65 & 67\\
 33 & \iso{Ge}{66}(p,$\gamma$)\iso{As}{67} &    20 &  66 & 66 & 67\\
 34 & \iso{As}{67}(p,$\gamma$)\iso{Se}{68} &   100 &  67 & 67 & 93-95,97-98\\
 35 & \iso{Se}{69}(p,$\gamma$)\iso{Br}{70} &     7 &  69 &  & 69,71-75\\
 36 & \iso{Br}{71}(p,$\gamma$)\iso{Kr}{72} &    90 &  71 & 71 & 72,98\\
 37 & \iso{Br}{72}(p,$\gamma$)\iso{Kr}{73} &     2 &  98 &  & 74,93-94,96-98\\
 38 & \iso{Kr}{73}(p,$\gamma$)\iso{Rb}{74} &     4 &  73 &  & 73,75\\
 39 & \iso{Rb}{75}(p,$\gamma$)\iso{Sr}{76} &    40 &  75 & 75 & \\
 40 & \iso{Y}{79}(p,$\gamma$)\iso{Zr}{80} &     6 &  79 &  & 79\\
 41 & \iso{Zr}{83}(p,$\gamma$)\iso{Nb}{84} &     3 &  83 &  & 83\\
 42 & \iso{Nb}{83}(p,$\gamma$)\iso{Mo}{84} &     2 &  83 &  & 83\\
 43 & \iso{Nb}{84}(p,$\gamma$)\iso{Mo}{85} &     3 &  84 &  & 84\\
 44 & \iso{Mo}{85}(p,$\gamma$)\iso{Tc}{86} &     3 &  85 &  & 85,90-98\\
 45 & \iso{Mo}{86}(p,$\gamma$)\iso{Tc}{87} &     7 &  86 &  & 86,90-92\\
 46 & \iso{Tc}{89}(p,$\gamma$)\iso{Ru}{90} &     3 &  89 &  & 89\\
 47 & \iso{Rh}{92}(p,$\gamma$)\iso{Pd}{93} &     3 &  98 &  & 91,93-98\\
\enddata
\tablenotetext{a}{Abundance ratio between up and down variations of the reaction rate}
\end{deluxetable*}

\end{document}